\documentclass[a4paper,12pt]{iopart}

\usepackage{topcapt}
\usepackage{amssymb}
\usepackage{amsthm}
\usepackage{amsfonts}
\usepackage{url}
\usepackage{breakurl}
\usepackage{rotating}
\usepackage{adjustbox}
\usepackage{longtable}
\usepackage{tikz}
\usepackage{tcolorbox}
\usepackage{algorithm}
\usepackage{algpseudocode}
\usepackage{pdflscape}
\usepackage{microtype}
\usepackage{caption}
\usepackage{hyperref}

\newcommand{\BE}{\begin{equation}}
\newcommand{\EE}{\end{equation}}

\newcommand{\RGf}{{\mathcal R}}

\newcommand{\Sig}{\textstyle{\sum}}
\renewcommand{\Sig}{\raisebox{-0.35ex}{\textrm{\large $\Sigma$}}}
\newcommand{\U}{\raisebox{0.15ex}{\tcbox[on line, arc=2pt,colback=black!10!white,boxrule=0.5pt, boxsep=0.25ex,left=0.5ex,right=0.5ex,top=0.05ex,bottom=0.05ex]{\scriptsize)}}
}
\renewcommand{\L}{\raisebox{0.15ex}{\tcbox[on line, arc=2pt,colback=black!10!white,boxrule=0.5pt, boxsep=0.25ex,left=0.5ex,right=0.5ex,top=0.05ex,bottom=0.05ex]{\scriptsize $($}}}
\newcommand{\E}{\raisebox{0.15ex}{\tcbox[on line, arc=2pt,colback=black!10!white,boxrule=0.5pt, boxsep=0.25ex,left=0.4ex,right=0.4ex,top=0.5ex,bottom=0.45ex]{\scriptsize $\circ$}}}
\newcommand{\F}{\raisebox{0.15ex}{\tcbox[on line, arc=2pt,colback=black!10!white,boxrule=0.5pt, boxsep=0.25ex,left=0.55ex,right=0.55ex,top=0.05ex,bottom=0.05ex]{\scriptsize $|$}}}
\newcommand{\FU}{\raisebox{0.15ex}{\tcbox[on line, arc=2pt,colback=red!25!white,boxrule=0.5pt, boxsep=0.25ex,left=0.5ex,right=0.5ex,top=0.05ex,bottom=0.05ex]{\scriptsize \bf )}}}
\renewcommand{\O}{\raisebox{0.15ex}{\tcbox[on line, arc=2pt,colback=black!10!white,boxrule=0.5pt, boxsep=0.25ex,left=0.4ex,right=0.4ex,top=0.5ex,bottom=0.45ex]{\scriptsize $\star$}}}
\newcommand{\UB}{\raisebox{0.15ex}{\tcbox[on line,arc=2pt,colback=blue!20!white,boxrule=0.5pt, boxsep=0.25ex,left=0.5ex,right=0.5ex,top=0.05ex,bottom=0.05ex]{\scriptsize)}}}
\newcommand{\LB}{\raisebox{0.15ex}{\tcbox[on line, arc=2pt,colback=blue!20!white,boxrule=0.5pt, boxsep=0.25ex,left=0.5ex,right=0.5ex,top=0.05ex,bottom=0.05ex]{\scriptsize $($}}}
\newcommand{\EB}{\raisebox{0.15ex}{\tcbox[on line, arc=2pt,colback=blue!20!white,boxrule=0.5pt, boxsep=0.25ex,left=0.4ex,right=0.4ex,top=0.5ex,bottom=0.45ex]{\scriptsize $\circ$}}}
\newcommand{\FB}{\raisebox{0.15ex}{\tcbox[on line, arc=2pt,colback=blue!20!white,boxrule=0.5pt, boxsep=0.25ex,left=0.55ex,right=0.55ex,top=0.05ex,bottom=0.05ex]{\scriptsize $|$}}}
 
\newcommand{\MZ}{{\mathcal M}^{(0)}}
\newcommand{\MO}{{\mathcal M}^{(1)}}
\newcommand{\C}{{\mathcal C}}

\theoremstyle{plain}
\renewcommand{\thetable}{\arabic{table}}

\newcommand{\seqnum}[1]{OEIS \href{http://oeis.org/#1}{\underline{#1}}}
\newcommand{\orcid}[1]{\address{ORCID ID: \href{http://orcid.org/#1}{#1}}}

\begin{document}

\title[SAWs crossing the square and hexagonal lattices]{Self-avoiding walks and polygons crossing a domain on the square and hexagonal lattices.}

\author{Anthony J Guttmann}
\address{School of Mathematics and Statistics,
The University of Melbourne,
Vic. 3010, Australia}
\ead{guttmann@unimelb.edu.au}
\orcid{0000-0003-2209-7192}

\author{Iwan Jensen}
\address{College of Science and Engineering, Flinders University at Tonsley,
GPO Box 2100, Adelaide, SA 5001, Australia}
\ead{iwan.jensen@gmail.com}
\orcid{0000-0001-6618-8470}

\begin{abstract}
We have analysed the recently extended series for the number of self-avoiding walks (SAWs) $C_L(1)$ that cross an $L \times L$ square between diagonally opposed corners. The number of such walks is known to grow as $\lambda_S^{L^2}.$ We have made more precise the estimate of $\lambda_S,$ based on additional series coefficients provided by several authors, and refined analysis techniques.  We estimate that $\lambda_S = 1.7445498 \pm 0.0000012.$ 
We have also studied the subdominant behaviour, and conjecture that $$ C_L(1) \sim \lambda_S^{L^2+bL+c}\cdot L^g,$$ where $b=-0.04354 \pm 0.0001,$ $c=0.5624 \pm 0.0005,$ and $g=0.000 \pm 0.005.$

We implemented a very efficient algorithm for enumerating paths on the square and hexagonal lattices making use of a minimal perfect hash function and in-place memory updating of the arrays for the counts of the number of paths.

Using this algorithm we extended and then analysed series for SAWs spanning the square lattice and  self-avoiding polygons (SAPs) crossing the square lattice. These are known to also grow as $\lambda_S^{L^2}.$ The sub-dominant term $\lambda^b$ is found to be the same as for SAWs crossing the square, while the exponent  $g = 1.75\pm 0.01$ for spanning SAWs and $g = -0.500 \pm 0.005$ for SAPs.

We have also studied the analogous problems on the hexagonal lattice, and generated series for a number of geometries. In particular, we study SAWs and SAPs crossing rhomboidal, triangular and square domains on the hexagonal lattice, as well as SAWs spanning a rhombus. 
We estimate that the analogous growth constant $\lambda_H=1.38724951 \pm 0.00000005,$ so an even more precise estimate than found for the square lattice.
We also give estimates of the sub-dominant terms.
\end{abstract}




\section{Introduction}
\label{sec:intro}

A {\em $n$-step self-avoiding walk} (SAW) ${\bf \omega}$ on a regular lattice is 
a sequence of {\em distinct} vertices $\omega_0, \omega_1,\ldots , \omega_n$ 
such that each vertex is a nearest neighbour of its predecessor. SAWs are
considered distinct up to translations of the starting point $\omega_0$.
If  $\omega_0$ and  $\omega_n$ are nearest-neighbours we can form
a closed  $(n+1)$-step self-avoiding polygon (SAP) by adding
an edge between the two end-points.

\begin{figure}
\begin{center}
\includegraphics{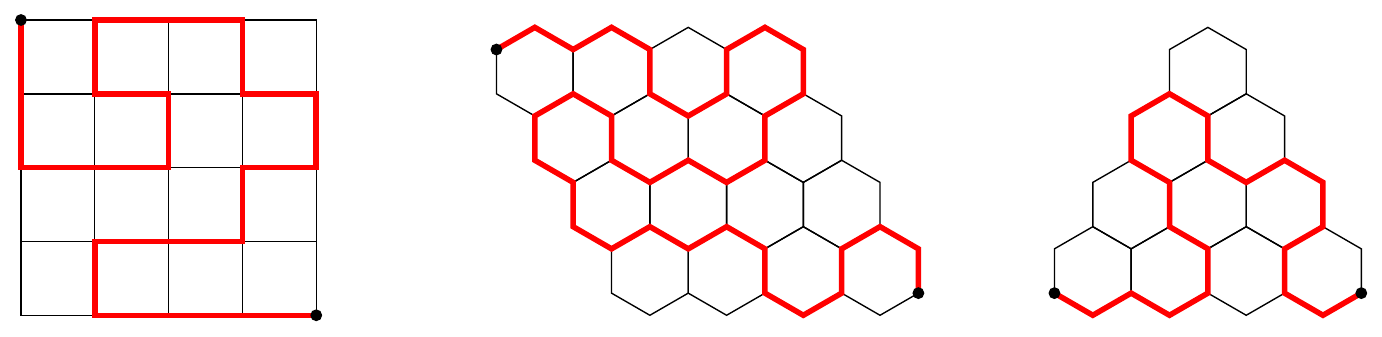}
\end{center}

\caption{\label{fig:domain}  A square domain and rhomboidal and triangular domains of size $n=4$  on the hexagonal lattice. Walks must extend between 
the points indicated by black circles as illustrated by the red walks.}
\end{figure}

We consider SAWs on an $L \times L$ square lattice, with the walks starting at the north-west corner $(0,L)$ and finishing at the south-east corner $(L,0),$ and constrained within the square (see the first diagram in \Fref{fig:domain}).
Clearly such walks vary in length from a minimum of $2L$ to a maximum of $L^2+2L$ (if $L$ is even).
Guttmann and Whittington \cite{GW90} computed the first 7 terms in 1990, then Bousquet-M\'elou, Guttmann and Jensen \cite{BGJ05}
 computed the terms up to $L=19$.  Iwashita  et al. \cite{IKM12}  computed the next two terms, $L=20$ and 21, R. Spaans computed three more terms, $L=22$ to 24, and  Iwashita  \cite{INK13} computed the terms for $L=25$ and 26. Details can be found in the On-line Encyclopaedia of Integer Sequences \cite{OEIS}, \seqnum{A007764}. Note that the listing in the OEIS runs from 1 to 27, which in our notation is $L=0$ to 26. 
 
 Recall that the number of SAWs in the bulk, $c_n$, grows exponentially with length $n$ as $\mu^n,$ where $\mu$ depends on the lattice. For the hexagonal lattice it is known \cite{DCS10} that $\mu = \sqrt{2+\sqrt{2}},$ while for the square lattice the growth constant $\mu$ has only been estimated numerically. The most precise estimate $\mu = 2.63815853032790(3)$ was obtained by Jacobsen, Scullard and Guttmann \cite{JSG16}.
 
We will be interested in the generating function $C_L(x) = \sum_{n \ge 2L} c_n x^n,$ where $c_n$  denotes the number of SAWs of length $n$ crossing the square from $(0,L)$ to $(L,0)$. Madras \cite{M95} proved that the limits $\mu_1(x):=\lim_{L \to \infty} C_L(x)^{1/L}$ and $\mu_2(x):=\lim_{L \to \infty} C_L(x)^{1/L^2}$ are well defined in ${\mathbb R} \cup \{+\infty\}.$
More precisely, Madras proved (i) $\mu_1(x)$ is finite for $0<x<1/\mu,$ and is infinite for $x > 1/\mu.$ Moreover, $0 < \mu_1(x) < 1$ for $0 < x < 1/\mu$ and $\mu_1(1/\mu) =1.$
(ii) $\mu_2(x)$ is finite for all $x>0.$ Moreover, $\mu_2(x)=1$ for $0 <x \le 1/\mu$  and $\mu_2(x) >1$ for $x > 1/\mu.$

The existence of the limit 
\BE \label{eq:CLlim}
\lim_{L \to \infty} C_{L}(1)^{1/L^2}=\lambda_S
\EE
was proved in  both \cite{AH78} and \cite{GW90} by different methods. 
In \cite{BGJ05} we estimated $\lambda_S = 1.744550 \pm 0.000005.$ Using the longer series now available, we have sharpened this to $\lambda_S = 1.7445498 \pm 0.0000012.$
We have also estimated the sub-dominant terms by finding precise numerical evidence for the asymptotic behaviour 
\BE \label{eq:CLas}
C_L(1) \sim \lambda_S^{L^2+bL+c}\cdot L^g,
\EE
 where $b=-0.04354 \pm 0.0001,$ $c=0.5624 \pm 0.0005,$ and $g=0.000 \pm 0.005,$ from which we conjecture that $g=0$, exactly.

For SAPs crossing a square we calculated the coefficients up to $L=26$ and then analysed the data for the first time.  The analysis clearly demonstrated that the two problems have the same growth constant. We conjecture that the subdominant term $\lambda^b$ is the same as for crossing SAWs, and that the corresponding exponent $g=-\frac12.$ For SAWs spanning a square we extended the known series up to $L=26$. This is a superset of SAWs crossing the square, as the origin can be any vertex on the left boundary and the end-point can be any vertex on the right boundary. In \cite{M95} it was proved that the two problems have the same growth constant and this is of course consistent with our analysis. We conjecture that the $\lambda^b$ term is the same as for crossing SAWs and that $g=\frac74$. 

We have also studied the analogous problems on the hexagonal lattice. We initially considered the problem on a square domain of the hexagonal lattice (see the last two diagrams in \Fref{fig:hexprob}), but this was soon found to be a rather unnatural domain, as the paths changed according as the size $L$ of the lattice was odd or even. A more natural domain is a rhombus, shown as the second diagram in \Fref{fig:domain}, or a triangular domain, shown as the third diagram in \Fref{fig:domain}.
We studied both self-avoiding walks (SAWs) and self-avoiding polygons (SAPs) in these three domains. For the triangular domain, we studied two cases, according as the path is forced to include the top vertex of the triangle or not. We also studied SAWs which {\em span} a rhombus of width $L.$

In Section~\ref{sec:algo} we give a detailed description of the new and very efficient algorithm we used to calculate the series for SAWs crossing a rhombus and briefly 
mention how to amend the algorithm to enumerate other problems such as SAPs. 
In Section~\ref{sec:ana} we give a brief description of the methods we used in our analysis of the series. Further details can be found in \ref{app:ratio} and \ref{app:da}.
Section~\ref{sec:wcas} contains a detailed analysis of the extended series for SAWs crossing a square with SAPs and spanning SAWs given a more cursory treatment. This is followed  in Section~\ref{sec:sqlat} by the results of our detailed asymptotic analysis of SAWs crossing rhomboidal and triangular domains with several other problems briefly mentioned.
Section~\ref{sec:conc} contains our conclusions and gives a summary of the estimates we have obtained.

\section{Algorithm to enumerate SAWs crossing a rhombus.}
\label{sec:algo}

The algorithm we use to count the number SAWs on domains of the hexagonal lattice builds on the  pioneering work of Enting \cite{IGE80} 
who enumerated square lattice self-avoiding polygons and extended by Conway, Enting and Guttmann \cite{CEG93} to enumerate square lattice SAWs. 
An algorithm for the enumeration of hexagonal SAWs was described in \cite{IJ06} and a detailed description of the general method can be found in \cite{EJ09}.

\subsection{Transfer matrix algorithm}
\label{sec:TM}

\begin{figure}
\begin{center}

\begin{minipage}[t]{0.55\textwidth} 
\centerline{\includegraphics[width=\textwidth,angle=0]{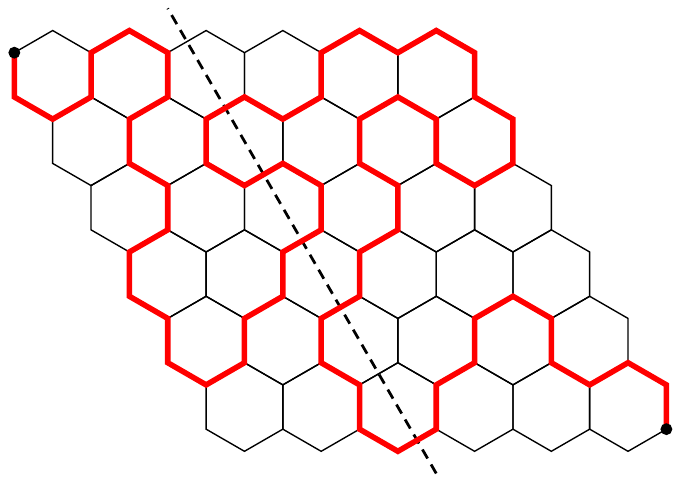} }
\end{minipage}
\hspace{0.1\textwidth}
\begin{minipage}[t]{0.25\textwidth} 
\centerline{\includegraphics[width=\textwidth,angle=0]{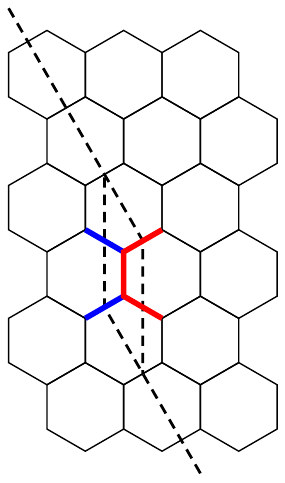}}
\end{minipage}
 \end{center}

\caption{\label{fig:TM}   An example (left panel) of a SAW crossing a rhomboidal domain of the hexagonal lattice intersected by a TM line. 
 The  basic TM move (right panel) in which the intersection is moved so as to add another two vertices
 and three edges to the section of the domain already visited. The states of the two blue `incoming' edges 
determines the type of update to apply while adding the three red edges to the visited section of the domain.}
\end{figure}

If we take an example of a SAW  crossing a rhombus and draw an line across the domain 
as shown in \Fref{fig:TM} we observe that the partial SAW to the left of the 
intersection consists of arcs connecting two edges on the intersection (we shall refer to 
these as arc-ends),  and a single edge that is not  connected to any other edge
on the intersection (we call this a free end). The free end is connected to the vertex in the upper left corner 
of the domain and the SAW must terminate in the lower right corner. 

We are not allowed to form closed loops, so two arc ends can only be joined 
if they belong to different arcs.  We must also ensure that the graphs we count have just 
a single component. To exclude arcs which close on themselves 
we label the occupied edges in such a way that we can easily determine 
whether or not two ends belong to the same arc. On two-dimensional 
lattices this can be done by relying on the fact that arcs 
can never intertwine. Each arc end is assigned a label
depending on whether it is the lower or upper end of an arc and these
labels can be viewed as balanced parenthesis. We shall refer to
the configuration along the intersection as a {\em signature}, denoted by $\Sig$, which
 can  be represented by a string of edge states, $\sigma_i$, where

\begin{equation}\label{eq:states}
\sigma_i  = \left\{ \begin{array}{rl}
\E &\;\;\; \mbox{empty edge},  \\ 
\L &\;\;\; \mbox{lower arc end}, \\
\U &\;\;\; \mbox{upper arc end}, \\
\F &\;\;\; \mbox{free end}. \\
\end{array} \right.
\end{equation}
\noindent
Take the SAW in \Fref{fig:TM} and consider the configuration 
associated with the partial SAW to the left of the line. Reading from bottom to top we find the  signature $\Sig = \L\E\U\F\L\U\E.$ 
Since all SAWs have to cross the rhombus it readily follows that any signature contains one free edge surrounded by a string of empty states and arc ends on either side (with the arc ends forming balanced parenthesis).

\begin{figure}
\begin{center}
\includegraphics[width=0.9\textwidth]{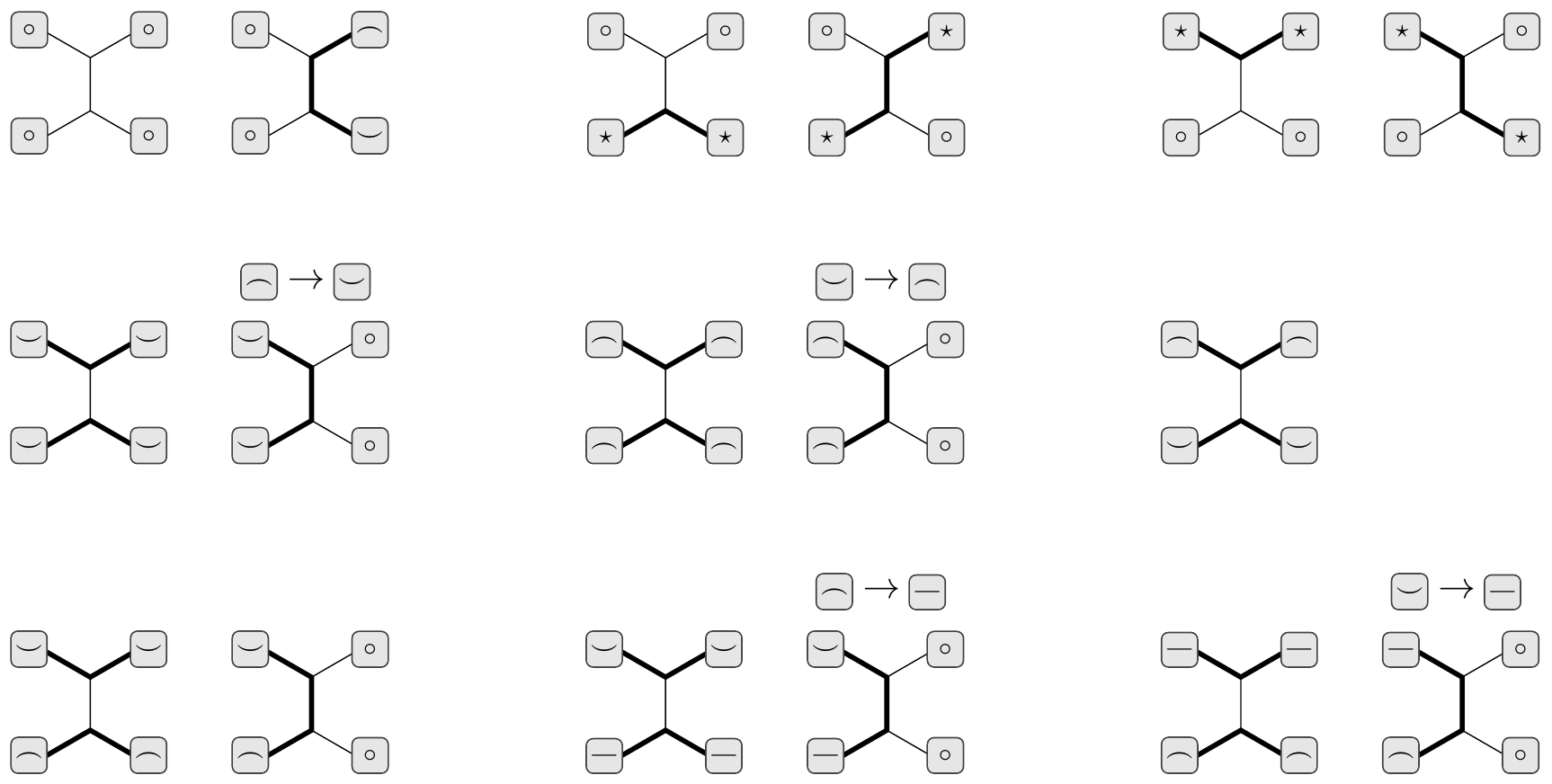}
\end{center}

\caption{\label{fig:TM-update} The  possible updates in a TM move with thin edges empty and thick edges occupied by  the walk. 
In the top row \O\ refers to any type of occupied edge \F, \L, or \U.
Relabelling of arc ends are indicated above the update, i.e., for the second transition in the second row two lower arc ends are
connected and the matching upper end is relabelled as a lower end.}
\end{figure}

For each signature $\Sig$ we simply count the number of partial SAWs, $\C(\Sig)$.
SAWs in a given domain of the hexagonal lattice are counted by moving the 
intersection so as to add  two vertices and three edges at a time, as illustrated in \Fref{fig:TM}.  
The updating of the counts $\C(\Sig)$ depends on the states of the
edges to the left of the new vertices. In \Fref{fig:TM-update} we display the possible local 
`input' states and the `output' states which arise as the kink in the boundary
is propagated by one step. Not all the possible local input states are displayed
since some are related by an obvious reflection symmetry  with straightforward changes
to the corresponding updating rules.  We shall refer to the signature before the move 
as the {\em source},  $\Sig_{\rm S}$, and a signature produced as a result of the move as a {\em target},  $\Sig_{\rm T}$. 
In all cases we see that the first update has the source appearing as a target as well. In the second row, last panel, we cannot connect the arc ends
since this would result in the formation of a cycle. 
Most of the updates are local and involve only the two edges in the kink, but some
of the updates involves a non-local transformation of the signature. This happens when we connect two
lower (upper) arc ends or a free end to a lower (upper) arc end. In these cases we need to locate a matching
arc end in the signature and relabel it accordingly. We illustrate these here:
\begin{eqnarray*}
\fl
 \qquad \L\U\E\F\LB\LB\E\L\U\E\L\U\U\U &\quad \to \quad &   \L\U\E\F\EB\EB\E\L\U\E\L\U\LB\U   \\[2mm]
 \fl
\qquad  \L\L\E\L\U\E\L\U\U\UB\FB\L\U\E &\quad \to \quad &   \FB\L\E\L\U\E\L\U\U\EB\EB\L\U\E   \\
\end{eqnarray*}
Two consecutive blue tiles indicate the edges that are involved in the update as per \Fref{fig:TM}, while
the isolated blue tile indicate the edge which has a change of state. In the first example we connect 
two lower arc ends (second row second update of \Fref{fig:TM-update}) and we then have to relabel the
upper arc end of the inner arc to a lower arc end as indicated. How do we find the matching arc end?
We start at the update position of the innermost arc and set a counter to 1, we then scan
to the right and increase the counter by 1 for every \L\ we encounter and decrease the counter by 1 for every \U; once
the counter records a value of 0 we have found the matching end. Similarly in the second example 
(illustrating the last update in row three of \Fref{fig:TM-update}) we connect an upper arc end to the
free end and we then have to locate the matching lower end of the arc and change the state from \L\ to \F. 

\subsection{Motzkin path representation of signatures}
\label{sec:Mrep}

It is possible to represent the signatures as  Motzkin paths, which are directed walks  from $(0,0)$ to $(n,0)$ 
in the first quadrant of the square lattice with step-set $\Omega = \{(1,0),(1,1),(1,-1)\}$,
see \seqnum{A001006} for numerous references to this classical combinatorial problem. We shall refer to the steps 
as horizontal, up and down steps,  respectively. The basic mapping from a signature to a Motzkin path is to map \E\ to horizontal steps,
\L\ to up steps, and \U\ to down steps. 

What about the free end? Since the walk has to cross the domain the free end  
can never be enclosed inside an arc and therefore splits the signature into two `halves' such that on either side of the free 
end one has a standard Motzkin path. Next we consider what happens to updates involving a free end and show that we don't need to explicitly keep track of the free end but can treat it as if it were an (excess) upper end arc, which we denote \FU. 

\begin{description}
\item[$\F\E:$] An input of $\F\E$ (or $\E\F$) produces outputs $\E\F$ and $\F\E$ which is the same as for an input $\FU\E$ with the free end remaining the excess \FU. 
\item[$\F\L:$] $\F\L \to \F\L$ is clearly the same as $\FU\L \to \FU\L.$  In the case $\F\L \to \E\E$ we are connecting 
a free end to a lower arc end and relabelling the matching upper end as free. However, this is equivalent to $\FU\L \to \E\E$ 
with the incoming \FU\ now moving to the position of the matching \U\ of the arc with no change of state  required.
\item[$\U\F:$] $\U\F\ \to \U\F$ is clearly the same as $\U\FU \to \U\FU.$ In the case $\U\F \to \E\E$ we are connecting 
a free end to an upper arc end and relabelling the matching lower end as free. However, this is equivalent to $\U\FU \to \E\E$ 
with the  incoming \FU\ now moving to the position of the matching \L\ of the arc and being relabelled as  \FU.
\item[$\L\F/\F\U:$] Not possible since free end would be enclosed inside an arc. 
\end{description}

\begin{figure}
\begin{center}
\includegraphics{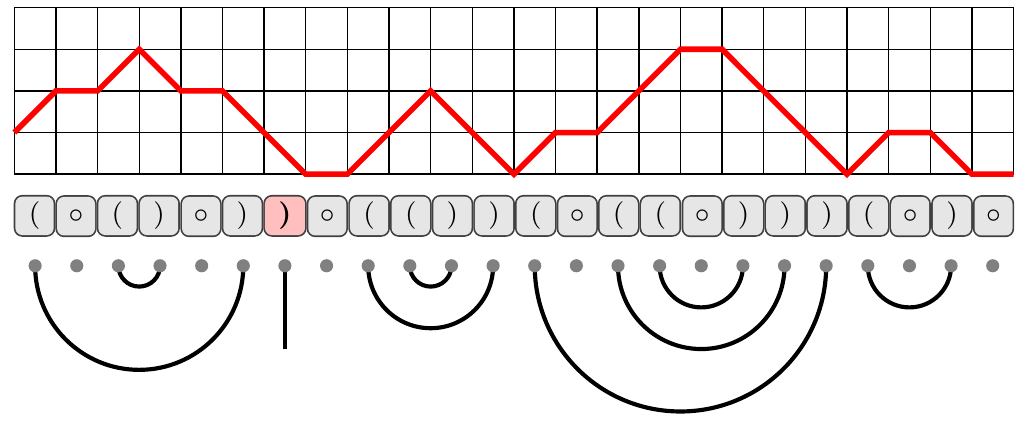}
\caption{\label{fig:Mrep} 
Illustration of the representation of an arc configuration along the TM intersection line (grey circles mark the intersection 
with an edge), the corresponding signature, and the corresponding Motzkin path starting at (0,1) and ending at (24,0). 
The free end can be represented by an excess upper arc end \FU\ and located as the first return of the Motzkin path to level 0.
}
\end{center}
\end{figure}

It now follows that the set of signatures can be represented as the set of Motzkin paths 
starting at height 1, i.e. at vertex $(0,1)$, and ending at $(L+1,0)$. This is another well known combinatorial problem
as evidenced by its low sequence number, \seqnum{A002026}.
Should we need to know the position of the free end (as it happens we don't for this problem) it is easy to find it 
as the excess \FU\ when looking from the first state in the signature. In Motzkin parlance the position of the free end 
corresponds to the {\em first} return of the path to height 0. The representation is illustrated in \Fref{fig:Mrep}.

\subsection{Minimal perfect hashing}
\label{sec:hash}

\begin{figure}
\begin{center}
\includegraphics{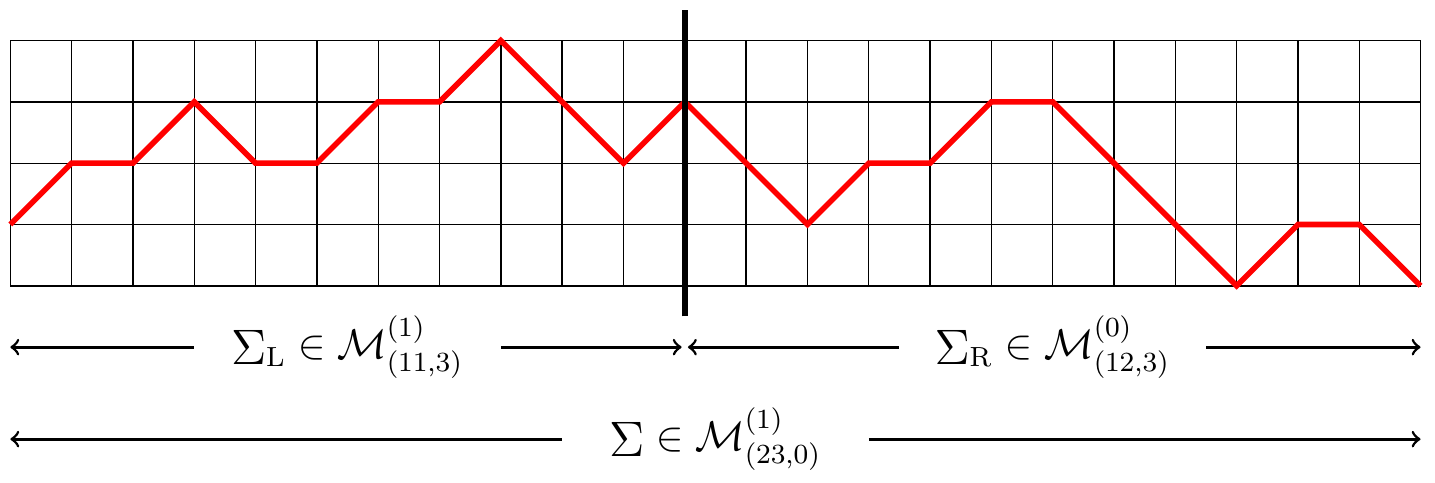}
\caption{\label{fig:hash} 
Illustration of the splitting of a Motzkin path representing a signature, $\Sig \in \MO_{(23,0)}$  ($L=22$), into 
left and right halves of height $h=3$, with  $\Sig_{\rm L} \in \MO_{(11,3)}$ and $\Sig_{\rm R} \in \MZ_{(12,3)}$. 
Note that the right Motzkin path has to be reversed, i.e., it starts at (23,0) and proceeds leftwards to (11,3).
}
\end{center}
\end{figure}

We implement the minimal perfect hashing scheme of Iwashita \etal \cite{INK13}. 
Let $\MZ_{(m,h)}$ be the set of $m$-step Motzkin paths starting at height 0 and ending at height $h$. 
Similarly, let $\MO_{(n,h)}$ be the set of $n$-step Motzkin paths starting a height 1 and ending at height $h$.
The total set of states is $\MO_{(L+1,0)}$ since there are $L+1$ edges along the TM intersection.
So we seek to construct a mapping $\Phi : \MO_{(L+1,0)} \to \{1,\ldots, |\MO_{(L+1,0)}|\}$. We implement this
as a sum of two functions 
\BE\label{eq:hash}
\Phi \left(\Sig\right)=\Phi_{\rm L} \left(\Sig_{\rm L}\right) + \Phi_{\rm R} \left(\Sig_{\rm R}\right),
\EE
where $\Sig_{\rm L}$ is the left part of the signature and  $\Sig_{\rm R}$ the right part. 
We divide the signature at the halfway point so that $\Sig_{\rm L}$ contains the
first $m=\lfloor (L+1)/2 \rfloor$ states of $\Sig,$ and  $\Sig_{\rm R}$ the remaining $n=L+1-m$ states.
We can view $\Sig$ as the concatenation of two Motzkin paths of height $h$ with $0\leq h \leq m$.
We then have that $\Sig_{\rm L} \in \MO_{(m,h)}$ and $\Sig_{\rm R} \in \MZ_{(n,h)}$, though the Motzkin path from  $\MZ_{(n,h)}$
has to be reversed, see \Fref{fig:hash}.

We divide the storage array for the counts into sections based on the height $h$ of the signatures $\Sig \in \MO_{(L+1,0)}$. 
Each $\Sig_{\rm L} \in \MO_{(m,h)}$ can be concatenated with any reversed path  from $\MZ_{(n,h)}$. 
The total number of paths of height $h$ is therefore $|\MO_{(m,h)}| \cdot |\MZ_{(n,h)}|$,  and this is the size of the section
of the storage array required to contain the counts for signatures of height $h$. Each main section
of the storage array is divided into subsections of size $ |\MZ_{(n,h)}|$ containing the counts
of the signatures with a particular left part  $\Sig_{\rm L}$. The paths in $\MO_{(m,h)}$ and
$\MZ_{(n,h)}$ are sorted in lexicographical order (using $\E < \L < \U$) so that paths can be assigned unique indices $I_{\rm L}$ 
and $I_{\rm R}$ (note there is separate index function for each $h$ and $L$). Define the number $b_h$ as
\begin{eqnarray*}
b_0&:=&0, \\
b_{h+1} &:=& b_h+|\MO_{(m,h)}| \cdot |\MZ_{(n,h)}|,
\end{eqnarray*}
then we define 
\begin{eqnarray*}
\Phi_{\rm L} \left(\Sig_{\rm L}\right) &=&b_h + (I_{\rm L} -1) \cdot |\MZ_{(n,h)}| \\
 \Phi_{\rm R} \left(\Sig_{\rm R}\right)&=& I_{\rm R}.
\end{eqnarray*}
$\Phi_{\rm L}$ tells us which subsection of the storage array to use 
and $\Phi_{\rm R}$ gives us the position within a given subsection.

\subsection{Data representation and storage}

The number of signatures  $|\MO_{(L+1,0)}|$ can be expressed in terms of Motzkin numbers $M_n=|\MZ_{(n,0)}|$ (\seqnum{A001006})
since $|\MO_{(L+1,0)}| = M_{L+2}-M_{L+1}$ (\seqnum{A002026}). The Motzkin numbers are given by the recurrence
\BE \label{eq:motz}
M_0=M_1=1, \quad (n+2)M_n= (2n+1)M_{n-1} + 3(n-1)M_{n-2}
\EE
and have the generating function 
\BE \label{eq:motzgf}
{\mathcal M}(z) = \sum_{n=0} M_nz^n = \big( 1 - z - \sqrt{1-2z-3z^2\,}\big) / (2z^2).
\EE
From this it follows immediately that $|\MO_{(L+1,0)}| \sim 3^{L}$ and this gives the asymptotic growth in the storage
required for the counts $\C(\Sig)$. This growth in storage is the main limitation on the maximum size $L_{\rm max}$ that
we can attain.	We therefore perform all calculations of the walk counts modulo several prime numbers $p_k$ which
yields remainders of $C_L(1)$  modulo $p_k$. The exact counts are then obtained from the remainders using the 
Chinese remainder theorem. We generally use primes of the form $p_k = 2^{62}-r_k$, such that $p_k$ are
the largest primes less than $2^{62}$. The counts $\C(\Sig)$ can therefore be stored in an array of 64-bit integers with
$\Phi(\Sig)$ giving the position where $\C(\Sig)$  is stored.

The signatures are represented as 64-bit integers with 2 bits required for each state, with $\E=00, \L=10$, and $\U=01$.
The left and right parts of a signature can then be represented by a 32-bit integer and the hash functions $\Phi_{\rm L}$
and $\Phi_{\rm R}$ can be coded directly as simple arrays or look-up tables. The total size of these two arrays is
about $2^{L+2}$ so insignificant compared to the storage needed for the counts. The integer representation of 
signatures also means that transformations between a signature $\Sig$ and its left and right parts $\Sig_{\rm L},\Sig_{\rm R}$
and from sources to targets can be done very efficiently using bit-wise manipulations.

\subsection{In-place memory updating}
\label{sec:inplace}

By controlling the order in which we access the signatures we can ensure that the counts can by updated in-place
without the need for any temporary storage. The way we order the signatures is by height and for given height in
lexicographical increasing order. Generally speaking, in-place updating is safe if a signature is updated only after 
it has been processed. The specific order of processing is controlled by the position at which we divide the signature into two
halves. Importantly this dividing position need not be the same as the one used to construct the hash function and
can be changed between iterations of the TM algorithm. The updates 
illustrated in \Fref{fig:TM-update} shows that processing a given source signature $\Sig_{\rm S}$
always give rise to the same signature (as a target). A signature 
mapping to itself results is no change to its count and hence nothing needs to be done and in-place updating is trivially safe. We now consider the updates in detail and show how in-place updating can be done safely.

\begin{description}
\item[$\E\E:$] Processing the signature   $\Sig_{\rm S}=\Sig_{\rm L}\E\E\Sig_{\rm R}$ leads to $\Sig_{\rm S}$ and the new signature 
$\Sig_{\rm T}=\Sig_{\rm L}\L\U\Sig_{\rm R}$.
Updating the count for $\Sig_{\rm T}$ is safe since $\Sig_{\rm T} $ does not give rise to any new target signatures (apart from itself) when processed.
\item[$\E\L/\L\E:$]  Processing $\Sig_1=\Sig_{\rm L}\E\L\Sig_{\rm R}$ leads to $\Sig_{\rm 1}$ and the signature 
$\Sig_2=\Sig_{\rm L}\L\E\Sig_{\rm R}$, while similarly processing $\Sig_2$ gives rise
to $\Sig_2$ and $\Sig_1$. In-place updating of the counts for $\Sig_1$ and $\Sig_2$ is safe provided they are updated simultaneously,
which is easily achieved.
\item[$\E\U/\U\E:$] Same as above.
\item[$\L\L:$] Processing the signature $\Sig_{\rm S}=\Sig_{\rm L}\L\L\U\U\Sig_{\rm R}$ leads to $\Sig_{\rm S}$ and  the new signature 
$\Sig_{\rm T}=\Sig_{\rm L}\E\E\L\U\Sig_{\rm R}$. Note that the matching upper arc ends \U\ need not be consecutive or next to \L\L. We now look at
 the four sites involved in the update and consider how the height of the signature at the dividing position changes. We have 

\begin{center}
{\renewcommand{\tabcolsep}{0.25mm}
\begin{tabular}{cccccccccccccccccccc}
 \vline & \L & \vline  & \L& \vline  & \U & \vline  & \U & \vline & \,\,\,$\to$\,\,\, &   \vline & \E& \vline  & \E& \vline  & \L & \vline  & \U & \vline  \\
  0 & &1  & & 2  & & 1  & & 0   & &   0 & &0  & & 0  & & 1  & & 0   \\
\end{tabular}
}
\end{center}

The possible dividing positions are indicated by vertical lines and the numbers below indicate the additional height of the
signature from the height of $\Sig_{\rm L}$. 

We see that $\Sig_{\rm T}$ is never higher than $\Sig_{\rm S}$ and when they have the same height $\Sig_{\rm T}$ is lexicographically smaller
than $\Sig_{\rm S}$. Hence in all cases we process $\Sig_{\rm T}$  before updating its count and in-place updating is therefore safe.

\item[$\U\U:$] Processing  $\Sig_{\rm S}=\Sig_{\rm L}\L\L\U\U\Sig_{\rm R}$ leads to $\Sig_{\rm S}$ and $\Sig_{\rm T}=\Sig_{\rm L}\L\U\E\E\Sig_{\rm R}$. We have 

\begin{center}
{\renewcommand{\tabcolsep}{0.25mm}
\begin{tabular}{cccccccccccccccccccc}
 \vline & \L & \vline  & \L& \vline  & \U & \vline  & \U & \vline & \,\,\,$\to$\,\,\, &   \vline & \L& \vline  & \U& \vline  & \E & \vline  & \E & \vline  \\
  0 & &1  & & 2  & & 1  & & 0   & &   0 & &1  & & 0  & & 0  & & 0   \\
\end{tabular}
}
\end{center}
 
 As for the case above in-place updating is  safe.

\item[$\L\U:$] No new signatures.

\item[$\U\L:$] Processing  $\Sig_{\rm S}=\Sig_{\rm L}\U\L\Sig_{\rm R}$ leads to $\Sig_{\rm S}$ and $\Sig_{\rm T}=\Sig_{\rm L}\E\E\Sig_{\rm R}$. We have 

\begin{center}
{\renewcommand{\tabcolsep}{0.25mm}
\begin{tabular}{cccccccccccccccccccc}
 \vline & \U & \vline  & \L& \vline  & \,\,\,$\to$\,\,\, &   \vline & \E & \vline  & \E & \vline  \\
  0 & & $\!\!\!-1$  & & 0   & &   0 & & 0  & & 0      \\
\end{tabular}
}
\end{center}
 
 In-place updating is  safe when the additional height is 0. There is a problem when the dividing position
 splits the signature between \U\ and \L. In that case $\Sig_{\rm T}$ is higher than $\Sig_{\rm S}$ and in-place updating is unsafe
 since the count of $\Sig_{\rm T}$ is updated before $\Sig_{\rm T}$ is processed.
\end{description}

The upshot of the above considerations is that in-place updating can be safely done provided the dividing position
never splits the signature between two edges involved in an update.  Thankfully we can easily avoid
 this from happening since we can change the dividing position so as to avoid such splits.

\begin{figure}
\begin{center}
\includegraphics{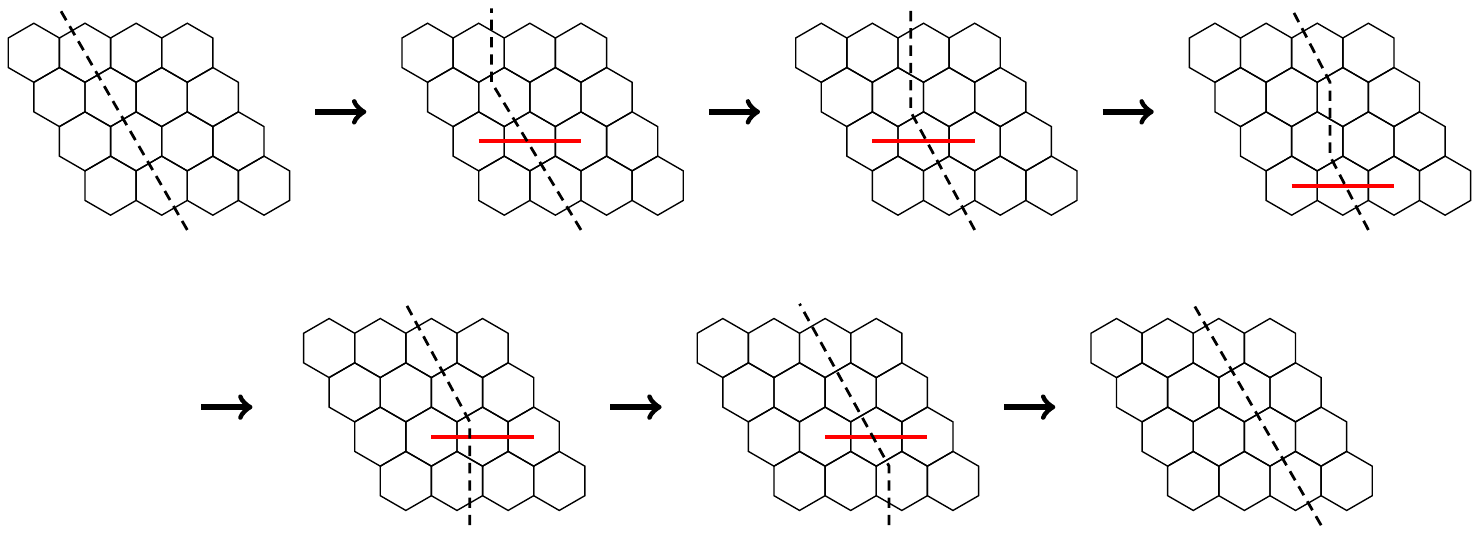}
\end{center}
\caption{\label{fig:TMevol} The TM moves used to add an extra column to a rhomboidal domain. The thick red line
shows the dividing position during an update.}
\end{figure}

\begin{algorithm}
\caption{\label{alg:main}
Calculate the number of SAWs crossing a rhombus of size $L$}\label{alg:rhombus} 
\begin{algorithmic}[1]
\State $l_h \gets \lfloor (L+1)/2 \rfloor$ 
\State $\Phi \gets$ {\sc ConstructHashFunction}($l_h$) 
\State $l_t \gets l_h -1$  \Comment{Upper signature divider}
\State $l_b \gets l_t -1$           \Comment{Lower signature divider}
\State $\C[k] \gets 0$ for all $1 \leq k \leq |\MO_{(L+1,0)}|$
\State $\C[\Phi(\E\E\cdots \E\U)] \gets 1$ \label{line:initsig} 
\For{$Row=0$ {\bf to}  $L$} \Comment{Build domain column-by-column}
\State $m_h \gets \min(l_t+1,L+1-l_t)$ \Comment{Max possible height} 
  \State  $\big\{\MO_{(l_t,h)},\MZ_{(L+1-l_t,h)}\big\} \gets$ {\sc ConstructSignatures}($l_t$) \Comment{$0 \leq h \leq m_h$}
   \For{$Col=L-1$ {\bf to}  $l_t$ {\bf by} $-1$} \label{line:fcl}  \Comment{Build top half of column}
    \For{$h$=0  {\bf to} $m_h$}   \Comment{Height of signatures}
    \ForAll{$\Sig_{\rm L} \in \MO_{(l_t,h)}$}  \Comment{Left signatures}
    \ForAll{$\Sig_{\rm R} \in \MZ_{(L+1-l_t,h)}$}  \Comment{Right signatures}
    \State $\Sig_{\rm S} \gets \Sig_{\rm L} \Sig_{\rm R} $    \Comment{Source signature}
    \State {\sc UpdateCounts}($\Sig_{\rm S}$)    \Comment{Process source signature}
    \EndFor
    \EndFor
    \EndFor
  \EndFor
  \State $m_h \gets \min(l_b+1,L+1-l_b)$ \Comment{Max possible height} 
  \State  $\big\{\MO_{(l_b,h)},\MZ_{(L+1-l_b,h)}\big\} \gets$ {\sc ConstructSignatures}($l_b$) 
  \State $Col \gets l_t-1$ \Comment{Add  unit cell to column}
    \For{$h$=0  {\bf to} $m_h$}   \Comment{Height of signatures}
    \ForAll{$\Sig_{\rm L} \in \MO_{(l_b,h)}$}  \Comment{Left signatures}
    \ForAll{$\Sig_{\rm R} \in \MZ_{(L+1-l_b,h)}$}  \Comment{Right signatures}
    \State $\Sig_{\rm S} \gets \Sig_{\rm L} \Sig_{\rm R} $    \Comment{Source signature}
    \State {\sc UpdateCounts}($\Sig_{\rm S}$)    \Comment{Process source signature}
    \EndFor
    \EndFor
    \EndFor

  \State $m_h \gets \min(l_t+1,L+1-l_t)$ \Comment{Max possible height} 
  \State  $\big\{\MO_{(l_t,h)},\MZ_{(L+1-l_t,h)}\big\} \gets$ {\sc ConstructSignatures}($l_t$) 
  \For{$Col=l_t-2$ {\bf to}  0 {\bf by} $-1$} \Comment{Build bottom half of column}
    \State {$\qquad \vdots$} \Comment{Repeat lines 11:--18:}
  \EndFor

\EndFor

\State  {\bf return} $\C[\Phi(\U\E\cdots \E\E)]$ \label{line:return} 
 \end{algorithmic}
\end{algorithm}
 
\begin{algorithm}
\caption{\label{alg:upd}
Update the counts of signatures}\label{alg:update} 
\begin{algorithmic}[1]
 \Procedure {UpdateCounts}{$\Sig_{\rm S}$}
\State $S \gets$ {\sc InputState}($\Sig_{\rm S}$)   \Comment{States of update edges}
\If{$S= \EB\EB$}
\State $\Sig_{\rm T} \gets$ {\sc ChangeSignature}($\Sig_{\rm S}$,\LB\UB)   \Comment{Insert new arc}
\State $\C[\Phi(\Sig_{\rm T})] \gets \C[\Phi(\Sig_{\rm T})]+\C[\Phi(\Sig_{\rm S})]$ \Comment{Update count of target}
\ElsIf{$S= \LB\EB$}
\State $\Sig_{\rm T} \gets$ {\sc ChangeSignature}($\Sig_{\rm S}$,\EB\LB)   
\State $\C[\Phi(\Sig_{\rm T})] \gets \C[\Phi(\Sig_{\rm T})]+\C[\Phi(\Sig_{\rm S})]$ \Comment{Update count of target}
\State $\C[\Phi(\Sig_{\rm S})] \gets \C[\Phi(\Sig_{\rm T})]$ \Comment{Simultaneous update of source}
\ElsIf{$S= \EB\LB$}
\State {\sc Null} \Comment{Do nothing. Processed in previous update}
\ElsIf{$S= \UB\EB$}
\State $\Sig_{\rm T} \gets$ {\sc ChangeSignature}($\Sig_{\rm S}$,\EB\UB)  
\State $\C[\Phi(\Sig_{\rm T})] \gets \C[\Phi(\Sig_{\rm T})]+\C[\Phi(\Sig_{\rm S})]$  \Comment{Update count of target}
\State $\C[\Phi(\Sig_{\rm S})] \gets \C[\Phi(\Sig_{\rm T})]$  \Comment{Simultaneous update of source}
\ElsIf{$S= \EB\UB$}
\State {\sc Null} \Comment{Do nothing. Processed in previous update}
\ElsIf{$S= \LB\LB$}
\State $\Sig_{\rm T} \gets$ {\sc RelabelSignature}($\Sig_{\rm S}$,\EB\EB,\LB)   \Comment{Connect arc ends and relabel}
\State $\C[\Phi(\Sig_{\rm T})] \gets \C[\Phi(\Sig_{\rm T})]+\C[\Phi(\Sig_{\rm S})]$ 
\ElsIf{$S= \LB\UB$}
\State {\sc Null} \Comment{Do nothing. No new signatures}
\ElsIf{$S= \UB\LB$}
\State $\Sig_{\rm T} \gets$ {\sc ChangeSignature}($\Sig_{\rm S}$,\EB\EB)   \Comment{Connect arc ends}
\State $\C[\Phi(\Sig_{\rm T})] \gets \C[\Phi(\Sig_{\rm T})]+\C[\Phi(\Sig_{\rm S})]$  
\ElsIf{$S= \UB\UB$}
\State $\Sig_{\rm T} \gets$ {\sc RelabelSignature}($\Sig_{\rm S}$,\EB\EB,\UB)   \Comment{Connect arc ends and relabel}
\State $\C[\Phi(\Sig_{\rm T})] \gets \C[\Phi(\Sig_{\rm T})]+\C[\Phi(\Sig_{\rm S})]$  
\EndIf
\EndProcedure
\end{algorithmic}
 \end{algorithm}

We are now ready to present our algorithm for counting SAWs crossing a rhombus. Algorithm~\ref{alg:rhombus} presents
pseudo code for the main body of our algorithm. First up we divide signatures into two halves at position $l_h$ and calculate
the corresponding hash function $\Phi$ or more specifically the two functions (look-up tables) $\Phi_{\rm L}$ and  $\Phi_{\rm R}$. The hash function $\Phi$ remains fixed throughout the entire calculation.
The value of $l_h$ determines {\em where} counts are stored in memory. Then we define two parameters $l_t$ and $l_b$ which determine the {\em order in which signatures are processed}. Next we initialise the counts to zero except for the signature with a free end at the top vertex. 
After this comes the main body of the algorithm where we build the rhombus column-by-column up to size $L$ with
each column built cell-by-cell as illustrated in \Fref{fig:TMevol}. Note that the first move, from panel 1 to 2, and the final move are `virtual'. These TM intersection moves add just a single edge and this means that all sources and targets are identical and hence no updating is actually required. These moves are included for illustrative purposes only.

We choose the parameters $l_t$ and $l_b < l_h$,  because memory access is crucial to the performance of the algorithm 
and we found that for $L$ large these choices resulted in the best performance. The reason for breaking the column 
construction into three separate pieces is  to allow in-place updating of the counts. During the first loop (at line \ref*{line:fcl}) the position of the divider $l_t$ is below 
the local states being changed by an update (see  \Fref{fig:TMevol}) so in-place updating can be done safely. Next we change the divider to
be at $l_b=l_t-1$, since otherwise the divider would lie between the two local states in the TM kink and as explained
above this would not be safe. We then change back to a divider at $l_t$ (which now lies above the local states of the update) 
and complete the column. Note that we could have set $l_b=l_t+1=l_h$ and completed the column with this divider, but as already stated using
dividers strictly less than $l_h$ is faster and the time taken by an extra call to  {\sc ConstructSignatures} is insignificant. The routine {\sc ConstructSignatures} generates the sets of left and right signatures using a simple back-tracking algorithm.
 
The updating rules for the counts of the signatures are given in Algorithm~\ref{alg:upd}.  {\sc InputState} simply extracts the states of the two input edges involved in the update. {\sc ChangeSignature} changes the states of the input edges to those indicated by the two blue tiles. {\sc RelabelSignature} changes the input states to empty states and finds and relabels the matching arc end in those updates where two arc ends are connected in a TM update.

\subsection{Parallelisation}
\label{sec:para}

The transfer-matrix algorithm is very well suited to parallel
computation. In previous work we implemented algorithms using the message passing interface (MPI) \cite{IJ03,EJ09} suited for distributed memory systems. 
For this work we used shared memory computers and hence
implemented the parallel algorithms using OpenMP which is somewhat
simpler but relies on the same basic ideas.
One of the main ways of achieving a good parallel algorithm using 
data decomposition is to try to find an invariant under the
operation of the updating rules. That is we seek to find some property
of the signature which
does not alter in a single iteration. There is such an invariant since any edge not
directly involved in the update cannot change from being 
empty to being occupied and vice versa (it may change, say, from state \U\ to \L). 
That is only the kink edges  can change their occupation status. This invariant
allows us to parallelise the algorithm in such a way
that we can do the calculation completely independently on each
core.  With the intersection straight (having no kinks) we distribute the
data across cores so that signatures with the same 
occupation pattern along the {\em lower} half of the intersection 
are processed by the same core.  We  then do the TM updates inserting 
the top-half of a new column.
This can be done {\em independently} by each core because the 
occupation pattern in the lower half remains unchanged.
When  reaching the half-way point we redistribute the data
so that configurations with the same occupation pattern along 
the {\em upper} half of the intersection are processed by  the same core and we then
do the TM update inserting the bottom-half of a new column. This is then
repeated column by column. 
 
\subsection{Changes needed to enumerate other hexagonal problems}
\label{sec:other}

The changes required to enumerate other types of configurations are mostly straightforward. To enumerate spanning SAWs we just need to change lines \ref*{line:initsig} and \ref*{line:return} in Algorithm \ref{alg:main}. At \ref*{line:initsig}  we need to initialise all signatures with just a single free end in some position (all other states empty) to have a count of one. This means a SAW can start in any position on the left side of the rhombus. Similarly at \ref*{line:return} we need to return the sum of the counts for signatures with just a single free end. 

To enumerate SAPs crossing a rhombus the main change to note is that we no longer have a free end and any signature can therefore be represented by a standard Motzkin path from $(0,0)$ to $(L+1,0)$. So the total set of signatures for this problem is $\MZ_{(L+1,0)}$.  Furthermore, we have that $\Sig_{\rm L} \in \MZ_{(m,h)}$ and $\Sig_{\rm R} \in \MZ_{(n,h)}$. Again we need to change lines \ref*{line:initsig}  and \ref*{line:return} of Algorithm \ref{alg:main}. Line \ref*{line:initsig}  is changed to: $\C[\Phi(\E\E\cdots \E\L\U)] \gets 1$. 
Line \ref*{line:return} is changed to: {\bf return} $\C[\Phi(\L\U\E\cdots \E\E)]$.  

Enumerations in a triangular domain just requires us to change the way in which the transfer matrix intersection is moved, that is, the moves for the rhombus TM calculation shown in \Fref{fig:TMevol} have to be changed appropriately. 
 
\subsection{Algorithm to enumerate square lattice problems}

The algorithm for enumerating walks crossing a square has been described in \cite{INK13},
and for this work we implemented our own version which we won't describe here other than to say that the main body is identical to Algorithm~\ref{alg:main}, but of course the updating rules are different, and Algorithm~\ref{alg:upd} must be amended accordingly. The interested reader can check out the actual code at our GitHub repository (see Section~\ref{sec:resource}).

\section{Series analysis} 
\label{sec:ana}

The method of series analysis  has, for many years, been a powerful tool in the study of a variety of problems in statistical mechanics, combinatorics, and other fields. In essence, the problem is the following: Given the first $N$ coefficients of the series expansion of some function, (where $N$ is typically as low as 5 or 6, or as high as 100,000 or more), determine the asymptotic form of the coefficients, subject to some underlying assumptions, or equivalently the nature of the singularity of the function.

A typical example is the generating function of self-avoiding walks (SAWs) in dimension two or three. This is believed to behave as 
\begin{equation}\label{generic}
F(z) = \sum_n c_n z^n \sim C \cdot (1 - z/z_c)^{-\gamma}.
\end{equation}
In this case, among regular two-dimensional lattices, the value of $z_c$ is only known for the hexagonal lattice \cite{DCS10}, while $\gamma=43/32$ \cite{BN82} is believed to be the correct exponent value for all two-dimensional lattices, but this has not been proved. 

The method of series analysis is used when one or more of the critical parameters is not known. For example, for the three-dimensional version of the above problems, none of the quantities $C,$ $z_c$ or $\gamma$ are known exactly.
 From the binomial theorem it follows from (\ref{generic}) that 
\begin{equation}\label{asymp1}
c_n \sim \frac{C }{\Gamma(\gamma)}\cdot  z_c^{-n} \cdot  n^{\gamma-1},
\end{equation}
where $a_n \sim b_n$ means that $\lim_{n \to \infty} a_n/b_n = 1.$
 Here $C, \,\, z_c, \,\, {\rm and} \,\, \gamma$ are referred to as the critical amplitude, the critical point (usually the radius of convergence) and the critical exponent, respectively. In combinatorics one often refers to the {\em growth constant $\mu=1/z_c,$} as the coefficients are dominated by the term $\mu^n.$

Obtaining these coefficients is typically a problem of exponential complexity, as is the case with our algorithm, described in Section~\ref{sec:algo}. The consequence is that usually fewer than 50 terms are known (and in some cases far fewer).

The standard methods of series analysis include the ratio method, described in \ref{app:ratio}, and the method of differential approximants, described in \ref{app:da}.
A relatively recent development has been the method of series extension \cite{G16}, described in \ref{app:pred}, in which differential approximants based on the exactly known terms is used to obtain a significant number of additional {\em approximate} terms. These approximate terms, if of sufficient accuracy, can then be used in the ratio method and its extensions to obtain more precise estimates of the various critical parameters.

In our analysis below we make use of all of these methods, but will just refer to them under the assumption that the material in the appendices has been understood.

\subsection{Methods of analysis} 
\label{sec:ana_met}

The existence of the limit (\ref{eq:CLlim}) and the more detailed asymptotic form (\ref{eq:CLas}), which we shall take for granted and
provide overwhelming numerical support for, suggests several methods of analysis that one can apply in order to estimate the growth
constant $\lambda$. For the first method (M1), we look at the quantity
\BE \label{eq:M1}
 \lambda_L := C_{L}(1)^{1/L^2} \sim \lambda.
\EE
While it has not been proved that the ratios $R_L := C_L(1)/C_{L-1}(1) \sim \lambda^{2L},$ it is almost certainly true, and we will assume it to be so in our analysis.  Given the expectation that $R_L \sim \lambda^{2L},$ for the second method (M2) we define the ratio-of-ratios 
\BE  \label{eq:M2}
{\mathcal C}_L := \frac{R_{L+1}}{R_L} = \frac{C_{L+1}(1) C_{L-1}(1) }{C_L(1)^2}.
\EE
From (\ref{eq:CLas}) it follows that 
\BE \label{eq:M2as}
{\mathcal C}_L = \lambda^2 \left ( 1 -\frac{g}{L^2} + O(L^{-3}) \right ).
\EE\
All of the sequences defined above will be analysed using ratio methods.

Next we  briefly describe three different methods  that we have used to estimate the parameters $b,$ $c$ and $g$  in the assumed asymptotic form (\ref{eq:CLas}).
In the first method (P1) we use our best estimate of $\lambda$ and form the sequence 
\BE \label{eq:P1}
d_L := C_L(1)/\lambda^{L^2} \sim \lambda^{bL+c}\cdot L^g.
\EE
This sequence, provided the assumed asymptotic form is correct, behaves as a typical power-law singularity, in which the coefficients grow as $a_n \sim C\cdot \alpha^n \cdot n^g,$ and can be analysed as such. With that notation, the growth constant $\alpha = \lambda^b,$ and the amplitude $C = \lambda^c.$ Of course, we have to use our estimated value of $\lambda.$

For the second method (P2) we fit to the assumed form by writing 
\BE \label{eq:P2} 
\log{d_L} \sim b\log(\lambda)L+ c\log(\lambda) + g \log{L}.
\EE
 We then use successive triples of data points $( \log{d_{k-1}},\log{d_{k}},\log{d_{k+1}}),$ with $k=2,3,\cdots, L_{\textrm{max}}-1,$ to obtain estimates of the parameters $b \log{\lambda},$ $c \log{\lambda},$ and $g.$  
 
 The third method (P3) makes use of the ratio ${\mathcal C}_L$  (\ref{eq:M2}). According to its asymptotic form (\ref{eq:M2as}), we can 
 fit the sequence $\{{\mathcal C}_L\} $ to $c_0 + c_2/L^2+c_3/L^3,$ so that $c_0$ should give estimators of $\lambda^2,$ and $c_2$ give estimators of $-g\lambda^2.$ 

If $C_L(1) \sim \lambda^{L^2}$, then the ratios $R_L=C_L(1)/C_{L-1}(1) \sim \lambda^{2L-1}$, so the exponent $\gamma$ in the canonical
form  (\ref{asymp1}) equals 1.  It follows that the corresponding function, $\RGf(z) := \sum_L R_Lz^L,$ will have a simple pole at the critical
point $z_c=1/\lambda^2$. If we include sub-dominant terms, so that $C_L(1) \sim \lambda^{L^2+bL+c} L^g$, then $R_L \sim \lambda^{2L-1+b}(1+O(1/L))$,
and all that has changed is the amplitude. The singularity is still a simple pole at $z_c=1/\lambda^2$. The series $\RGf(z)$ can therefore be analysed using 
differential approximants to obtain an estimate for $\lambda$. 
 
Since  $\RGf(z)$ has a simple pole this suggests two other ways to estimate $\lambda$. Firstly, one can simply form Pad\'e approximants, that
is set $P_{m,n}(z) := P_m(z)/Q_n(z)$, where $P_m(z)$ and $Q_n(z)$  are polynomials of degree $m$ and $n$, respectively, chosen so the first 
$n+m+1$ terms in the Taylor expansion of $P_{m,n}(z)$ coincide with those of  $\RGf(z)$. The first real zero of $Q_n(z)$ will then
provide an estimate of $1/\lambda^2$. 

The second method is a little more speculative and novel and we are not entirely sure of its validity. 
We force the differential approximants to have a singularity at a critical point $\widehat{z}_c$ close to the expected true value $z_c=1/\lambda^2$.
This is done by forming {\em biased}   differential approximants as outlined in~\ref{app:bda} and the associated critical exponent is calculated.
Many  biased   differential approximants are formed  for each value of the biasing critical point $\widehat{z}_c$ and the average critical exponent calculated.
One can then conjecture that the value of $\widehat{z}_c$ for which the average critical exponent attains the value $-1$ provides a reasonable estimate
for $1/\lambda^2$.

\section{Walks and polygons in a square.}
\label{sec:sqlat}

\begin{figure}[ht!] 
\centerline{\includegraphics[width=0.9\textwidth,angle=0]{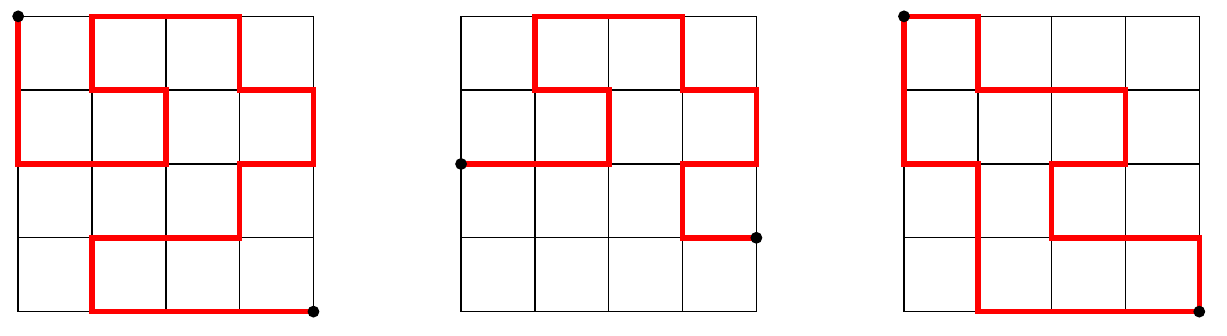} }
\caption{Classes of walk and polygon configurations investigated on the square lattice.} 
\label{fig:sqconf}
\end{figure}

In this section we analyse walks and polygons crossing a square domain of the square lattice, using the techniques just discussed. We study three different variants of the problem, namely 
SAWs crossing or spanning  a square and SAPs crossing a square. These are shown in \Fref{fig:sqconf}.

\subsection{Walks crossing a square}
\label{sec:wcas}

Firstly, we apply method M1 (\ref{eq:M1}) to the analysis of the series for walks crossing a square. For want of greater knowledge about the sub-dominant asymptotic terms we simply extrapolate $\lambda_L$ against $1/L.$ Recall that we only have  27 terms. We therefore use the method of series extension, mentioned above and described in \ref{app:pred}, to extend the sequence of {\em ratios} $R_L = C_L(1)/C_{L-1}(1),$ and this sequence is then used to extend the
$C_L(1)$ series. In this way we obtained 20 additional {\em approximate} coefficients. These are given in \Tref{tab:CLex}.

\begin{table}[htp]

\centering
\begin{tabular}{cc}
\hline
$L$ & $C_L(1)$ estimates. \\
\hline
27 & 1.092762277820988255238897693624593273299$\times 10^{176}$\\
28 & 2.092263800732296637339584460940199207179$\times 10^{189}$\\
29 & 1.219188494943327773136239385657818116903$\times 10^{203}$\\
30 & 2.162167627691293760665426155350775028513$\times 10^{217}$\\
31 & 1.167003905184619653378731561256980927898$\times 10^{232}$\\
32 & 1.916990667670442255801047617746147903033$\times 10^{247}$\\
33 & 9.583688332141159129759056552823132225046$\times 10^{262}$\\
34 & 1.458178102419213554003374021702217439866$\times 10^{279}$\\
35 & 6.752333021793147034314988105916341545574$\times 10^{295}$\\
36 & 9.516180772478135635389490590804240161152$\times 10^{312}$\\
37 & 4.081663288146408412423849764027291063947$\times 10^{330}$\\
38 & 5.328162506991801337436456805173755617688$\times 10^{348}$\\
39 & 2.116818597440340726855200831821163701531$\times 10^{367}$\\
40 & 2.559504109272639104369198989850180317538$\times 10^{386}$\\
41 & 9.418767710224918432123841878087214586086$\times 10^{405}$\\
42 & 1.054869066038373202559187284758968442477$\times 10^{426}$\\
43 & 3.595581533556538000636173781717640795638$\times 10^{446}$\\
44 & 3.729975451051537109220327069642553666508$\times 10^{467}$\\
45 & 1.177630435162076031609822304850879989404$\times 10^{489}$\\
46 & 1.131562339582151957192359485190854061339$\times 10^{511}$\\
 \hline
\end{tabular}
\caption{Estimated coefficients $C_L(1).$ }
\label{tab:CLex}
\end{table}%

\begin{figure}[ht!] 
\includegraphics[width=0.95\textwidth,angle=0]{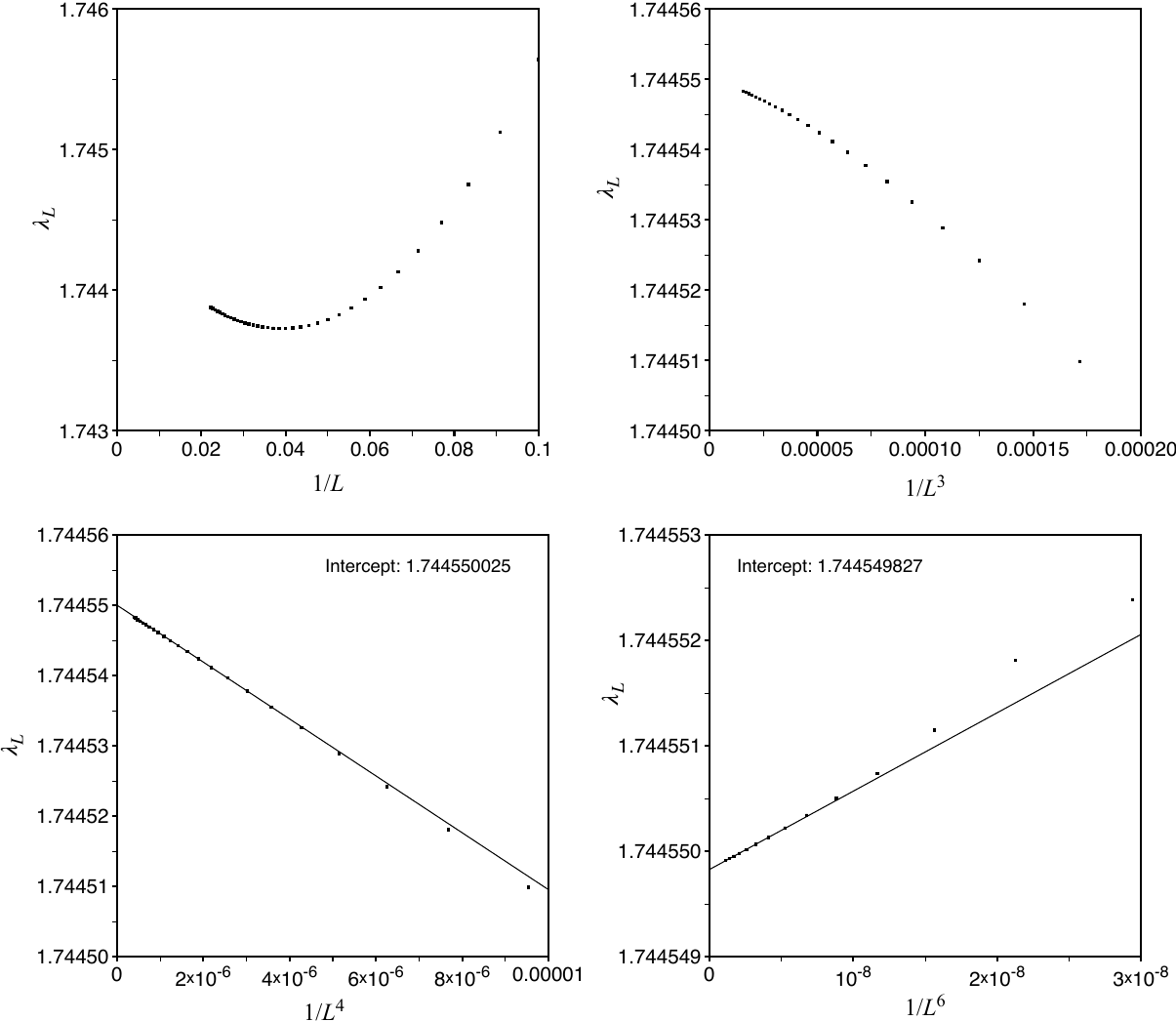}
\caption{\label{fig:wcas-M1} The first panel shows $\lambda_L$ plotted against $1/L$. The second and third panels
show plots of the estimator  $\lambda_L$  using a quadratic correction term against $1/L^3$ and $1/L^4$, respectively. 
The fourth panel is a plot of the estimator  $\lambda_L$  using the correction term  $c_1/L + c_2/L^2+c_4/L^4$.} 
\end{figure}

We show a plot of $\lambda_L$ against $1/L$ in the top-left panel of \Fref{fig:wcas-M1}, and it is seen to be quite well converged, and can visually be extrapolated to $\lambda_S \approx 1.7442.$ 
It is reasonable to assume that the curvature is due to the presence of higher-order terms, such as $1/L^2,$ $1/L^3$ etc. In the top-right panel of \Fref{fig:wcas-M1} we show values of the estimator of $\lambda_S$ assuming $\lambda_L$ converges to $\lambda_S$ with correction term $c_1/L + c_2/L^2,$ plotted against $1/L^3$, and we estimate $\lambda_S \approx 1.74455.$  There is still considerable curvature in this plot and we therefore tried plotting against $1/L^4$ instead, as shown in the bottom-left panel  of \Fref{fig:wcas-M1}, and in this case the plot appears linear. The straight line is a simple linear fit to the data which intercepts the $y$-axis at
$\lambda_L = 1.74550025$ and we therefore conclude that $\lambda_S \approx 1.744550.$ This analysis indicates that the $1/L^3$ correction term is absent or at least has a very small amplitude. Finally in the bottom-right panel of  \Fref{fig:wcas-M1} we plot the estimator of $\lambda_S$ assuming $\lambda_L$ converges with correction terms $c_1/L + c_2/L^2+c_4/L^4,$ plotted against $1/L^6.$ For this plot we have used only the first 4 of the 20 extra approximate coefficients, as using more than this produces some ripples in the plot, indicating that the approximate coefficients are insufficiently precise for such an extreme extrapolation. The linear fit has an intercept at $\lambda_L = 1.745549827$ and hence we estimate  $\lambda_S \approx 1.7445498.$

\begin{figure}[ht!] 
\includegraphics[width=0.95\textwidth,angle=0]{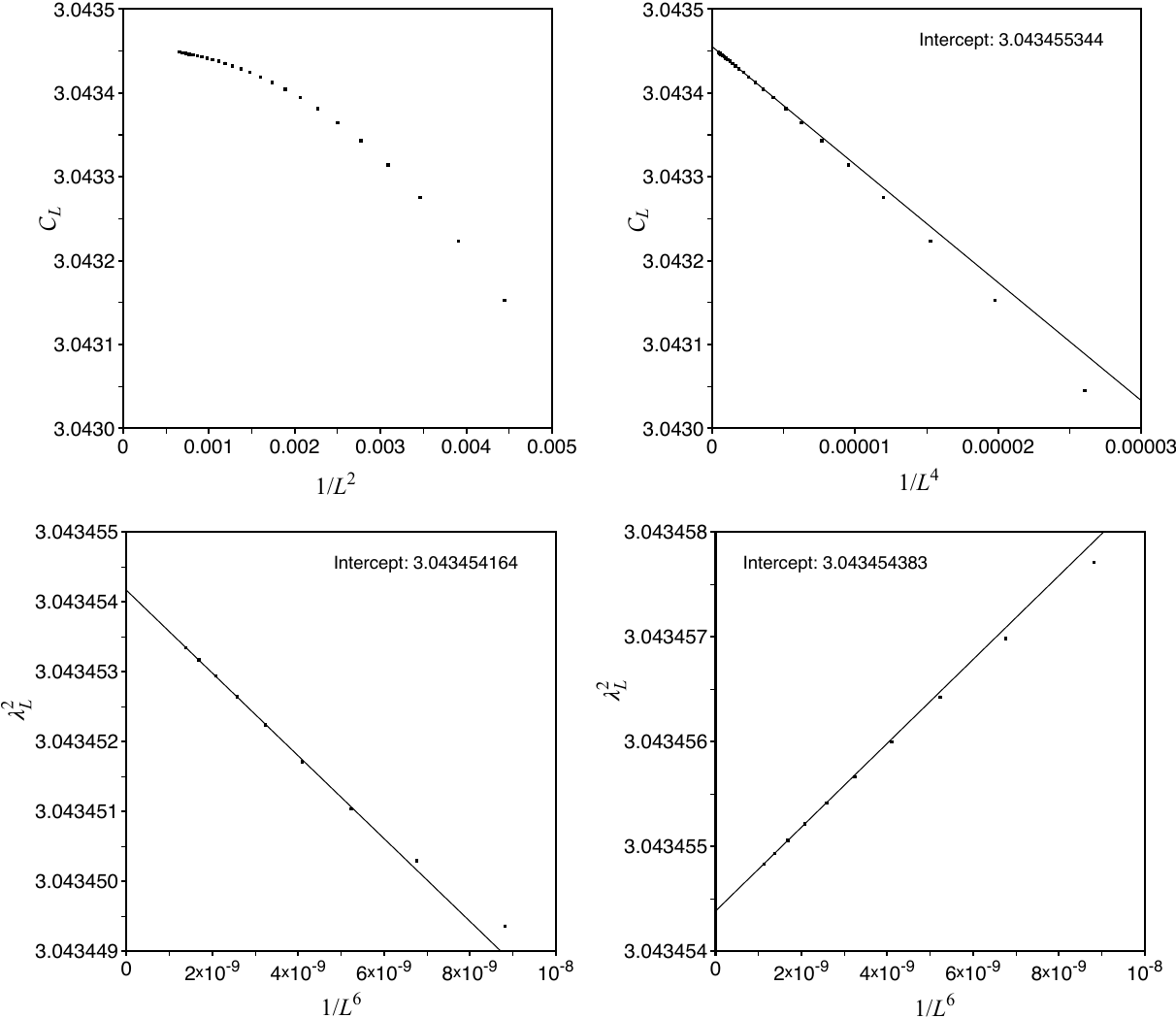} 
 \caption{\label{fig:wcas-M2} The top two panels show plots of ${\mathcal C}_L$  plotted against $1/L^2$ and $1/L^4$, respectively.
 The bottom panels show plots of the estimators $\lambda_L^2$ assuming correction terms $c_4/L^4$ (left panel) and  $c_2/L^2+c_4/L^4$ (right panel).} 
\end{figure}

\begin{figure}[ht!] 
\includegraphics[width=0.95\textwidth,angle=0]{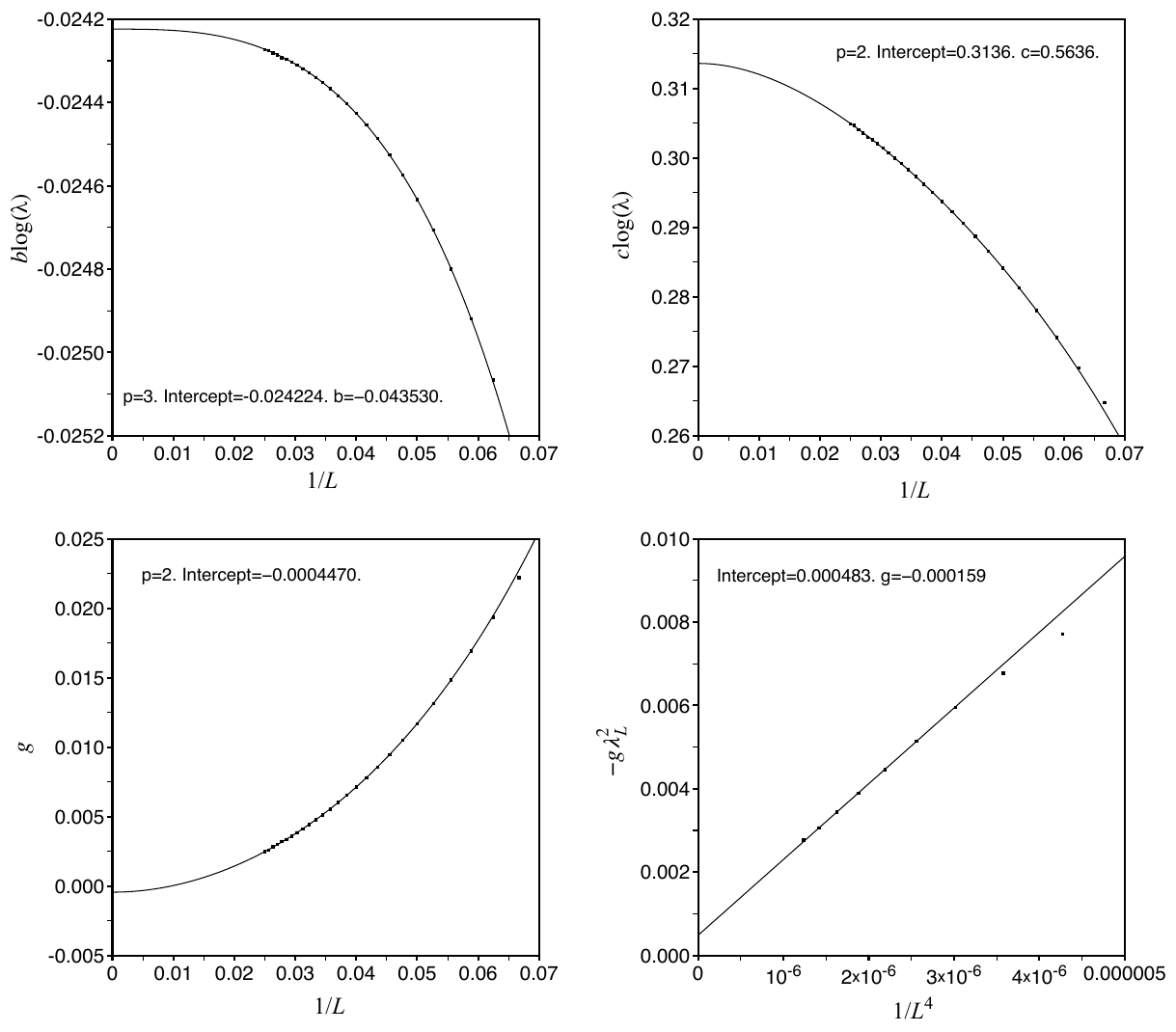}
\caption{\label{fig:wcas-para} Estimators of $b \log \lambda_S$, $c \log \lambda_S$, and $g$ from method P2 plotted against $1/L$,
and the estimator $-g\lambda_L^2$ from method P3 plotted against $1/L^4$.} 
\end{figure}

Next, we apply method M2 (\ref{eq:M2}) to the analysis of $C_L(1)$.
We show a plot of the ratios ${\mathcal C}_L \sim \lambda_L^2$  against $1/L^2$ in the top-left panel of \Fref{fig:wcas-M2}. It is seen to display considerable curvature, but can be visually extrapolated to $\lambda_S^2 \approx 3.04345$. In fact the curvature in the plot is suggestive of quadratic behaviour which would mean that ${\mathcal C}_L$  depends on $1/L^4$.  A plot of ${\mathcal C}_L$ against $1/L^4$ is shown in the top-right panel of \Fref{fig:wcas-M2} and we do indeed see a nice linear plot. The linear fit has intercept at $\lambda_L^2=3.043455344$ from which we estimate that $\lambda_S \approx 1.744550.$
As above, we now directly include powers of $1/L$ in the extrapolation. In the bottom-left panel of \Fref{fig:wcas-M2} we show the estimator of $\lambda_S^2$ assuming $\lambda_L^2$ converges with correction term $c_4/L^4,$ plotted against $1/L^6$ (intercept at 3.043454383).  Then in the bottom-right panel of \Fref{fig:wcas-M2} we show the estimator of $\lambda_S^2$ assuming $\lambda_L^2$ converges with correction terms $c_2/L^2+c_4/L^4,$ plotted against $1/L^6$ (intercept at 3.043454164). For these plot we have used only the first 4  approximate coefficients, for similar reasons to those given above.
The two extrapolated values of $\lambda_L^2$ are in excellent agreement and we obtain the precise estimate  $\lambda_S \approx 1.74454985.$
The clear indication from this analysis is that the parameter $g=0$. To further examine this point we plot in the last panel of \Fref{fig:wcas-para} the values of $c_2\sim -g\lambda_L^2$ from the analysis with correction terms $c_2/L^2+c_4/L^4$. Clearly the value of this parameter is very small and entirely consistent with $g=0$.

\begin{figure}[ht!] 
\includegraphics[width=0.95\textwidth,angle=0]{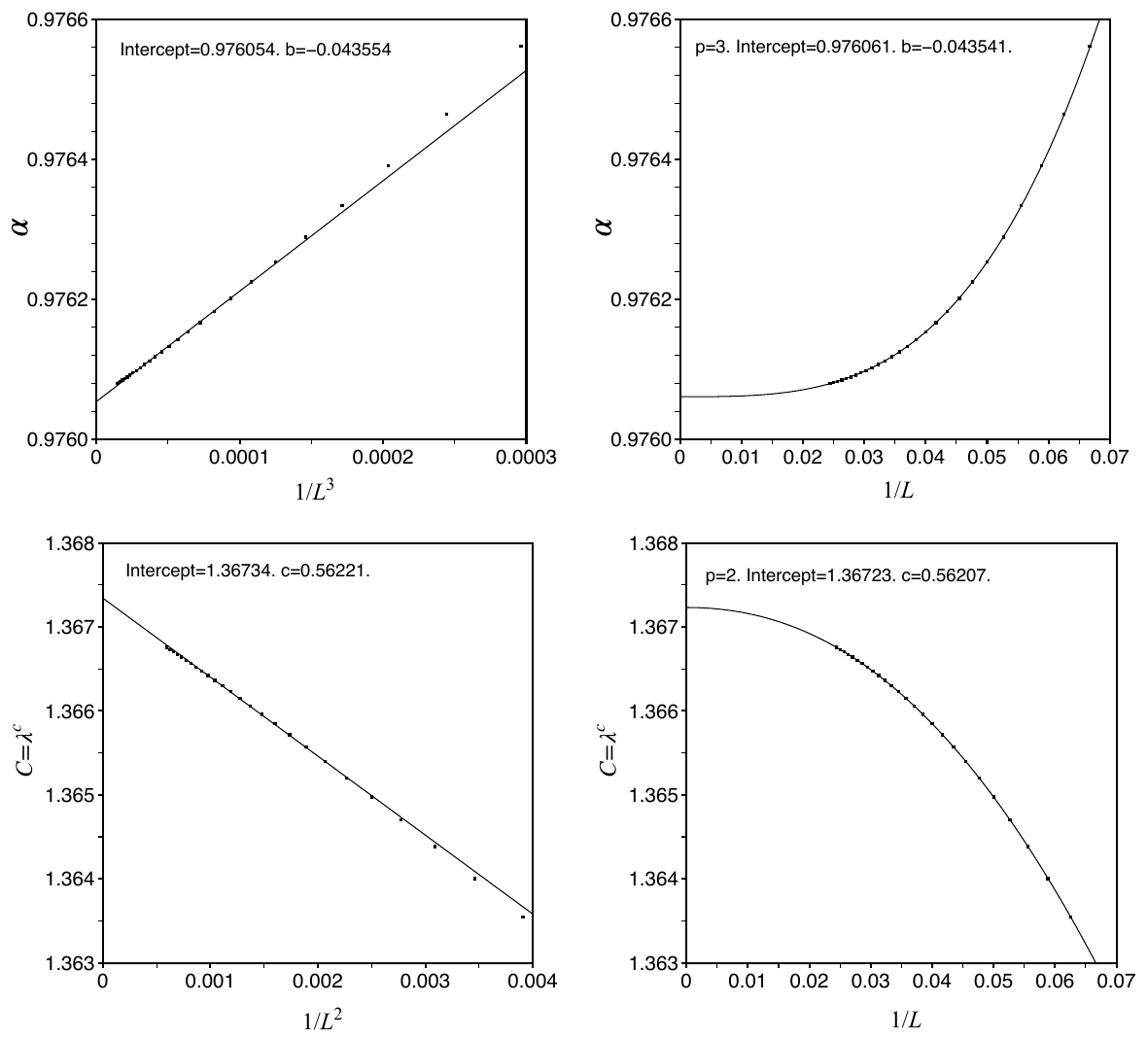}
\caption{\label{fig:wcas-P1} 
Ratios $d_L/d_{L-1} \sim \alpha$ plotted against $1/L^3$ (top-left) and the amplitude $C=\lambda^c$ plotted against $1/L^2$ (bottom-left). The panels on the right display the same data plotted against $1/L$ with the solid curve being a least-squares fit.} 
 \end{figure}

Next we will attempt to estimate the parameters $b,$ $c$ and $g$  in the assumed asymptotic form (\ref{eq:CLas}) by the  methods described in Section~\ref{sec:ana_met}. 
Using our best estimate of $\lambda_S=1.7445498,$ we first form the sequence $d_L = C_L(1)/\lambda_S^{L^2}.$  
Usually, ratios are plotted against $1/L,$ and the gradient of the linear plot gives a measure of the exponent $g.$ The ratio plot displays considerable curvature when plotted against $1/L,$ becoming approximately linear only when plotted against $1/L^3$ as shown in the in the top-left panel of \Fref{fig:wcas-P1}. This suggests that the coefficient of $1/L$ in the asymptotic expansion of the expression for the ratios is zero, or at least very small, that is $g\approx 0$. We estimate from this plot that $\alpha = 0.97605 \pm 0.00001,$ so that $b=\log \alpha/\log \lambda_S= -0.04355 \pm 0.00001.$ From the plot it is clear that there is some residual curvature. 

Next we performed a least-squares fit of the data to the form $c_0+c_3/L^3+c_4/L^4+c_5/L^5$ using the data-points from $L=20$ up to $L=35$ (we display the data from $L=15$ to 41) and the resulting plot is shown in the top-right panel. We estimate from this plot that $\alpha = 0.976061 \pm 0.000005,$ so that $b=\log \alpha/\log \lambda_S= -0.04354 \pm 0.00001.$
Finally, we use this latter value of $\alpha$ to estimate the value of the parameter $c,$ or equivalently, the amplitude $C,$ by observing that $d_L/\lambda_S^{bL} \sim \lambda_S^c \cdot L^g.$
We have argued that $g \approx 0,$ so that $d_L/\lambda_S^{bL} \sim \lambda_S^c.$ In the bottom-left panel of \Fref{fig:wcas-P1} we show a plot of the estimator of $C=\lambda_S^c$ plotted against $1/L^2$, from which we estimate $C=\lambda_S^c = 1.3673 \pm 0.001,$ or $c=0.5622 \pm 0.0005.$ As before we next did a least-squares fit of the data, but now to the form $c_0+c_2/L^2+c_3/L^3+c_4/L^4$, which we display in the bottom-right panel. We estimate  $C=\lambda_S^c = 1.36723 \pm 0.0001,$ or $c=0.56207 \pm 0.00005.$

We now turn to the second method P2 to obtain estimates of the parameters $b \log{\lambda_S},$ $c \log{\lambda_S},$ and $g.$ As was the case above, these estimators have a lot of curvature when plotted against $1/L$. Hence we plotted against integer powers $p$ of $1/L$ until we found a value for which approximate linearity was achieved and we then performed a least-squares fit to the data to the form $c_0+c_p/L^p+c_{p+1}/L^{p+1}+c_{p+2}/L^{p+2}$. 
Plots of these against $1/L$ are shown in the first three panels of  \Fref{fig:wcas-para}.
From these plots, we estimate $b\log{\lambda_S} = -0.02422 \pm 0.00002,$ or $b = -0.04353 \pm 0.00002,$ $c \log{\lambda_S} = 0.314 \pm 0.001,$ or $c = 0.564 \pm 0.002,$ and $g \approx 0.$ The agreement between the two methods is excellent and well within quoted confidence limits. 

Accordingly we conclude that our analysis has provided overwhelming numerical evidence that the conjectured asymptotic form 
$$ C_L(1) \sim \lambda_S^{L^2+bL+c}\cdot L^g,$$ 
is correct. For SAWs crossing a square we estimate (conservatively) that the parameters take the values $\lambda_S = 1.7445498 \pm 0.0000012,$  $b=-0.04354 \pm 0.0001,$ $c=0.5624 \pm 0.001,$ and $g=0.000 \pm 0.005.$

\begin{table}
\begin{center}
\begin{tabular}{|lll|}
 \hline
 \multicolumn{1}{|c}{$L$} &
 \multicolumn{1}{c}{Singularity} &
 \multicolumn{1}{c|}{Exponent} \\
 \hline
0 & 0.3285739(14)& $-0.99989(27)$   \\
 \hline
1  & 0.32857478(95)& $-1.00010(18)$ \\
  \hline
2        & 0.32857481(99)& $-1.00012(20)$\\
 \hline
3        & 0.3285745(10)& $-1.00007(21)$ \\
 \hline
4   & 0.3285746(24)& $-1.00004(42)$ \\
\hline
5   & 0.3285730(49)& $-0.9998(10)$ \\
 \hline
6   & 0.3285745(18)& $-1.00004(38)$ \\
 \hline
 \end{tabular}
 \end{center}
 \caption{\label{tab:da}
 Estimates of the singularity and exponent of the sequence for the ratios of walks crossing a square series. The estimates are from third order differential approximants with various degrees $L$ of the inhomogeneous polynomial.}
 \end{table}

Next we use differential approximants to analyse the series $\RGf(z)=\sum R_L z^L.$ We show the results of the analysis, using 3rd order differential approximants (and of course only the exactly known 27 terms) in \Tref{tab:da}. From this we estimate the radius of convergence as $z_c=1/\lambda_S^2=0.3285735 \pm 0.000001,$ which gives $\lambda_S = 1.744551 \pm 0.000003.$ This is slightly less precise than the ratio methods. The estimates for the critical exponent are clearly supportive of  $\RGf(z)$ having a simple pole adding even more evidence
to the validity of the assumed asymptotic form. 

We also tried using Pad\'e approximants to estimate $\lambda_S$. In \Tref{tab:pade} we list some estimates of $\lambda_S$ obtained
from $P_{m,n}(z)$  Pad\'e approximants to  $\RGf(z)$ by calculating  the real roots of the denominator polynomial $Q_n(z)$, 
finding the smallest positive root to obtain an estimate of   $\lambda_S$.
It is clear that this method works just fine but it is, perhaps not surprisingly, at least an order of magnitude less accurate than differential approximants
let alone the ratio methods. Hence we shall not consider this method  or  differential approximants any further. 

\begin{table}
\begin{center}
\begin{tabular}{|cc|cc|cc|}
 \hline
 $(n,m)$ & Root & $(n,m)$ & Root & $(n,m)$ & Root  \\ \hline
(8,8) &    1.7445242860 & (8,10) &    1.7445439450 & (8,12) &    1.7445454270 \\
 (10,8) &    1.7445415440 & (10,10) &    1.7445441380 & (10,12) &    1.7445488060 \\
 (12,10) &    1.7445497750 & (12,12) &    1.7445487670 & (12,14) &     1.7445489710 \\
 (13,11) &    1.7445488890 & (13,13) &    1.7445491150 &  (14,12) &    1.7445491730 \\
 \hline
 \end{tabular}
 \end{center}
 \caption{\label{tab:pade}
 Estimates of $\lambda_S$ obtained from Pad\'e approximants.}
 \end{table}
 
\begin{figure}[ht!] 
\centerline{\includegraphics[width=\textwidth,angle=0]{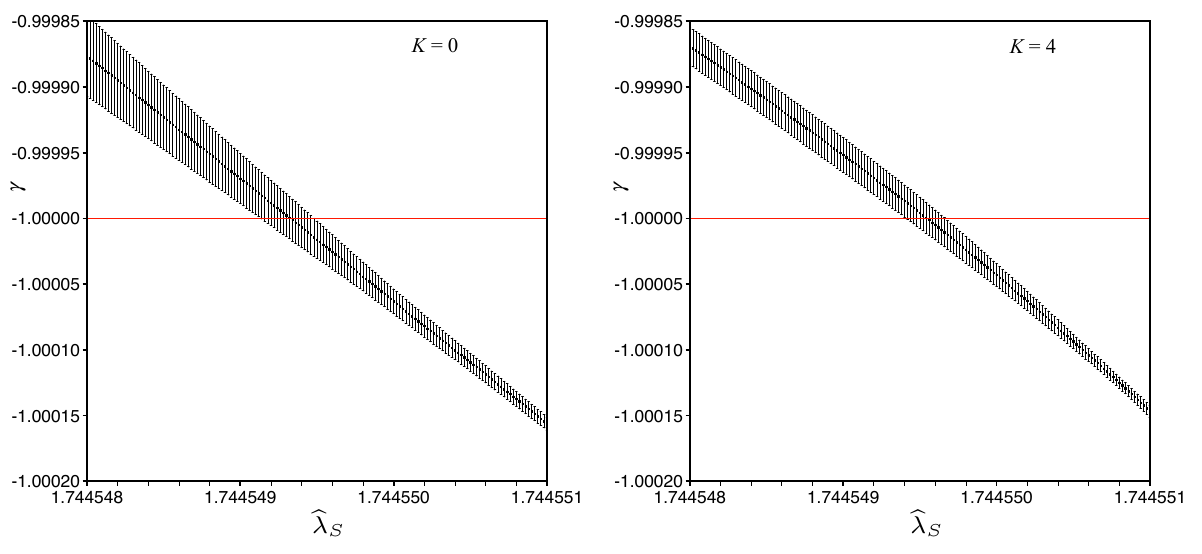} }
\caption{\label{fig:wcas-bda} Biased estimators for the critical exponents $\gamma$ of $\RGf(z)$ plotted against the
biasing value $\widehat{\lambda}_S$.} 
\end{figure}

 Finally, we turn to the analysis of $\RGf(z)$ using biased differential approximants (see \ref{app:bda}). 
 We pick a biasing value $\widehat{\lambda}_S$ and  force the differential approximants to have a singularity of order 1 at $z_c=1/\widehat{\lambda}_S^2$. 
 We calculate many ($> 100$) 3rd order biased differential approximants with an inhomogeneous polynomial of degree $K$, such that the number of required terms of the approximants $N \geq 22$. 
 Each approximant in turn provides us with an estimate of the critical exponent $\gamma$, which we confidently conjecture has the value $-1$. 
 From all of these $\gamma$ estimates we discard the outlying 10\% on either side. The remaining estimates are used to calculate the mean and standard deviation. 
 This procedure is then repeated for different values of $\widehat{\lambda}_S$ so as to cover the full range  of values within our estimated range $\lambda_S = 1.7445498 \pm 0.0000012.$ 
 In \Fref{fig:wcas-bda} we show a plot of the $\gamma$ estimates (with error-bars) as a function of $\widehat{\lambda}_S$ for the two cases where
 the degree of the inhomogeneous polynomial is 0 and 4, respectively. We notice that the curve of exponent estimates intersects $\gamma=-1$ over a very narrow range (smaller than the error estimate on  $\lambda_S$). Obviously it is very tempting to try and use this to provide an even more precise
 estimate of  $\lambda_S$. One may say that $\lambda_S$ could be estimated from the crossing with an error given by the width of the range over
 which error-bars on the exponent estimates overlap with  $\gamma=-1$ (or perhaps a factor of two or three times this range). However, this
 is a very new method and we are not yet confident that it is a valid method for obtaining more accurate estimates of critical points in cases where
 the exponent is known exactly. In particular we have no real idea of how to confidently estimate an error-bar. All we are willing to say at
 the moment is that it appears to be a promising method that warrants further detailed investigation.

\subsection{Polygons crossing a square}
\label{sec:pcas}

Using the algorithm described in Section~\ref{sec:algo} we calculated
$P_L(1)$ to lattice size $L=26$ and we then used the method of series extension to obtain a further 30 approximate terms.  

\begin{figure}[ht!] 
 \includegraphics[width=0.95\textwidth,angle=0]{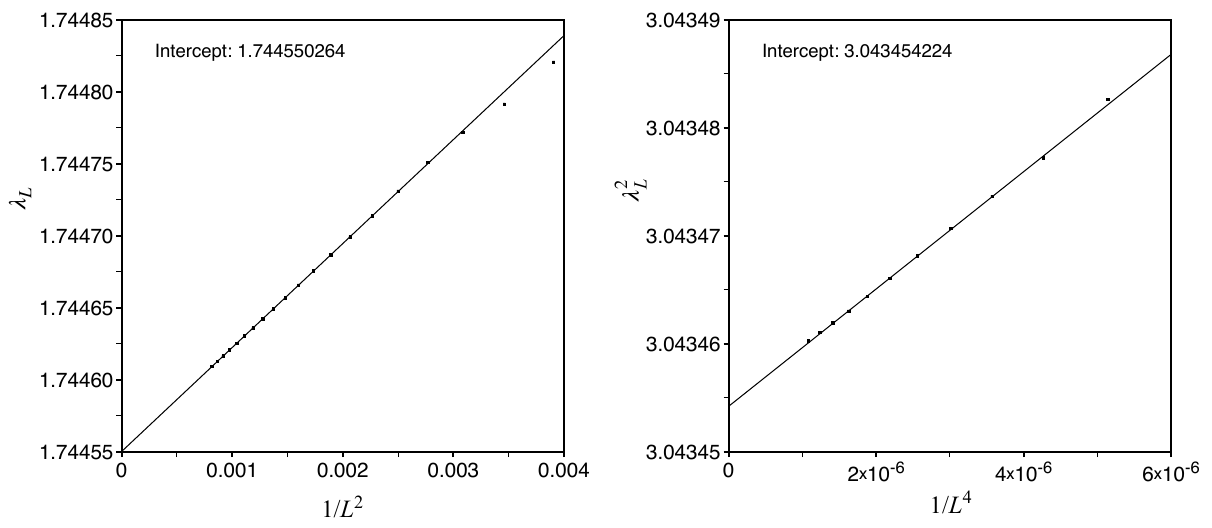}
\caption{\label{fig:pcas-lambda} Estimators of $\lambda_S$ from method M1 plotted against $1/L^2$ and $\lambda_S^2$ from method M2 plotted against $1/L^4$. } 
\end{figure}

We first estimated $\lambda_S$ by method M1, that is extrapolating the sequence $\lambda_L = P_{L}(1)^{1/L^2}$ against $1/L.$ There was some curvature in the plot, so we extrapolated against $c_0+c_1/L+c_2/L^2+c_3/L^3$. In this case the estimates appear fairly straight when plotted against $1/L^2$ as shown in the left panel of \Fref{fig:pcas-lambda}.   From this plot we estimate that
 $\lambda_S= 1.744550\pm 0.000005$. 
 Next we used method M2, that is we looked at the ratio of ratios. We extrapolated against  $c_0+c_2/L^2+c_3/L^3,$ and plotted this against $1/L^4$  as shown in the right panel of \Fref{fig:pcas-lambda}.
   This allowed us to  estimate $\lambda_S^2=3.043454\pm 0.000003,$  and hence  $\lambda_S=1.7445498\pm 0.0000008,$ in agreement with the previous analysis.
     
\begin{figure}[ht!] 
 \includegraphics[width=0.95\textwidth,angle=0]{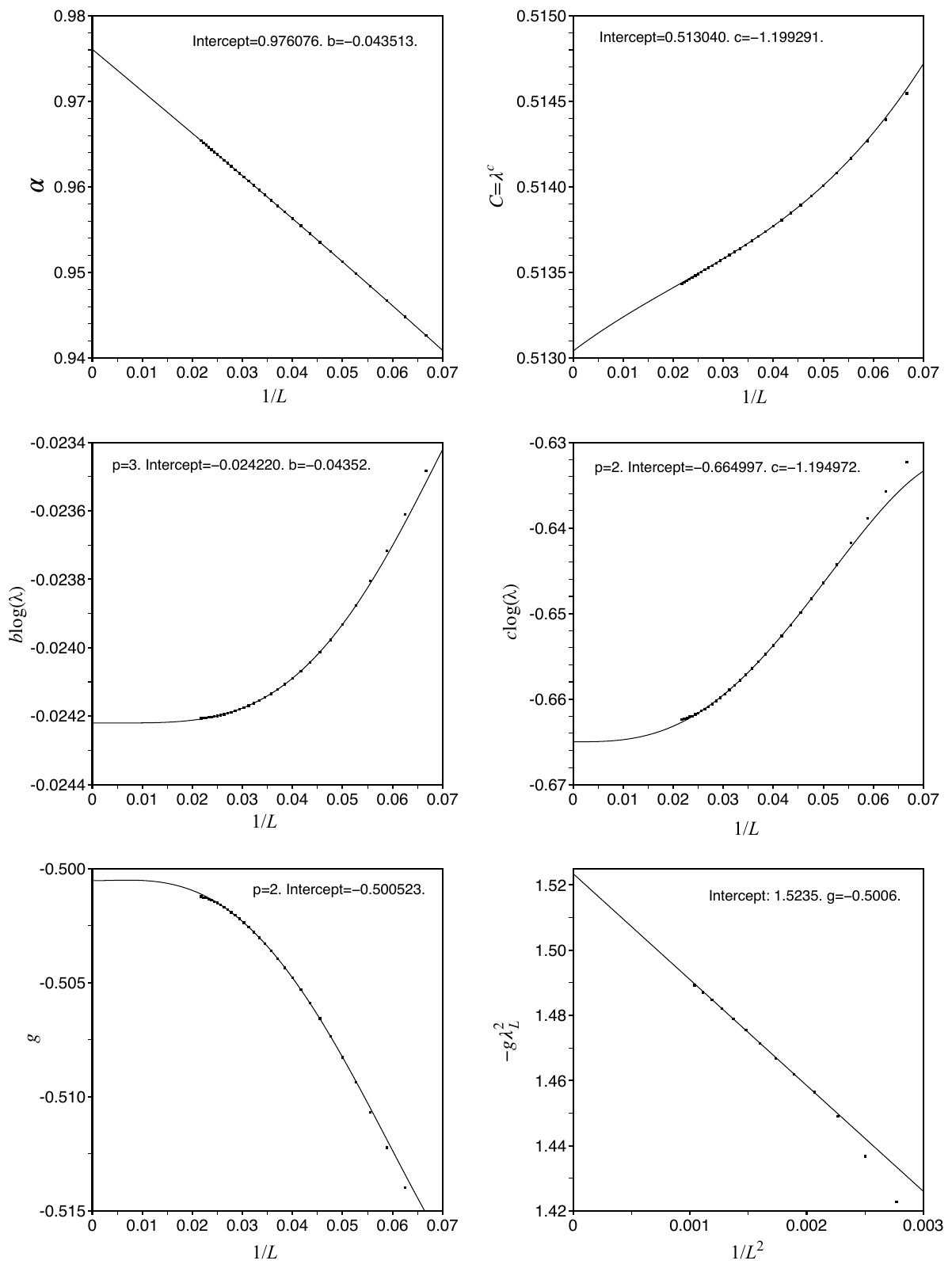}
\caption{\label{fig:pcas-para} Plots of the estimators for parameters $\alpha$ and
$C=\lambda_S^c$ from method P1, $b \log \lambda_S$, $c \log \lambda_S$, and $g$ from method P2
and the estimator $-g\lambda_L^2$ from method P3 for polygons crossing a square.}
\end{figure}

We estimated the values of the sub-dominant terms by method P2 and  we also estimated $g$ by method P3. The various plots are shown in \Fref{fig:pcas-para}.   In this way we estimate $b\log{\lambda_S} \approx -0.02422,$ or $b \approx -0.04352,$ $c \log{\lambda_S} \approx -0.6665$ or $c \approx -1.195,$ and $g \approx -0.5005$. From the estimate 
 $-g\lambda_S^2 \approx 1.5235$ we get $g \approx -0.5006$. We conjecture with some confidence that $g=-\frac12$, exactly. Using our best estimate for $\lambda_S$ and our conjecture for the exact value of $g$ we then turned to method P1. The plot of the estimator for $\alpha$ is close to linear against $1/L$, but to account for small correction we used a least-squares cubic fit in $1/L$ (solid curve) and found from the intercept that $\alpha = \lambda_S^b \approx 0.9761$ and hence $b\approx -0.04351$. We next make use of the intercept value from the $\alpha$-plot to estimate $c$. We look at the quantity $C=\lambda_S^c \sim d_L/(\alpha^L L^g)$, plot it against $1/L$, and use a cubic least-square fit to estimate the intercept $C\approx 0.5130$ and
 hence $c\approx -1.199$. The parameter estimates from the various method are in good agreement and clearly it seems that $b$ has the same value as for walks crossing a square.
    
\subsection{Walks spanning a square}
\label{sec:wcas-span}

We calculated $C_L(1)$ to lattice size $L=26$ and we then used the method of series extension to obtain a further 30 approximate terms. The plots used to estimate the parameters of this model are shown in \ref{app:numana} \Fref{fig:wcas-span}. We estimate $\lambda_S \approx 1.74455$ from method M1 using a fourth degree estimator and $\lambda_S^2 \approx 3.043455$ (and hence  $\lambda_S \approx 1.744550$) from method M2 fitting to a cubic polynomial. We estimated the values of the sub-dominant terms by method P2, and we also estimated $g$ by method P3. We estimate $b \approx -0.0435,$  $c \approx 0.603,$ and $g \approx 1.74$. From the estimate 
 $-g\lambda_S^2 \approx -5.33$ we get $g \approx 1.75$. It seems reasonable to  conjecture that $g=\frac74$, exactly. We finally used this value of $g$ in method P1 and we found $b\approx -0.0435$ and  $c\approx 0.4088$. Our estimates for $b$ and $g$ are in agreement and $b$ again has the same value as for walks crossing a square, but there is quite a variation in our estimates of $c.$

\section{Walks crossing a domain of the hexagonal lattice.}
\label{sec:wcah}

\begin{figure}[ht!] 
\centerline{\includegraphics[width=0.9\textwidth,angle=0]{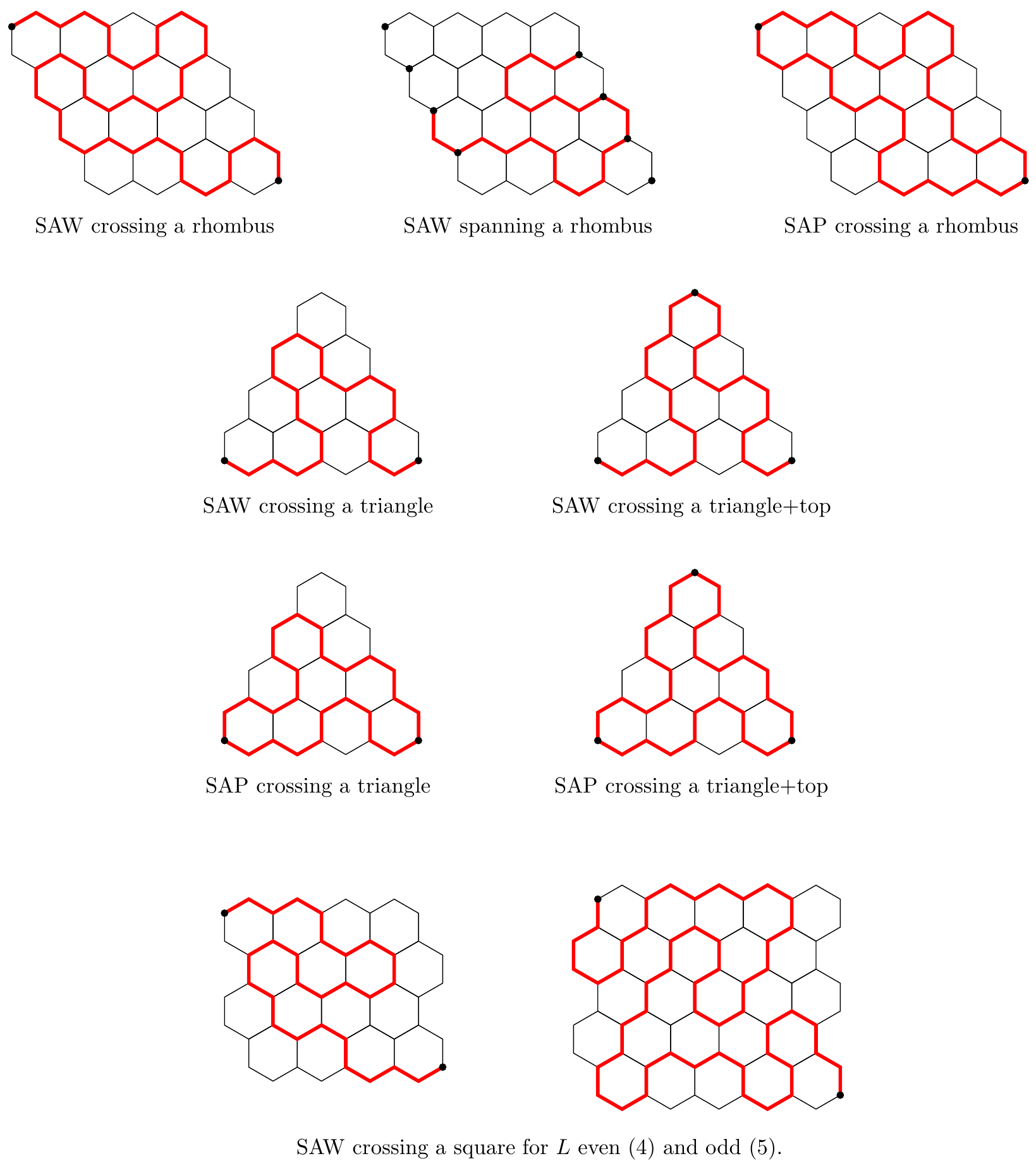} }
\caption{Classes of walk and polygon configurations investigated.} 
\label{fig:hexprob}
\end{figure}

In this section we study walks and polygons crossing a specified domain of the hexagonal lattice. We study several different variants of the problem, including both SAWs and SAPs, on triangular, rhomboidal domains and and SAWs on square domains. These are illustrated in \Fref{fig:hexprob}. We expect that the number of walks $C_L(1)$ for all these cases will have the asymptotic form (\ref{eq:CLas}). More specifically
the number of walks should have dominant asymptotic growth determined by 
$$C_L(1) \sim \kappa^{\textrm{\# vertices in domain}}.$$ 
The number vertices in a  triangular domain of size $L$ is  $L^2+4L+1$, while there are $2L^2+4L-1$ vertices in the rhomboidal and square domains.
Hence, we expect there to be a common growth constant $\lambda_H$ such that
 $\kappa = \lambda_H$ for the triangular domain and $\kappa=  \lambda_H^2$ for the rhomboidal and square domains. The other parameters,
 $b$, $c$ and $g$, in the asymptotic form (\ref{eq:CLas}) may differ from problem to problem.

In all cases we used the method of series extension to obtain further {\em approximate} terms from the known exact terms by using differential approximants to predict further coefficients, as described in \ref{app:pred}. How many further terms can be obtained varies from problem to problem and in each case we take all predicted coefficients whose spread among estimates (as measured by 1 standard deviation) is less then 1 part in $10^5.$ In this way we expect the least accurate coefficients to be accurate to around 1 part in $10^5.$
As a consequence, we expect that simple ratio plots will be smooth and indistinguishable from those obtained by the exact coefficients. However when we use more elaborate calculations, such as extrapolating against a polynomial in $1/L,$ that operation magnifies the errors. This is made manifest by smooth plots starting to display irregularities. Accordingly, we cut off such values, and don't use these less accurate coefficients in those plots.
To be more specific, if we extend a series by, say, 60 terms, we will use them all in a ratio plot, but when fitting to say, $c_0+c_1/L+c_2/L^2+c_3/L^3$, we may only use the first 30 extra coefficients. Method M2 is particularly sensitive and we could often only make use of as few as 4 of the approximate terms.

 \subsection{Self-avoiding walks crossing a triangle.}

\begin{figure}[ht!] 
\includegraphics[width=0.95\textwidth,angle=0]{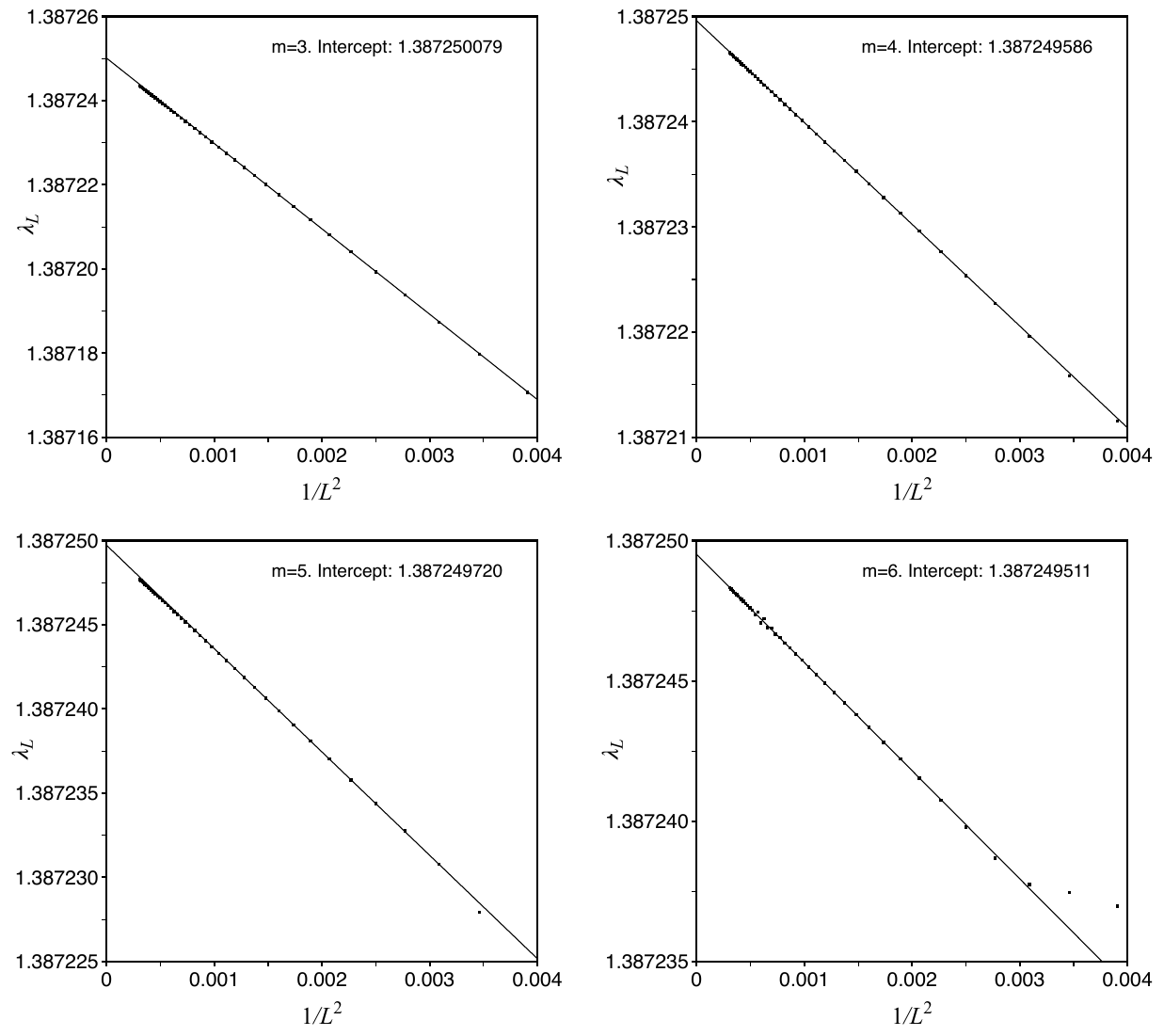}
\caption{\label{fig:wcah-tri-M1} Estimators of $\lambda_H$ from method M1 
when fitting against polynomials in $1/L$ of degree $m=3$ to 6 for SAWs crossing a triangle.} 
\end{figure}

\begin{figure}[ht!] 
\centerline{\includegraphics[width=0.95\textwidth,angle=0]{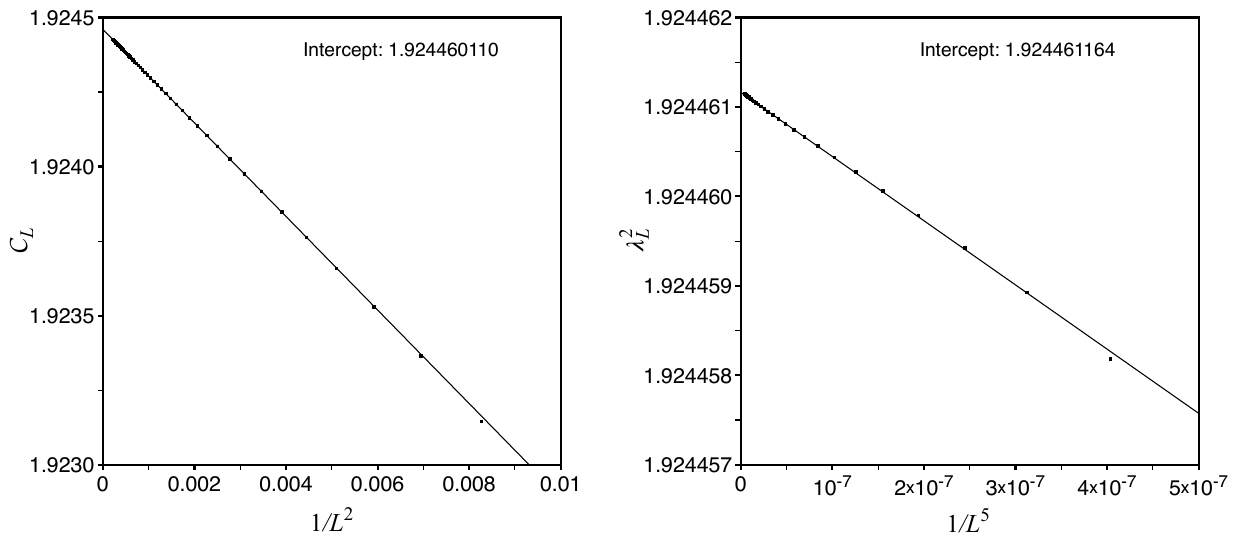} }
\caption{${\mathcal C}_L$ plotted against $1/L^2$ and the estimator $\lambda_L^2$ from method M2 with a cubic fit plotted against $1/L^5$ for SAWs crossing a triangle.} 
\label{fig:wcah-tri-M2}
\end{figure}

 The paths we are counting are shown in \Fref{fig:hexprob}. Using the algorithm described in Section~\ref{sec:algo} we calculated
$C_L(1)$ to lattice size $L=27$ and we then used the method of series extension to obtain a further 60 terms.   
We first estimated $\lambda_H$ by method M1, that is extrapolating the sequence $\lambda_L = C_{L}(1)^{1/L^2}$ against $1/L.$ There was some curvature in the plot, so we extrapolated against $c_0+c_1/L+\cdots+c_m/L^m,$  which allowed us to make a rather precise estimate,  
 $\lambda_H= 1.3872495\pm 0.0000005$. We show, in \Fref{fig:wcah-tri-M1}, just how well-converged this data is.
  We next considered  the sequence $\{{\mathcal C}_L\}$ which plotted against $1/L^2$  is an almost straight line. We then fitted the sequence to $c_0 + c_2/L^2+c_3/L^3.$ This gave exceptionally good apparent precision, allowing for a very precise estimate.
 We estimate $\lambda_H^2 = 1.9244612 \pm 0.0000002$, or $\lambda_H=1.38724951 \pm 0.00000001$. The plots are shown in \Fref{fig:wcah-tri-M2}.

\begin{figure}[ht!] 
\centerline{\includegraphics[width=0.95\textwidth,angle=0]{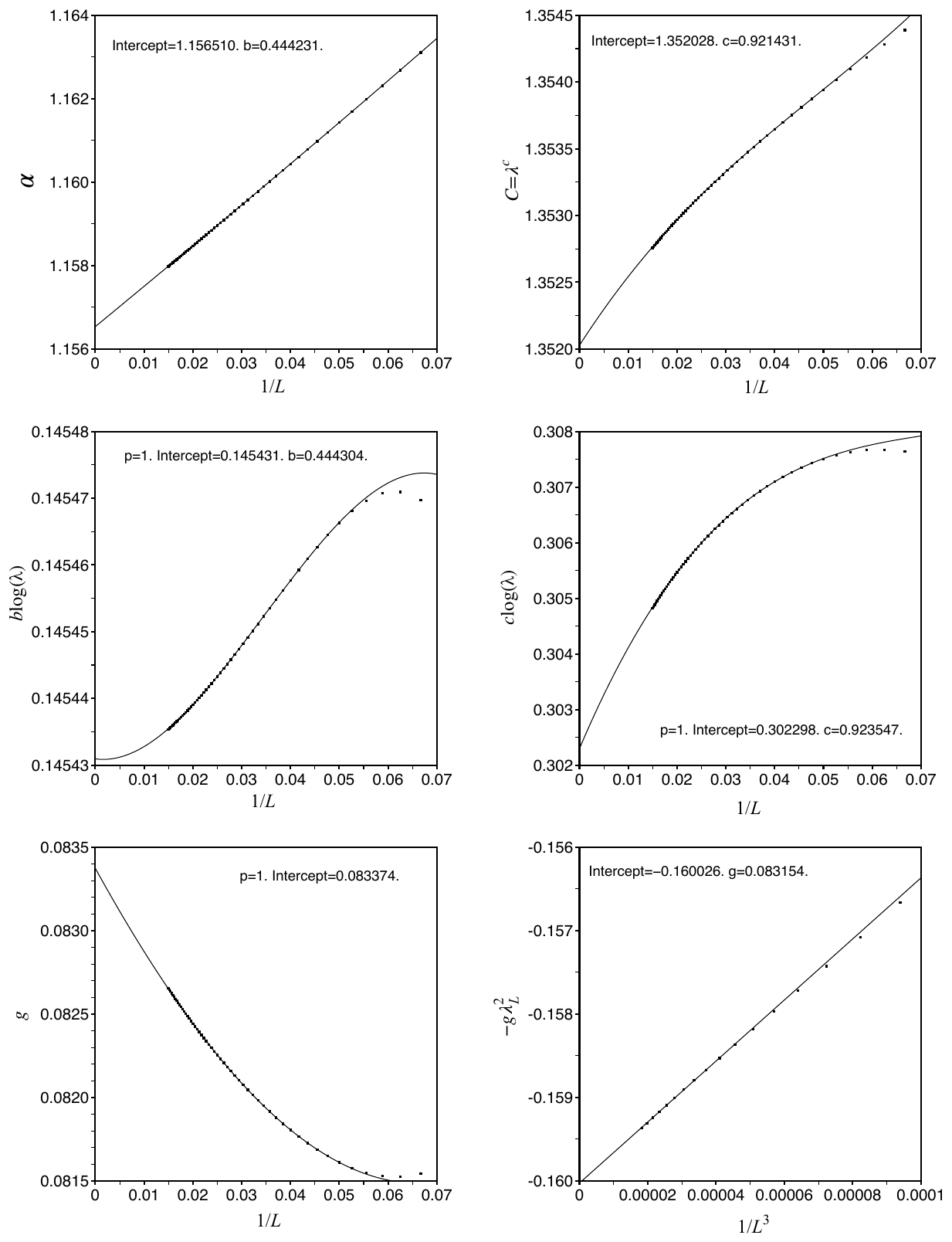} }
\caption{Estimators of $\alpha$ and $C=\lambda^c$ from method P1, $b \log \lambda_S$, $c \log \lambda_S$, and $g$ from method P2,
and the estimator $-g\lambda_L^2$ from method P3 for SAWs crossing a triangle.} 
\label{fig:wcah-tri-para}
\end{figure}

We estimated the values of the sub-dominant terms by method P2, fitting successive coefficients to $$\log{d_L} \sim b\log(\lambda_H)L + c\log(\lambda_H) + g \log{L},$$ 
and we estimated $-g\lambda_H^2$ from the cubic fit to  the sequence $\{{\mathcal C}_L\}$. The relevant plots are shown in \Fref{fig:wcah-tri-para}.
In this way we estimate $b \approx 0.4443,$  $c \approx 0.924,$ $g \approx 0.0834,$ and   $-g\lambda_H^2 \approx -0.1602,$ so $g \approx 0.0832,$  which is suggestive of the exact fraction $1/12.$ This exponent value was then used in method P1 from which we estimate $b\approx 0.4442$ and $c\approx 0.9214$ in good agreement with the results of method P2.
 
\begin{figure}[ht!] 
\centerline{\includegraphics[width=\textwidth,angle=0]{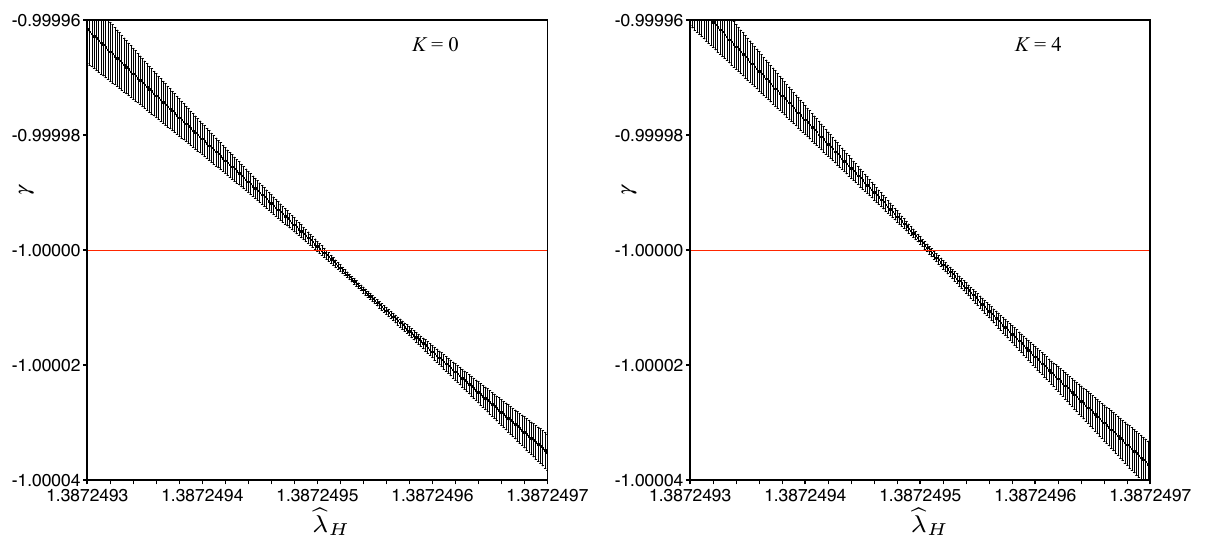} }
\caption{\label{fig:wcah-tri-bda} Biased estimates for the critical exponent $\gamma$ of $\RGf(z)$ plotted against the
biasing value $\widehat{\lambda}_H$ for SAWs crossing a triangle.} 
\end{figure}

 Finally we display in \Fref{fig:wcah-tri-bda} the results from a biased differential approximant analysis of  $\RGf(z)$. The biased estimates of $\gamma$ cross the value $-1$ in a very narrow range very close to our estimate $\lambda_H\approx 1.38724951$
 from the previous analysis.

We therefore conclude that for SAWs crossing a triangular domain of the hexagonal lattice we have found very firm numerical evidence that the conjectured asymptotic form~(\ref{eq:CLas}) is correct and we estimate that the parameters have the values $\lambda_H=1.38724951 \pm 0.00000005,$ $b=0.4443 \pm 0.001$, $c=0.923 \pm 0.005$, and $g=0.0833 \pm 0.0005$, where possibly $g=1/12$ exactly.

 \subsection{Self-avoiding walks crossing a rhombus.}
  
\begin{figure}[ht!] 
\includegraphics[width=0.95\textwidth,angle=0]{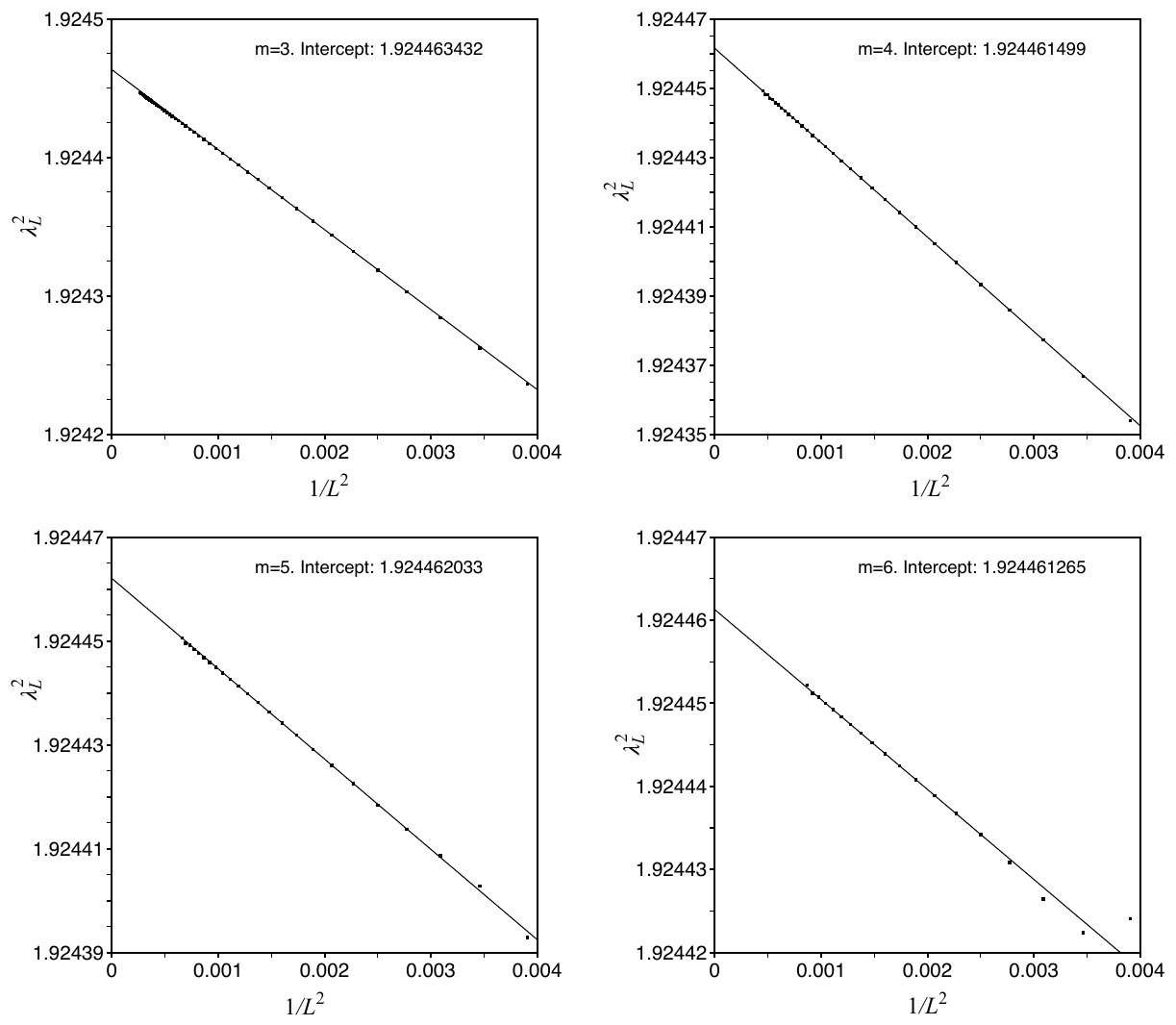}
\caption{\label{fig:wcah-rhom-M1} Estimators of $\lambda_H$ from method M1 
when fitting against polynomials in $1/L$ of degree $m=3$ to 6 for SAWs crossing a rhombus.} 
\end{figure}

\begin{figure}[ht!] 
\centerline{\includegraphics[width=0.95\textwidth,angle=0]{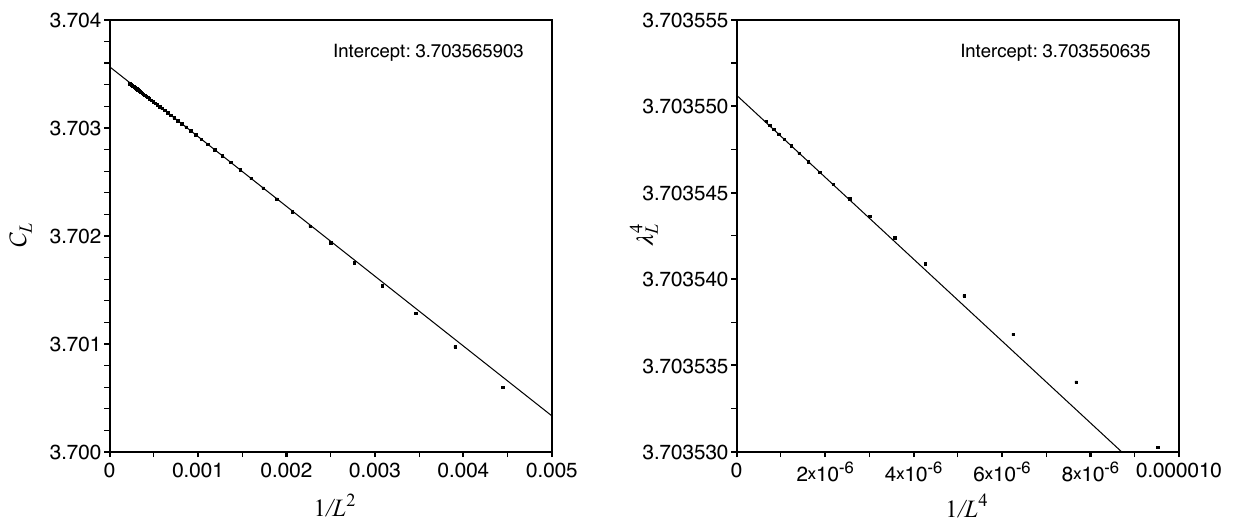} }
\caption{${\mathcal C}_L$ plotted against $1/L^2$ and the estimator $\lambda_L^4$ from method M2 with a cubic fit plotted against $1/L^4$ for SAWs crossing a rhombus.} 
\label{fig:wcah-rhom-M2}
\end{figure}

The paths we are counting are shown in \Fref{fig:hexprob}. We calculated
$C_L(1)$ to lattice size $L=26$ and then extended this sequence by a further 50 terms.   We first estimated $\lambda_H^2$ using method M1 by extrapolating against $c_0+c_1/L+\cdots+c_m/L^m,$  as shown in \Fref{fig:wcah-rhom-M1} for $m=3$ to 6.
From this we estimate that  $\lambda_H^2= 1.924461\pm 0.000002$, or 
$\lambda_H= 1.3872494\pm 0.000008$. Next we used method M2 to obtain an estimate for 
$\lambda_H^4$. In \Fref{fig:wcah-rhom-M2} we show a plot of ${\mathcal C}_L$  plotted against $1/L^2$  and the estimates obtained by fitting the sequence $\{{\mathcal C}_L\}$  to $c_0 + c_2/L^2+c_3/L^3.$  
 We estimate $\lambda_H^4 = 3.7035506 \pm 0.0000006$, or $\lambda_H=1.38724948 \pm 0.00000006$. These estimates for $\lambda_H$ are consistent with the estimate obtained above for the triangular domain.

\begin{figure}[ht!] 
\centerline{\includegraphics[width=0.95\textwidth,angle=0]{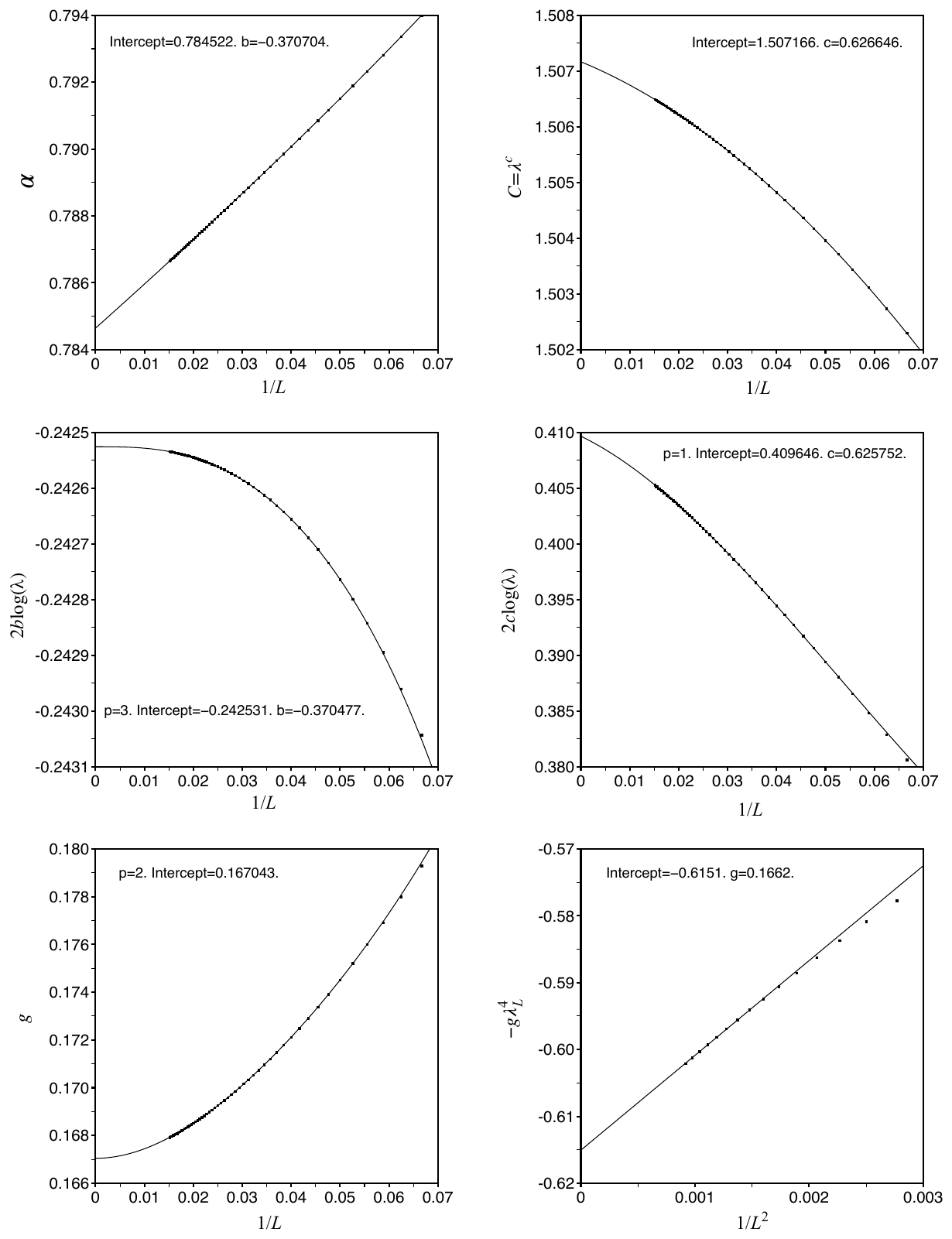} }
\caption{Estimators of $\alpha$ and $C=\lambda^c$ from method P1, $b \log \lambda_S$, $c \log \lambda_S$, and $g$ from method P2,
and the estimator $-g\lambda_L^2$ from method P3 for SAWs crossing a rhombus.} 
\label{fig:wcah-rhom-para}
\end{figure}

We estimated the values of the sub-dominant terms by method P2, fitting successive coefficients to $$\log{d_L} \sim 2b\log(\lambda_H)L + 2c\log(\lambda_H) + g \log{L},$$ 
and we estimated $-g\lambda^2$ from the cubic fit to  the sequence $\{{\mathcal C}_L\}$. The relevant plots are shown in \Fref{fig:wcah-rhom-para}.
  We estimate $b \approx -0.3705,$ $c \approx 0.6258,$ $g \approx 0.167,$ and   $-g\lambda_H^2 \approx -0.615,$ so $g \approx 0.166,$  which is suggestive of the exact fraction $1/6.$ From method P1 we then obtained the estimates $b\approx -0.3707$
and $c\approx 0.6266.$ 

\begin{figure}[ht!] 
\centerline{\includegraphics[width=\textwidth,angle=0]{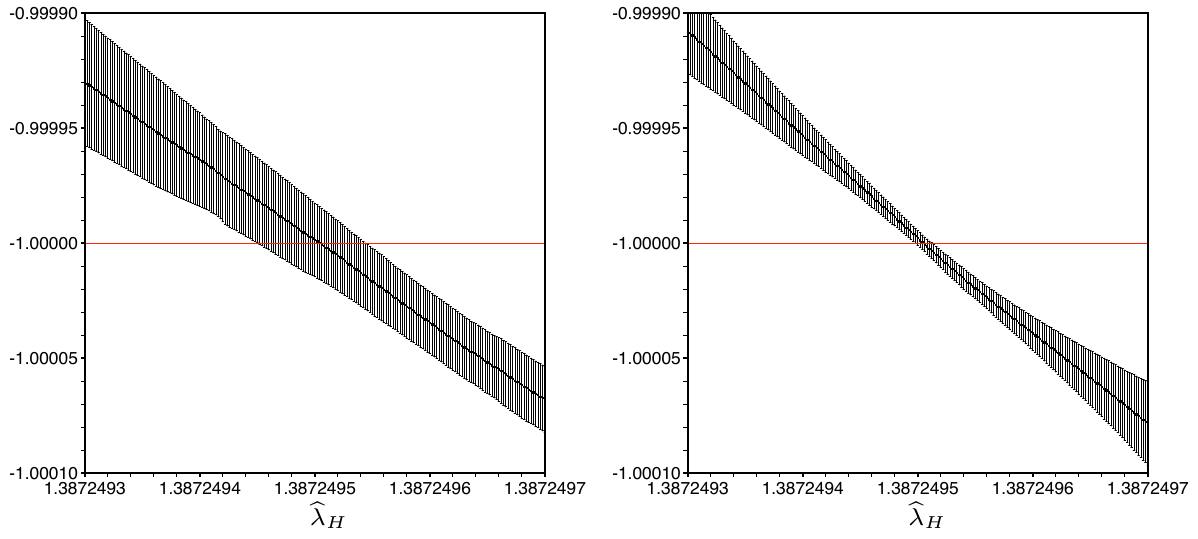} }
\caption{\label{fig:wcah-rhom-bda} 
Biased estimates for the critical exponent $\gamma$ of $\RGf(z)$ plotted against the
biasing value $\widehat{\lambda}_H$ for SAWs crossing a rhombus. }
\end{figure}

 Finally we display in \Fref{fig:wcah-rhom-bda} the results from a biased differential approximant analysis of  $\RGf(z)$. Once again we see that the biased estimates of $\gamma$ cross the value $-1$ in a narrow range contained within our best estimate $\lambda_H\approx 1.38724951 \pm 0.00000005$.

Hence, SAWs crossing a rhomboidal domain of the hexagonal lattice follows the conjectured asymptotic form~(\ref{eq:CLas}) with growth constant $\lambda_H^2$ 
and sub-dominant parameters $b=-0.3706 \pm 0.0005$, $c=0.6262 \pm 0.001$, and $g=0.167 \pm 0.002$, where possibly $g=1/6$.

\subsection{SAWs spanning a rhombus}

We have series to lattice size $L=26$ and we managed to obtain a further 37 approximate terms. Method M1 with a degree six polynomial extrapolation allowed us to make the estimate $\lambda_H^2 = 1.92446 \pm 0.00003$, or $\lambda_H=1.38725\pm 0.00001$.
From method M2 and P3 with a cubic fit we  estimate $\lambda_H^4 = 3.703551 \pm 0.000005,$ or $\lambda_H= 1.3872495  \pm 0.0000005,$ and $-g\lambda_H^4 \approx -6.18,$ so $g \approx 1.67.$  From method P2 we estimate $b \approx -0.3705,$  $c \approx 1.44,$ and $g \approx 1.675,$ in precise agreement with the estimate from  method P3. 
We suggest that perhaps $g=\frac53$. Method P1 then yielded the estimates
$b\approx -0.3706$ in agreement with the previous estimate and $c\approx 1.56$ somewhat large but still consistent with the estimate from method P2.
The plots can be seen in \Fref{fig:wcah-rhom-span}.

 \subsection{Polygons crossing a rhombus}
 
 We calculated series to lattice size $L=26$ and extended by 60 approximate terms. 
 Method M1 yielded the estimate $\lambda_H^2 = 1.92446 \pm 0.00001$ ($\lambda_H = 1.387249 \pm 0.000005$ ) and method M2 and P3
  $\lambda_H^4 = 3.7035515 \pm 0.0000015$ ($\lambda_H = 1.38724956 \pm 0.00000015$) and $-g\lambda_H^2 \approx 2.157,$ so $g \approx -0.583$, where  in each method we used a cubic extrapolation of the sequence.
This is clear evidence that the growth parameter $\lambda_H$ for polygons is the same as for SAWs, which is to be expected.    
 Method P2 was again used to estimate the values of the sub-dominant terms  and we estimate $b \approx -0.3705,$ $c \approx -1.0543,$ and $g \approx -0.583,$ in  agreement with the estimate from method P3. We hazard the guess that $g=-7/12$, exactly. From method P1 we then estimated $b\approx -0.3705$ and $c\approx -1.0529$. The plots can be seen in \Fref{fig:wcah-rhom-sap}.

 \subsection{Self-avoiding walks crossing a triangle and passing through the top vertex.}

We calculated  $C_L(1)$ to lattice size $L=26$ and extended the series by a further 60  approximate terms. We estimated $\lambda_H,$ by method M1 where an extrapolation of degree six allowed us to  estimate  $\lambda_H = 1.3872495 \pm 0.0000005.$ 
 Using method M2 we estimated $\lambda_H^2 = 1.9244611 \pm 0.0000001$ ($\lambda_H = 1.38724947 \pm 0.00000005$) and $-g\lambda_H^2 \approx -0.160,$ so $g \approx 0.0831,$ in 
 agreement with the value found for SAWs crossing a triangle.
 We estimated the values of the sub-dominant parameters by method P2 and   we found $b \approx 0.4443,$  $c \approx -1.7861,$ and $g \approx 0.0833,$ in good agreement with the estimate given immediately above, and suggestive of an exact fraction $1/12.$ 
 Method P1 yielded the estimates $b\approx 0.4442$ and $c\approx -1.7891$ in agreement with the previous estimates. The plots can be seen in \Fref{fig:wcah-tri-top}.

 \subsection{Polygons in a triangle.}
 
 We define $P_L(1)$ as the number of polygons in a triangular domain passing through two of the three corner vertices as shown in \Fref{fig:hexprob}.
 We calculated series to lattice size $L=26$ and obtained a further 50 approximate terms. 
 We first estimated $\lambda_H,$ by method M1 and found $\lambda_H = 1.387245 \pm 0.000002.$  Method M2 gave good apparent precision, giving  us the precise estimate $\lambda_H^2 =1.924461 \pm 0.000001$ ($\lambda_H = 1.3872494 \pm 0.0000004$) and from method P3 we found $-g\lambda_H^2 \approx 1.282,$ so $g \approx -0.666,$  which is  very suggestive of the exact fraction $-2/3.$ We estimated the values of the sub-dominant terms from method P2 and  found  $b \approx 0.4443,$  $c \approx -1.394,$ and $g \approx -0.666,$ in total agreement with the estimate of $g$ given immediately above. Method P1 gave $b\approx 0.4443$ and $c\approx -1.380$ in good agreement with the previous analysis. The plots can be seen in \Fref{fig:wcah-tri-sap}.
  
 \subsection{Polygons in a triangle passing through the top vertex.}
 
 We define $P_L(1)$ as the number of polygons in a triangular domain passing through all three corner vertices of the domain as illustrated in \Fref{fig:hexprob}. The series was calculated to lattice size $L=26 $ and extended by  40 further approximate terms. Method M1 yielded  the 
 estimate $\lambda_H = 1.38725 \pm 0.00001$ and method M2 gave $\lambda_H^2 = 1.924461 \pm 0.000002$ ($\lambda_H = 1.38725 \pm 0.00001$) while method P3 gave $-g\lambda_H \approx 1.284,$ so $g \approx -0.667$.  Method P2  resulted in the estimates  $b \approx 0.4443,$  $c \approx -4.106,$ and $g \approx -0.667$. We conjecture $g=-2/3$ exactly. Method P1 resulted in the estimates $b\approx 0.4443$ and $c\approx -4.091$ in agreement with the previous results. The plots can be seen in \Fref{fig:wcah-tri-sap-top}.

\subsection{SAWs crossing a square}

The paths we are counting are shown in \Fref{fig:hexprob}.  We calculated series to lattice size $L=24$ and we extended the series by a
further  25 approximate terms.  
A consequence of the lattice geometry is that different paths had to be counted according as the lattice size $L$ was odd or even, as shown in \Fref{fig:hexprob}.   This induced a period-2 oscillation in the ratios and other parameters. To accommodate this we redefined the ratios as the square-root of the ratio of alternate terms. That is to say, the ratio $r_L = \sqrt{C_L(1)/C_{L-2}(1)}.$
Similarly, when attempting to  extrapolate the sequence $\lambda_L = C_{L}(1)^{1/L^2}$ against a polynomial in $1/L$ we used tuples of alternate terms, rather than successive terms. Even after this adjustment the estimates of $\lambda_L$ showed some parity 
effects. Hence we decided to look at the average of consecutive terms, that is, $(\lambda_L+\lambda_{L-1})/2$. Similar changes were made for all the other parameter estimators.
The resulting plots are shown in \Fref{fig:wcah-square}

This allowed us to make the precise estimate $\lambda_H^2 = 1.924461 \pm 0.000005$ ($\lambda_H = 1.387249 \pm 0.000003$) from a cubic fit to the
sequence $\{C_{L}(1)^{1/L^2}\}$.
  When we fitted the sequence $\{{\mathcal C}_L\} $ to $c_0 + c_2/L^2+c_3/L^3,$  strong period-2 oscillations required a redefinition, so we defined  $${\mathcal C}^{\textrm{\scriptsize *}}_{L} := \left( \frac{C_{L-2}(1)\cdot C_{L+2}(1)}{C_L(1)^2}\right )^{1/4}.$$ As was the case for WCAS
  this redefined sequence of ratios showed linearity when plotted against $1/L^4$ hence suggesting that $g=0$ in this case as well. We therefore extrapolated the new sequence against $c_0 + c_4/L^4,$ and from a plot against $1/L^7$ we made the estimate  $\lambda_H^4 = 3.7035505 \pm 0.0000015,$ or $\lambda_H =  1.3872495  \pm 0.0000002$.  
 
 We also estimated the values of the sub-dominant terms by method P2 appropriately
 altered to deal with parity effects. In that way we estimated  $b \approx -0.3765,$ $c \approx 0.736,$ and $g \approx 0.003,$ in agreement with $g=0$.
Finally, we used also method P3 to estimate $g$. Here we fitted the sequence 
$\{{\mathcal C}^{\textrm{\scriptsize *}}_{L}\}$ to $c_0 + c_2/L^2+ c_4/L^4,$ 
so that $c_2$ becomes an estimator for $-g\lambda^4_H$ and from the plot we estimate
that $-g\lambda^4_H\approx -0.00240$ which again is consistent with the conjecture that $g=0$ exactly.

\section{Conclusion}
\label{sec:conc}
 For SAWs crossing a square on the square lattice, we conjecture that $$ C_L(1) \sim \lambda_S^{L^2+bL+c}\cdot L^g,$$ where $\lambda_S = 1.7445498 \pm 0.0000012,$ $b=-0.04354 \pm 0.0001,$
 $c=0.5624 \pm 0.0005,$ and $g=0.000 \pm 0.005.$
 
 For SAWs crossing a closed, connected, convex region on the hexagonal lattice we similarly conjecture $C_L(1) \sim \lambda_H^{L^2},$ where our best estimate of $\lambda_H=1.38724951 \pm 0.00000005.$
 For a number of combinatorial problems associated with SAWs on the hexagonal lattice, the growth constant is either known or conjectured. We have not been able to even guess a potential algebraic expression for $\lambda_H$ that is remotely plausible.
 
 We show in \Tref{tab:results} our estimates of the parameters $b,$ $c,$ and $g$ for the various geometries and path types we have studied, as well as the conjectured exact values of the exponent $g.$ 
 For the hexagonal lattice, it is seen that the parameter $b$ takes one of two values. The value $b \approx -0.3705$ is associated with the rhomboidal geometry, while the value $b\approx 0.4444$ is associated with the triangular geometry. All the exponents $g$ appear to be multiples of $1/12.$

 \begin{table}
 \caption{\label{tab:results} Estimates of the parameters $b,$ $c,$ and $g$ when fitting to the assumed asymptotic form $C_L(1) \sim \lambda_S^{L^2+bL+c}\cdot L^g,$ for the square lattice, $C_L(1) \sim \lambda_H^{2(L^2+bL+c)}\cdot L^g,$ for the hexagonal lattice on non-triangular domains, and $C_L(1) \sim \lambda_H^{L^2+bL+c}\cdot L^g,$ on triangular domains.}
\begin{center}
\begin{tabular}{|l|r|r|rr|}
 \hline \hline
 Geometry and lattice & \multicolumn{1}{c|}{$b$} & \multicolumn{1}{c|}{$c$}
 & \multicolumn{2}{c|}{$g$ and conjecture}\\
 \hline
 \hline
{\bf  Square lattice} &&&&\\

SAWs crossing a square &$-0.04354$ & 0.5624 & 0 & 0\\
SAWs spanning a square &$-0.04354$ & 0.5 & 1.75 & 7/4 \\
SAPs crossing a square &$-0.04354$ & $-1.197$ & $-0.5000$ & $-1/2$ \\
 \hline
 \hline
{\bf  Hexagonal lattice} &&&& \\ 
 SAWs crossing a rhombus & $-0.3706$ & 0.626 & 0.167 & 1/6 \\
 SAWs spanning a rhombus & $-0.3704$ & 1.78& 1.667 & 5/3 \\
 SAPS crossing a rhombus & $-0.3705$ & $-1.052$ & $-0.583$ & $-7/12$ \\
 SAWs crossing a triangle& 0.4443 & 0.923 & 0.0833 & 1/12 \\
 SAWs crossing a triangle $+$ top vertex & 0.4443 & $-1.787$ & 0.0833 & 1/12 \\
 SAPs crossing a triangle & 0.4444 & $-1.387$ & $-0.666$ & $-2/3$ \\
SAPs crossing a triangle $+$ top vertex & 0.4443 & $-4.10$ & $-0.667$ & $-2/3$ \\
 SAWs crossing a square & $-0.3765$ & 0.736 & 0.003 & 0 \\
 
 \hline \hline

 \end{tabular}
 \end{center}
 \end{table}

\section{Resources}
\label{sec:resource}

The enumeration data and extended series for all problems studied in this paper,
some Maple worksheets used for the asymptotic analysis and  some of the source code used to calculate the exact coefficients can be found at our GitHub repository \url{https://github.com/IwanJensen/Self-avoiding-walks-and-polygons/tree/WCAS(H)}.
 
\section*{Acknowledgements}
We would to thank Nathan Clisby for many conversations about
the implementation of the perfect hashing algorithm which greatly assisted our efforts and for several helpful
suggestions about problems worth studying and for a careful reading of an early version of the manuscript.
AJG wishes to thank the ARC Centre of Excellence for Mathematical and Statistical Frontiers (ACEMS) for support.

\section*{References}


\newpage
\appendix

\section{Ratio Method}
\label{app:ratio}
 
The ratio method was perhaps the earliest systematic
method of series analysis employed,
and is still the most useful method when only a small number of terms are known. Given a series $\sum c_n z^n,$ which behaves as in eqn. (\ref{generic}), it is assumed that $\lim_{n \to \infty} c_n/c_{n-1}$ exists and is equal to the growth constant. For some combinatorial sequences such as classical pattern-avoiding permutations of length up to 5, this has been proved by Atapour and Madras \cite{AM14}.

From eqn. (\ref{asymp1}),
it follows that the {\it ratio} of successive terms
\begin{equation} \label{ratios}
r_n = \frac{c_n}{c_{n-1}}=\frac{1}{z_c}\left (1 + \frac{\gamma -1}{n} + {\rm o}\left (\frac{1}{n}\right )\right ).
\end{equation}
 It is then natural to plot the successive ratios $r_n$ against $1/n.$
If the correction terms ${\rm o}(\frac{1}{n})$ can be ignored\footnote{For a purely algebraic singularity eqn. (\ref{generic}), with no confluent terms, the correction term will be ${\rm O}(\frac{1}{n^2}).$}, such a plot will be linear,
with gradient $\frac{\gamma-1}{z_c},$ and intercept $\mu=1/z_c$ at $1/n = 0.$

Linear intercepts $l_n$ eliminate the $O\left ( \frac{1}{n} \right )$ term in eqn. (\ref{ratios}), so in the case of a pure power-law singularity, one has
$$l_n := nr_n - (n-1)r_{n-1} = \mu \left (1+ \frac{c}{n^2} + O\left (\frac{1}{n^3} \right ) \right ).$$

Various refinements of the method can be readily derived. If the critical point
is known exactly, it follows from eqn. (\ref{ratios}) that estimators of the exponent
$\gamma$ are given by
$$ \gamma_n: = n(z_c\cdot r_n-1)+1 = \gamma +  {\rm o}(1).$$

If the critical point is not known exactly, one can still estimate the exponent $\gamma.$ From eqn. (\ref{ratios}) it follows that 
\BE \label{eq:exp}
\delta_n := 1+n^2\left ( 1-\frac{r_n}{r_{n-1}} \right )= \gamma +  {\rm o}(1).
\EE

Similarly, if the exponent $\gamma$ is known, estimators of the growth constant  $\mu$
are given by  $$\mu_n =  \frac{n r_n}{n+\gamma-1}= \mu + {\rm o}(1/n).$$

\section {Differential approximants}
\label{app:da}

The generating
functions  of some problems in enumerative combinatorics are sometimes algebraic, such as that for $Av(1342)$ pattern-avoiding permutations, sometimes D-finite, such as with $Av(12345)$ pattern-avoiding permutations,
sometimes differentially algebraic, and sometimes transcendentally transcendental.
The not infrequent occurrence of D-finite solutions was the origin of the method of {\em differential approximants}, a very successful method of series analysis for analysing power-law singularities \cite{G89}.

The basic idea is to approximate a generating function $F(z)$ by solutions
of differential equations with polynomial coefficients. That is to say, by D-finite ODEs. The singular behaviour
of such ODEs is  well documented
(see e.g. \cite{Forsyth02,Ince27}), and the singular points and
exponents are readily calculated from the ODE. 

The key point for series analysis is that even if {\em globally} the function is not describable by a solution
of such a linear ODE (as is frequently the case) one expects that
{\em locally,} in the
vicinity of the (physical) critical points, the generating
function is still well-approximated by a solution of a linear ODE, when the singularity is a generic power law (\ref{generic}).

An $M^{th}$-order differential approximant (DA) to a function $F(z)$  is formed by matching
the coefficients in the polynomials $Q_k(z)$ and $P(z)$ of degree $N_k$ and $K$, respectively,
so that the formal solution of the $M^{th}$-order inhomogeneous ordinary differential equation
\BE \label{eq:ana_DA}
\sum_{k=0}^M Q_{k}(z)\left(z\frac{{\rm d}}{{\rm d}z}\right)^k \tilde{F}(z) = P(z)
\EE
agrees with the first $N=K+\sum_k (N_k+1)$ series coefficients of $F(z)$. 

Constructing such ODEs only involves
solving systems of linear equations. The function
$\tilde{F}(z)$ thus agrees with the power series expansion of the (generally unknown)
function $F(z)$ up to the first $N$ series expansion coefficients.
We normalise the DA by setting $Q_M(0)=1,$ thus leaving us with $N$ rather
than $N+1$ unknown coefficients to find. The choice of the differential operator $z\frac{{\rm d}}{{\rm d}z}$ in (\ref{eq:ana_DA}) forces the origin to be a regular singular point. The reason for this choice is that most lattice models with holonomic solutions, for example, the free-energy of the two-dimensional Ising model, possess this property. However this is not an essential choice.

From the theory of ODEs, the singularities of $\tilde{F}(z)$ are approximated by zeros
$z_i, \,\, i=1, \ldots , N_M$ of $Q_M(z),$ and the
associated critical exponents $\gamma_i$ are estimated from the indicial equation. If there is only a single root at $z_i$  this is just
\BE \label{eq:ana_indeq1}
\gamma_i=M-1-\frac{Q_{M-1}(z_i)}{z_iQ_M ' (z_i)}.
\EE
Estimates of the critical amplitude $C$ are rather more difficult to make, involving the integration of the differential approximant. For that reason the simple ratio method approach to estimating critical amplitudes is often used, whenever possible taking into account higher-order asymptotic terms \cite{GJ09}.

Details as to which approximants should be used and how the estimates from many approximants are averaged to give a single estimate are given in \cite{GJ09}. Examples of the application of the method can be found in \cite{G15}. In that work, and in this, we reject so-called {\em defective} approximants, typically those that have a spurious singularity closer to the origin than the radius of convergence as estimated from the bulk of the approximants. Another  method sometimes used is to reject outlying approximants, as judged from a histogram of the location of the critical point (i.e. the radius of convergence) given by the DAs. It is usually the case that such distributions are bell-shaped and rather symmetrical, so rejecting approximants beyond two or three standard deviations is a fairly natural thing to do.

\subsection{Biased differential approximants} \label{app:bda}

If the critical point $z_c$ is known exactly (or very accurately) one may try to obtain improved numerical
estimates for the exponents by forcing the differential equation (\ref{eq:ana_DA})  to have a singular point at $z_c$, that is 
one may look at {\em biased differential approximants}.   In \cite{Jensen16} we developed a new method in which we form 
biased approximants by multiplying the derivatives in (\ref{eq:ana_DA}) by appropriate ``biasing polynomials''. 
This allows us to bias in such a manner that the singularity at $z_c$ is of order $q\leq K$. Let

\begin{equation}\label{eq:ana_bias}
F_k(z) = \left(z\frac{{\rm d}}{{\rm d}z}\right)^k F(z) \quad \textrm{and} \quad G_k(z) = (1-z/z_c)^{q_k} F_k(z),
\end{equation}
where $q_k=\max(q+k-M,0)$.  With this definition we have that $G_k= (1-z/z_c)^qF_k(z)$, while subsequent
lower order derivatives have ``biasing polynomials'' of degree decreasing in steps of 1 (until 0). 
Then we form biased differential approximants (BDA) such that
 \begin{equation} \label{eq:BDA}
P(z)+\sum_{k=0}^M \widehat{Q}_{k}(z)G_k(z)   = O(z^{N+1}).
\end{equation}
For biased approximants the degree of the polynomial multiplying the $k$'th derivative still have degree 
$N_k$ such that the degrees of $\widehat{Q}_{k}(x)= N_k-q_k$ and the number of unknown coefficients  is 
$\widehat{N}=K+1 +  \sum_k (N_k-q_k+1).$

\section{Coefficient prediction}
\label{app:pred}
In \cite{G16} we showed that the ratio method and the method of differential approximants  work serendipitously together in many cases, even when one has stretched exponential behaviour, in which case neither method works particularly well in unmodified form. 

To be more precise, the method of differential approximants (DAs)  produces ODEs which, by construction, have solutions whose series expansions agree term by term with the known coefficients used in their construction. Clearly, such ODEs implicitly define {\em all}  coefficients in the generating function, but if $N$ terms are used in the construction of the ODE, all terms of order $z^{N}$ and beyond will be approximate, unless the exact ODE is discovered, in which case the problem is solved, without recourse to approximate methods.

It is useful to construct a number of DAs that use all available coefficients, and then use these to predict subsequent coefficients. Not surprisingly, if this is done for a large number of approximants, it is found that the predicted coefficients of the term of order $z^n,$ where $n > N,$ agree for the first $k(n)$ digits, where $k$ is a decreasing function of $n.$ We take as the predicted coefficients the mean of those produced by the various DAs, with outliers excluded, and as a measure of accuracy we take the number of digits for which the predicted coefficients agree, or the standard deviation. These two measures of uncertainty are usually in reasonable agreement.

Now it makes no logical sense to use the approximate coefficients as input to the method of differential approximants, as we have used the DAs to obtain these coefficients. However there is no logical objection to using the ({\em approximate}) predicted coefficients as input to the ratio method. Indeed, as the ratio method, in its most primitive form, looks at a graphical plot of the ratios, an accuracy of 1 part in $10^4$ or $10^5$ is sufficient, as errors of this magnitude are graphically unobservable. 

Recall that, in the ratio method one looks at {\em ratios} of successive coefficients. We find that the ratios of the approximate coefficients are predicted with even greater precision than the coefficients themselves by the method of DAs. That is to say, while a particular coefficient and its successor might be predicted with an accuracy of 1 part in $10^p$ for some value of $p$, the {\em ratio} of these successive coefficients is frequently  predicted with significantly greater accuracy (the precision being typically improved by a factor varying between 2 and 20).

The DAs use all the information in the coefficients, and are sensitive to even quite small errors in the coefficients. As an example, in a recent study of some self-avoiding walk series, an error was detected in the eighteenth significant digit in a new coefficient, as the DAs were much better converged without the last, new, coefficient\footnote{Given 69 terms of the square-lattice self-avoiding walk series, the 70th term is predicted by 4th order ODEs to be $4190893020903935057 \times 10^{12}.$ The actual coefficient is $4190893020903935054619120005916,$ which differs in the nineteenth digit. An error in the eighteenth digit was thus discovered during development. Several other less dramatic examples are known where lower-order errors have been discovered by this means.}.  The DAs also require high numerical precision in their calculation. In favourable circumstances, they can give remarkably precise estimates of critical points and critical exponents, by which we mean up to or even beyond 20 significant digits in some cases. Surprisingly perhaps, this can be the case even when the underlying ODE is not D-finite. Of course, the singularity must be of the assumed power-law form.

Ratio methods, and direct fitting methods, by contrast are much more robust. The sort of small error that affects the convergence of DAs would not affect the behaviour of the ratios, or their extrapolants, and would thus be invisible to them. As a consequence, approximate coefficients are just as good as the correct coefficients in such applications, provided they are accurate enough. We re-emphasise that, in the generic situation (\ref{generic}), ratio type methods will rarely give the level of precision in estimating critical parameters that DAs can give. By contrast, the behaviour of ratios can more clearly reveal features of the asymptotics, such as the fact that a singularity is not of power-law type. This is revealed, for example, by curvature of the ratio plots \cite{G15}.

As an example, consider the OGF for $Av(12453)$ PAPs (see OEIS \cite{OEIS} A116485). This is known to order $x^{38}.$ Let us take the coefficients to order $x^{16}$ and use the method of series extension described above to predict the next 22 ratios, so that we can compare them to the exact ratios. The results, based on 3rd order differential approximants, are shown in \Tref{tab:serpred}. For the first predicted ratio, $r_{18},$ the discrepancy is in the 10th significant digit. For the last predicted ratio, $r_{39}$, the error is in the 5th significant digit. This level of precision is perfectly adequate for ratio analysis.

\begin{table}[htbp!]
   \begin{center}
   \topcaption{Ratios $r_{18}$ to $r_{39}$ actual and predicted from the coefficients of $Av(12453),$ with percentage error shown.}
   \begin{tabular}{|l|l|l|} \hline
   Predicted ratios & Actual ratios & Percentage error\\ \hline
10.654655347& 10.65465504& $4.78 \times 10^{-7}$\\ 
10.828226522& 10.82822539&$1.04 \times 10^{-5}$\\
10.986854456& 10.98685140&$2.79 \times 10^{-5}$\\
11.132386843&11.13238007&$4.78 \times 10^{-5}$\\
11.266382111&11.26636895&$6.08 \times 10^{-5}$\\
11.390163118&11.39013998&$2.03 \times 10^{-4}$\\
11.504857930&11.50482182&$3.14 \times 10^{-4}$\\
11.611441483&11.61138359&$4.99 \times 10^{-4}$\\
11.710743155&11.71066190&$6.94 \times 10^{-4}$\\
11.803496856&11.80338255&$9.68 \times 10^{-4}$\\
11.890333733&11.89017822&$1.31 \times 10^{-3}$\\
12.048402545&12.04814337&$2.15 \times 10^{-3}$\\
12.120553112&12.12022972&$2.67 \times 10^{-3}$\\
12.188650126& 12.18824275&$3.34 \times 10^{-3}$\\
12.252994715&12.25252103&$3.87 \times 10^{-3}$\\
12.313939194&12.31336663&$4.65 \times 10^{-3}$\\
12.371707700&12.37104982&$5.32 \times 10^{-3}$\\
12.426619450&12.42581319&$6.49 \times 10^{-3}$\\
12.478784843&12.47787509&$7.29 \times 10^{-3}$\\
12.528486946&12.52743256&$8.41 \times 10^{-3}$\\
\hline
      \end{tabular}
   \label{tab:serpred}
   \end{center}
\end{table}

In practice we find that the more exact terms we know, the greater is the number of predicted terms, or ratios that can be predicted.

\section{Enumeration data}
\label{app:data}

\begin{landscape}
\thispagestyle{empty}
 \begin{table}
{\tiny
\begin{tabular}{ll}
\hline \hline
$L$ &  $C_L(1)$ \\ \hline

1  &  8 \\ 
2  &  95 \\ 
3  &  2320 \\ 
4  &  154259 \\ 
5  &  30549774 \\ 
6  &  17777600753 \\ 
7  &  30283708455564 \\ 
8  &  152480475641255213 \\ 
9  &  2287842813828061810244 \\ 
10  &  102744826737618542833764649 \\ 
11  &  13848270995235582268846758977770 \\ 
12  &  5613766870113075134552249300590982081 \\ 
13  &  6856324633418315229580098999727214234534626 \\ 
14  &  25264653780547704599613926971040640439380254497299 \\ 
15  &  281194924965510769640501069703642937039678809002355743600 \\ 
16  &  9461739046646537749639494171503923182753987897972167546351180871 \\ 
17  &  963236702020101408274810653629921860636656580683490560257709270360444788 \\ 
18  &  296872411379358777499142156584947972393781613934413706389772635139720532797697401 \\ 
19  &  277150300263332125727926989254635730407844207233646123561354535935393720183262709640734296 \\ 
20  &  784096265647396811778105941874438158236581845146768685766318151014460448963606598066808194055196391 \\ 
21  &  6725180841063080568765785521839331530600623203136984200976765831832263641839818443238635675098039099764477094 \\ 
22  &  174931600296771588816418921915331826961754552793606147578780164414287627531728146738237058492693139833462228085357900931 \\ 
23  &  13803603811254425104633152972993523761617439474917134222103400574517678544806707098426335287312812055811653812588064999045835964788 \\ 
24  &  3305148095303296700320144368689162420653300006202515254218029114864900324594717492699469036772928641130210225812040363529280982602180021971501 \\ 
25  &  2401952907672357993462515287034263569296810854353779576930606173996736217445824082165189903808422564905380792093559726262352646493540668423785540986958618 \\ 
26  &  5299107129769378506953456534389910056797761651575284727185718491325361350700020349029824701741232397885203596081145956599662000200438618578333326328843424545244470913 \\ 
\hline \hline
\end{tabular}
}
\caption{Number of SAWs spanning a square.} 
\end{table}

\begin{table}
{\tiny
\begin{tabular}{ll}
\hline \hline
$L$ &  $P_L(1)$ \\ \hline
1  &  1 \\ 
2  &  3 \\ 
3  &  42 \\ 
4  &  1799 \\ 
5  &  232094 \\ 
6  &  92617031 \\ 
7  &  115156685746 \\ 
8  &  442641690778179 \\ 
9  &  5224287477491915786 \\ 
10  &  188825256606226776728029 \\ 
11  &  20879416139356164466643759334 \\ 
12  &  7057757437924198729598570424130207 \\ 
13  &  7287699030020917172151307665469211016474 \\ 
14  &  22973720258279267139936821063450448822110219653 \\ 
15  &  220999541336018343231658363621596453585823579325485544 \\ 
16  &  6485093759718494344865537501691711476194821918864090506157759 \\ 
17  &  580338710138214792049192419944468721379579881619954352303395183377868 \\ 
18  &  158337812302865122325340454524668159260049140429114314750279637797162731935795 \\ 
19  &  131686133943477323496319974983490271815302632940624543675717883251973010678492927145164 \\ 
20  &  333791921301450408656424393731824932225524914478794139217214328043764176667483451057387588939581 \\ 
21  &  2578284699331238205287505462049410591202075986811965490195831413301513664200106021200647983846671319023376 \\ 
22  &  60681018617202345518945611945550350166677922700280807250098751378055134432090527432691244436095874301999620318286417 \\ 
23  &  4351075330271556361458913058062785859178198294438374222572342619944855786600569691060003274678918267681800765492921507255444030 \\ 
24  &  950435810029045769123624069823361419361696021093908583594503116871962553684187067543456237120685036487281955076387155649492347583783912357 \\ 
25  &  632407534045235304278897181408229137621456170029071955419475667299649868819055002950538251352503131227737994021488874040396259243064748293794055487122 \\ 
26  &  1281707896370751708653066922805265028882836851074044433082078379196572742914435468007626647333767206265847516495713522985546806840650483671342846200191630108286969 \\ 

\hline \hline
\end{tabular}
}
\caption{Number of SAPs crossing a square.} 
\end{table}

\end{landscape}

\begin{landscape}
\thispagestyle{empty}
 \begin{table}
{\tiny
\begin{tabular}{ll}
\hline \hline
$L$ &  $C_L(1)$ \\ \hline
1  &  2 \\ 
2  &  14 \\ 
3  &  316 \\ 
4  &  25092 \\ 
5  &  7374480 \\ 
6  &  8029311942 \\ 
7  &  32223151155864 \\ 
8  &  476605408516689238 \\ 
9  &  26016526700583361056456 \\ 
10  &  5246595079903462547245876694 \\ 
11  &  3911053741699230141571030313824664 \\ 
12  &  10780907768757190963361134040036893772360 \\ 
13  &  109919900687141309301630828947780890728732496678 \\ 
14  &  4146148169372563020871034877194447551275644544417216784 \\ 
15  &  578668580332775727107695799371628560927178835729875790606922120 \\ 
16  &  298872860145313265329322304090348192097227121631333193254451061450023212 \\ 
17  &  571292892753639610811496925540653319819009464854621261888736201676638277892860364 \\ 
18  &  4041877636548925601268934261053439777968614414770138847482643177563162891499826990868686710 \\ 
19  &  105849680445660298017662516167192274494877530131095615720184731073055676134641221548956561836515847160 \\ 
20  &  10261319175888813072109344281334022257660847729142398797395911985785352481803270582806576593011349057648597629702 \\ 
21  &  3682522861496742274013714098245794929775776187625314598131060860173699707921073898860823021154156312609385314847082364604336 \\ 
22  &  4892542075116215747349775890169094456449789602921450060431267745393588411359934920766964621175270271453676206611892541512628195569791000 \\ 
23  &  24065022635991318624332037902196644133241139050298562673155834537019737576953755257676129977770681689706285684286956275152985138134931365323467344612 \\ 
24  &  438242218832195088801894111132005025739819100831104898458347287148981323073167259406443353675374485580614409526121217079214285336977046690201810752933696772629332 \\ 
25  &  29548150764354051108986653372266838516881491330935777859166128849296753863671730114976395779815333178178360613565695537261400270819437732521145740229031786418466822328155210774 \\ 
26  &  7376409612724881246275082273655527171437045694901336339786650436361148933382241819470915896534342159863922343241461009706359183544468912989538551447112728993641722959010706705865911047130282 \\ 
\hline \hline
\end{tabular}
}
\caption{Number of SAWs crossing a rhomboidal domain of the hexagonal lattice.} 
\end{table}

\begin{table}
{\tiny
\begin{tabular}{ll}
\hline \hline
$L$ &  $C_L(1)$ \\ \hline
1  &  2 \\ 
2  &  50 \\ 
3  &  2256 \\ 
4  &  292006 \\ 
5  &  124394172 \\ 
6  &  182189852062 \\ 
7  &  937116505296162 \\ 
8  &  17167376550995687961 \\ 
9  &  1130911800993488803731078 \\ 
10  &  269650395624478266477331223678 \\ 
11  &  233772496350603982679550385266064014 \\ 
12  &  739330863241806743025423160490836132227125 \\ 
13  &  8551000409049037000098287028025432585191736309022 \\ 
14  &  362378501157171575915086740862352731989136965188978227480 \\ 
15  &  56355164888885592354051749345529297798069126440063716209024866536 \\ 
16  &  32200232301152973892060847293393239105831802930525492217459523426803019578 \\ 
17  &  67665662468515970834966508500944029204762050650693102413477819738278462353187499568 \\ 
18  &  523379303813002076273464810690096008845689319359263297454993915567005968614237413818526075604 \\ 
19  &  14910759530495548949623554019916848509888902630562528597658761833271654794704911307455162596911430188758 \\ 
20  &  1565552766529028680644951163416182891619237381422347104413735417263587336683056846547108571311383043144417391243521 \\ 
21  &  606084065190103550545340197138093542241444175895240820944713102199472484407589836051473378294437969118344993099132112988392420 \\ 
22  &  865520866516174852434302085316123704413013184383267585803777199791333436065070398518809805939514793360791184628285127191002600593996365368 \\ 
23  &  4560992075553129850922927762995312993575376533697147813417446333497150777368818523814441027249715528879466279022480571252085818753070411144055018204876 \\ 
24  &  88718729299059562850997307819335993122314801341423818394843508516540284335394844383553598429174621313143090882997603769027577322213964104279601294505719147199388248 \\ 
25  &  6371850587510704465849294714166605694358498327959158268953676175064720102088107446652503452442438514332661839511396033679344990952004170416492446601538481306356885482155377886638 \\ 
26  &  1690113361272638089564412600147895085451226125158697875455343806307988763259201392451007110767935854996504549679554397321246918091898802222542458839160812960021404785245887985513225511176452673 \\ 
\hline \hline
\end{tabular}
}
\caption{Number of SAWs spanning a rhomboidal domain of the hexagonal lattice.} 
\end{table}

\end{landscape}

\begin{landscape}
\thispagestyle{empty}
 \begin{table}
{\tiny
\begin{tabular}{ll}
\hline \hline
$L$ &  $P_L(1)$ \\ \hline
1  &  1 \\ 
2  &  3 \\ 
3  &  48 \\ 
4  &  3126 \\ 
5  &  775842 \\ 
6  &  727870836 \\ 
7  &  2575728525240 \\ 
8  &  34244061451559094 \\ 
9  &  1703999058661009145746 \\ 
10  &  316543880488539946466963896 \\ 
11  &  219157996022284922702859434801868 \\ 
12  &  564858713948847373563461482383973674774 \\ 
13  &  5415142061627863782256892670635702203299498106 \\ 
14  &  192965908859455255222444585453472066280402031983076676 \\ 
15  &  25546198443752201604792021828520875111113011948793636471115986 \\ 
16  &  12559327077982128401344048554297110314066721517873014754182697036556596 \\ 
17  &  22922091883660814526614648049302957461020819783000936029058072227670397210344452 \\ 
18  &  155263447483572551766390960410624837560693086531157173552843098568401053874449090345960952 \\ 
19  &  3902210830303866544089288909268585128297599458763876386711662682968145520021314838767588755204876552 \\ 
20  &  363826279944404033043481454750498594828384686156661907448081883116767989915064170563158680799058587038093669970 \\ 
21  &  125819288614038800635491456373348096978155582443835316985068491781106711591018731348747689905774023677560616130526519792788 \\ 
22  &  161364676308721043071062335667640391981582844992152210690103275445701605800992646736829901291623422443479273978505601685944958319502908 \\ 
23  &  767404723000807383687986740560681434496809093061630495379302826095788965371427398181191526510671680461233928816145587981072980847709460390319231092 \\ 
24  &  13531589671078110974893162463064998617986779218655081753249240837076015118491730281579616308320287573428293648926750890064755844067619542837352365219373425001182 \\ 
25  &  884587740071585897410736339557658407744835896491278031519338391577738398470263456602417012949952825706708843546792281536393617902202119150298209236054002295717190761721586136 \\ 
26  &  214370491395195888234645652662533040178049645573231037100079862232102374051311163431171231717044978758387398235068063926962934021314526635561030249606476912450218466192900542978818780384974 \\ 
\hline \hline
\end{tabular}
}
\caption{Number of SAPs crossing a rhomboidal domain of the hexagonal lattice.} 
\end{table}
\end{landscape}

\begin{landscape}
\thispagestyle{empty}
 \begin{table}
{\small
\begin{tabular}{ll}
\hline \hline
$L$ &  $C_L(1)$ \\ \hline
1  &  2 \\ 
2  &  7 \\ 
3  &  44 \\ 
4  &  515 \\ 
5  &  11500 \\ 
6  &  493704 \\ 
7  &  40751496 \\ 
8  &  6463642330 \\ 
9  &  1970190022696 \\ 
10  &  1154437344815284 \\ 
11  &  1300686960810345198 \\ 
12  &  2818300749120970598426 \\ 
13  &  11745284697899678209887246 \\ 
14  &  94153940687296424300453605522 \\ 
15  &  1451915619132744566900848537333082 \\ 
16  &  43072062058620235613855525243039798546 \\ 
17  &  2458218787430131938141065342199631011888808 \\ 
18  &  269917990612156037679955033913220231218482526540 \\ 
19  &  57022048161016261704452967864058833682099233234074924 \\ 
20  &  23177397882827812987656054354088621630193659021408496092114 \\ 
21  &  18126208865601871898868235390674787298375068592505362074324218782 \\ 
22  &  27275828087021466037231281803108531532614036012259410718518383677989994 \\ 
23  &  78974101601865877096497572762267816542675600879070694217812459537275320667130 \\ 
24  &  439980515324228439963646464930268543060978419686632840124513851873692354257184355418 \\ 
25  &  4716606546189621488078969490297265985170243927748285792380749595975920915553704131199964610 \\ 
26  &  97292222614020401528875654356525325735532995996523907301076613477132417329484579095044048258220716 \\ 
27  &  3861740982967126791934974463996504445993431827647538470677158069324943832308988274731817887045190314942500 \\  
\hline \hline
\end{tabular}
}
\caption{Number of SAWs crossing a triangular domain of the hexagonal lattice.} 
\end{table}
\end{landscape}

\begin{landscape}
\thispagestyle{empty}

\begin{table}
{\small
\begin{tabular}{ll}
\hline \hline
$L$ &  $C_L(1)$ \\ \hline
1  &  1 \\ 
2  &  3 \\ 
3  &  18 \\ 
4  &  210 \\ 
5  &  4716 \\ 
6  &  203130 \\ 
7  &  16781528 \\ 
8  &  2661898722 \\ 
9  &  811337884328 \\ 
10  &  475395297020430 \\ 
11  &  535618774376758222 \\ 
12  &  1160567857061063474508 \\ 
13  &  4836675324919658534327348 \\ 
14  &  38772333263059858336182467950 \\ 
15  &  597894854584620490267288203881970 \\ 
16  &  17736956492510173648327596231133813426 \\ 
17  &  1012287723222402775005385313973408357507928 \\ 
18  &  111151484863070215708849728284201214059413569272 \\ 
19  &  23481522343431693736560242087640111797935241906792060 \\ 
20  &  9544388601505664173784379076794209212239937007395941459026 \\ 
21  &  7464322880925069857683897811600948880215514557439627560911154272 \\ 
22  &  11232110875321164747567467659828479928446150234247426811308149074039470 \\ 
23  &  32521317511278850216940549112361104580618379635763819229016915699625133297104 \\ 
24  &  181182764336015552734273130240200423605997687829676784582391379637383247087868602758 \\ 
25  &  1942285584539983234933331010286728144642773519634047277599154248174196516886152824735901816 \\ 
26  &  40064669298138196682088095071796367265068180648770697785528635200087726423296089992305061500566756 \\ 
\hline \hline
\end{tabular}
}
\caption{Number of SAWs crossing a triangular domain of the hexagonal lattice and including the top vertex.} 
\end{table}

\end{landscape}

\begin{landscape}
\thispagestyle{empty}
 \begin{table}
{\small
\begin{tabular}{ll}
\hline \hline
$L$ &  $P_L(1)$ \\ \hline
 1  &  1 \\ 
2  &  2 \\ 
3  &  9 \\ 
4  &  85 \\ 
5  &  1605 \\ 
6  &  59896 \\ 
7  &  4392639 \\ 
8  &  629739138 \\ 
9  &  175745776816 \\ 
10  &  95207239875508 \\ 
11  &  99934927799315359 \\ 
12  &  202993550188918062298 \\ 
13  &  797200289814680588454420 \\ 
14  &  6048794511036987586252009778 \\ 
15  &  88623124229469033988344357343229 \\ 
16  &  2506168305598107863294101582119745559 \\ 
17  &  136742066892485673488096591777101574684341 \\ 
18  &  14391095306419863125025082539141317797920679808 \\ 
19  &  2920637571762330449794165953013715565926946586966972 \\ 
20  &  1142780121652579092442989213824129363529214905674607409456 \\ 
21  &  861928813419640412952428304528142087056944927600343349249100770 \\ 
22  &  1252960133060510490994725871202276919994651077934833437111933731780232 \\ 
23  &  3509963453723621942826513300378279853247659026894598196945505524358307547596 \\ 
24  &  18945984524072416973165104755335799616808372006565339168062614482119446796495592941 \\ 
25  &  197032077332349626704638536077733550874900563736415557346148448949082140805149991012506724 \\ 
26  &  3947507851539205775146388396017001015202508590957965919271768932077125446293950595857281240459716 \\ 
\hline \hline
\end{tabular}
}
\caption{Number of SAPs crossing a triangular domain of the hexagonal lattice.} 
\end{table}
\end{landscape}

\begin{landscape}
\thispagestyle{empty}

\begin{table}
{\small
\begin{tabular}{ll}
\hline \hline
$L$ &  $P_L(1)$ \\ \hline
1  &  1 \\ 
2  &  1 \\ 
3  &  4 \\ 
4  &  36 \\ 
5  &  666 \\ 
6  &  24696 \\ 
7  &  1808820 \\ 
8  &  259300148 \\ 
9  &  72369408510 \\ 
10  &  39205936157880 \\ 
11  &  41152969216872016 \\ 
12  &  83592236529606631688 \\ 
13  &  328284931491454739745904 \\ 
14  &  2490876950205850778116435156 \\ 
15  &  36494758452603010620499864088198 \\ 
16  &  1032033208911845667821292289616451218 \\ 
17  &  56310006747344597198073248186075772148180 \\ 
18  &  5926213428826485611611313527823854932071080074 \\ 
19  &  1202710510511720770819662867223620040669484274841448 \\ 
20  &  470593707331440145848250079430318880733169905225241510182 \\ 
21  &  354939911811827613400027738254513445185773676790950877558157556 \\ 
22  &  515965532286678291640886325718842923532551840839177342378988626653078 \\ 
23  &  1445393283922054883637378235832608861381031003585207142018132021675532043232 \\ 
24  &  7801904249270681046277482881424254681239226301915609070185058428520166740304455480 \\ 
25  &  81137266805100512823257637730776600977600011085064069900554442194897045916216667639237206 \\ 
26  &  1625572861413431635691529107338978659074358348381654539274841326821635464880471185029440059346822 \\ 
\hline \hline
\end{tabular}
}
\caption{Number of SAPs crossing a triangular domain of the hexagonal lattice and including top vertex.} 
\end{table}

\end{landscape}

\begin{landscape}
\thispagestyle{empty}

\begin{table}
{\tiny
\begin{tabular}{ll}
\hline \hline
$L$ &  $C_L(1)$ \\ \hline
1 &  2 \\
2 &  14 \\
3 &  264 \\
4 & 21512 \\
5 &  5663596 \\
6 &  6478476233 \\ 
7 &  23432328776346 \\
8 &  365121393771314359 \\
9  & 18039965927005597824652 \\
10 & 3847346539490622663060402802 \\
11 & 2604549807872636495439504536518768 \\
12 & 7613280873970130888072912524910312775000 \\
13 & 70659728324509466176595292882340210105184200002 \\
14 & 2831956810062815172946024396329723966506233510418891138 \\
15 & 360424703055912928274223706157781269084968015495478379832577374 \\
16 & 198097258016637755765939369950089310341388296845374445597477414443215248 \\
17 & 345765524783138086318892247783650000160221384394056330912454668222835230637412672 \\
18 & 2606338884649187506543399082354962241036644807771353337217794306196421868029243067294778048 \\
19 & 62392663751835087636515340004811611674555874089327041316405089409127243514061643853154930821350090724 \\
20 & 6450407172867437933486941949195444800686042090585344770339862190617805415359568631117548255651795966774975729408 \\
21 & 2117885679287759638663389972562580414723520464095542413611859724509315283169986698668888918984500033887174419453417031840108 \\
22 & 3003111631506205594200550519402977342109069309804916734448892038737633365710727044910318753595074453487711557977426845717574589858142229 \\
23 & 13524071180124614895872809797043935746289243109268223573969721018213938124696255997302291968299972324147755286357424784253377182127831016598030435648 \\
24 &  263025838000002506267728179467786825301378641433800017689615980657976366151408147401274046080321228727574309853794611789046332691442399607521892487626528327350411 \\
\hline \hline
\end{tabular}
}
\caption{Number of SAWs crossing a square domain of the hexagonal lattice.} 
\end{table}

\end{landscape}

\section{Supplementary numerical analysis}
\label{app:numana}

\begin{figure}[ht!] 
\begin{center}
\includegraphics[width=0.725\textwidth,angle=0]{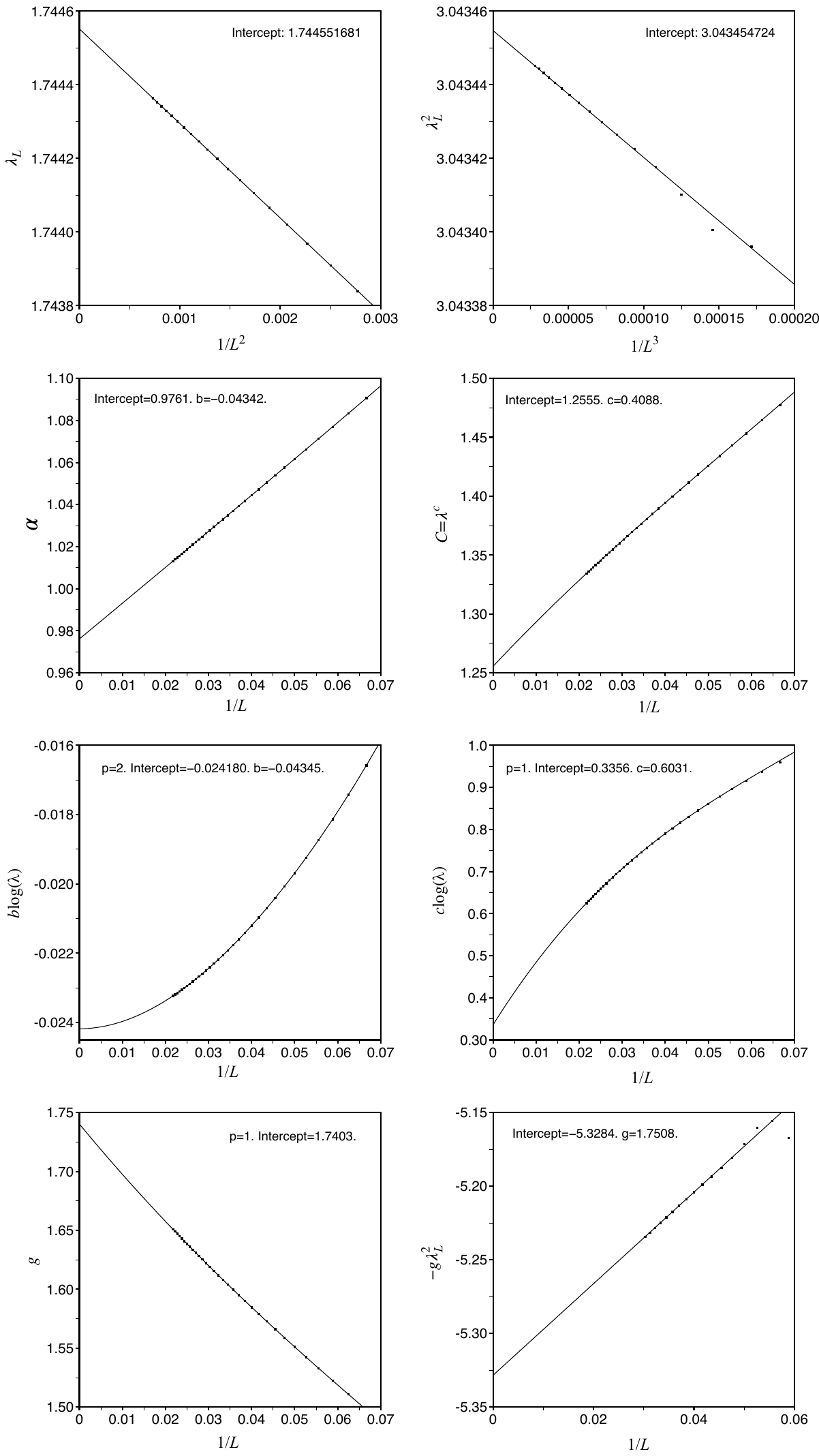} 
 \caption{\label{fig:wcas-span}  
 Plots of the various estimators used in the analysis of the data for SAWs spanning a square.
 }
\end{center}
\end{figure}

\begin{figure}[ht!]
\begin{center}
\includegraphics[width=0.78\textwidth,angle=0]{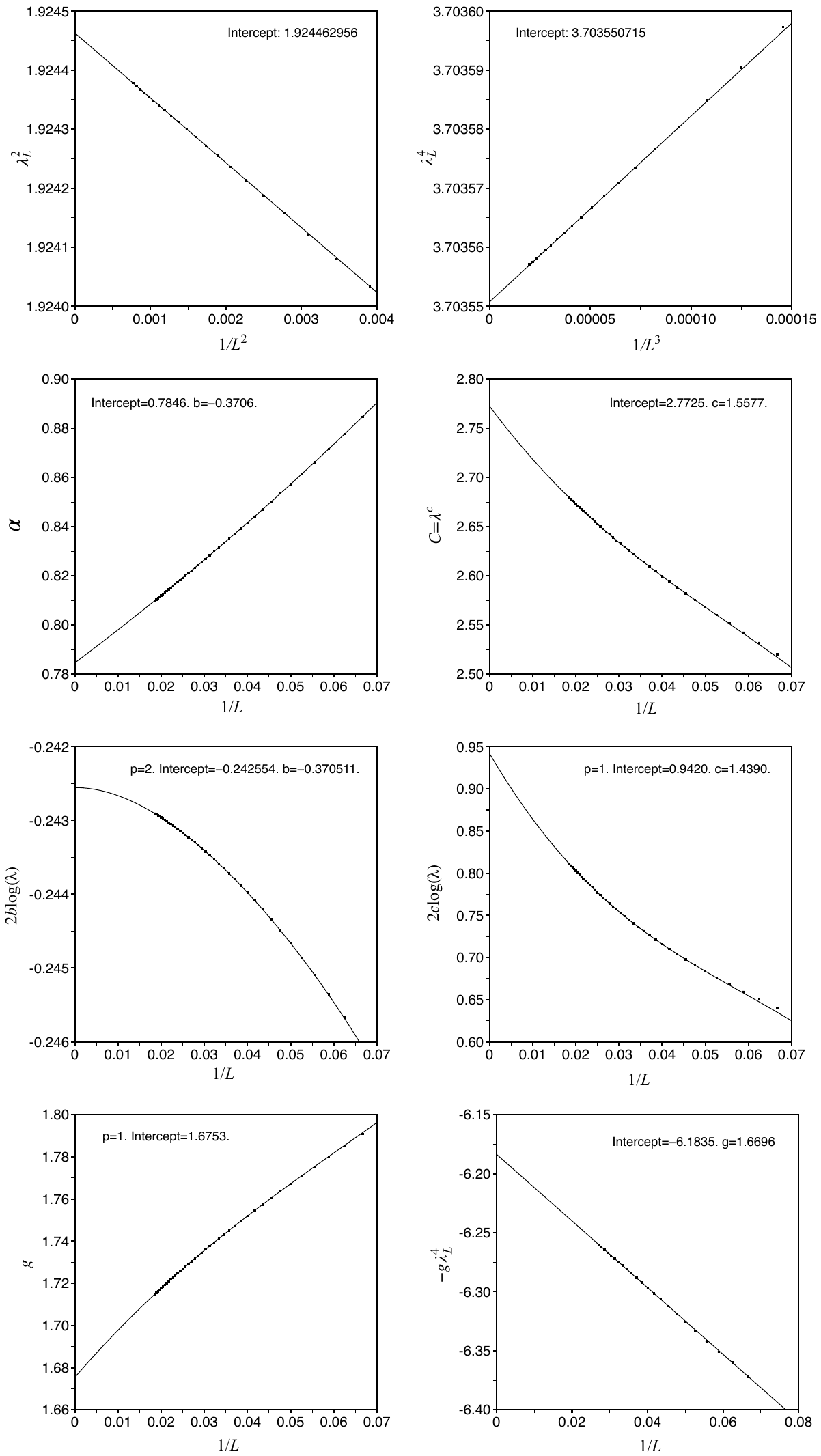} 
 \caption{\label{fig:wcah-rhom-span}  
 Plots of the various estimators used in the analysis of the data for SAWs spanning a rhomboidal domain of the hexagonal lattice.
 }
\end{center}
\end{figure}

\begin{figure}[ht!] 
\begin{center}
\includegraphics[width=0.78\textwidth,angle=0]{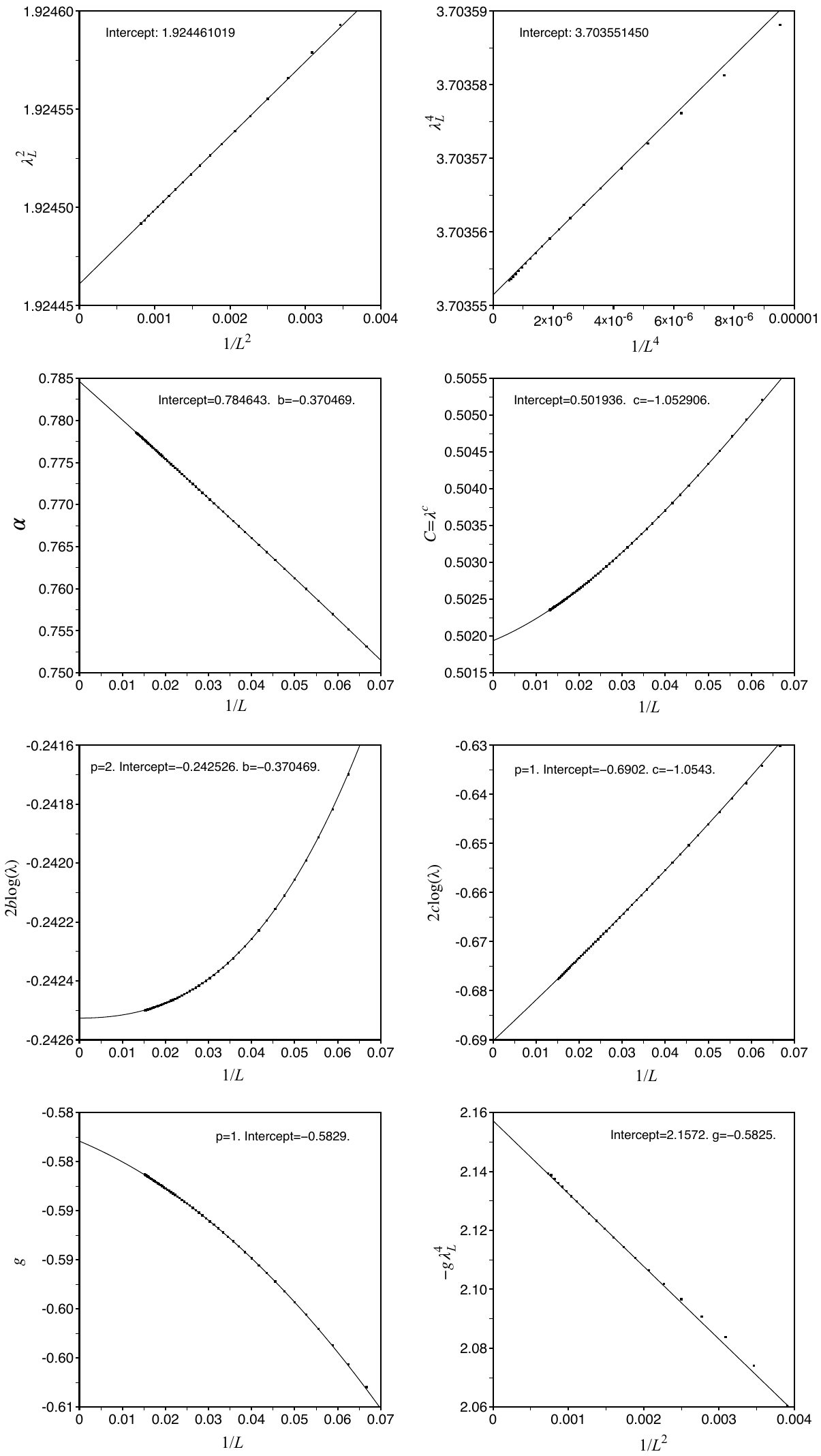} 
 \caption{\label{fig:wcah-rhom-sap}  
 Plots of the various estimators used in the analysis of the data for SAPs crossing a rhomboidal domain of the hexagonal lattice.
 }
 \end{center}
\end{figure}

\begin{figure}[ht!] 
\begin{center}
\includegraphics[width=0.78\textwidth,angle=0]{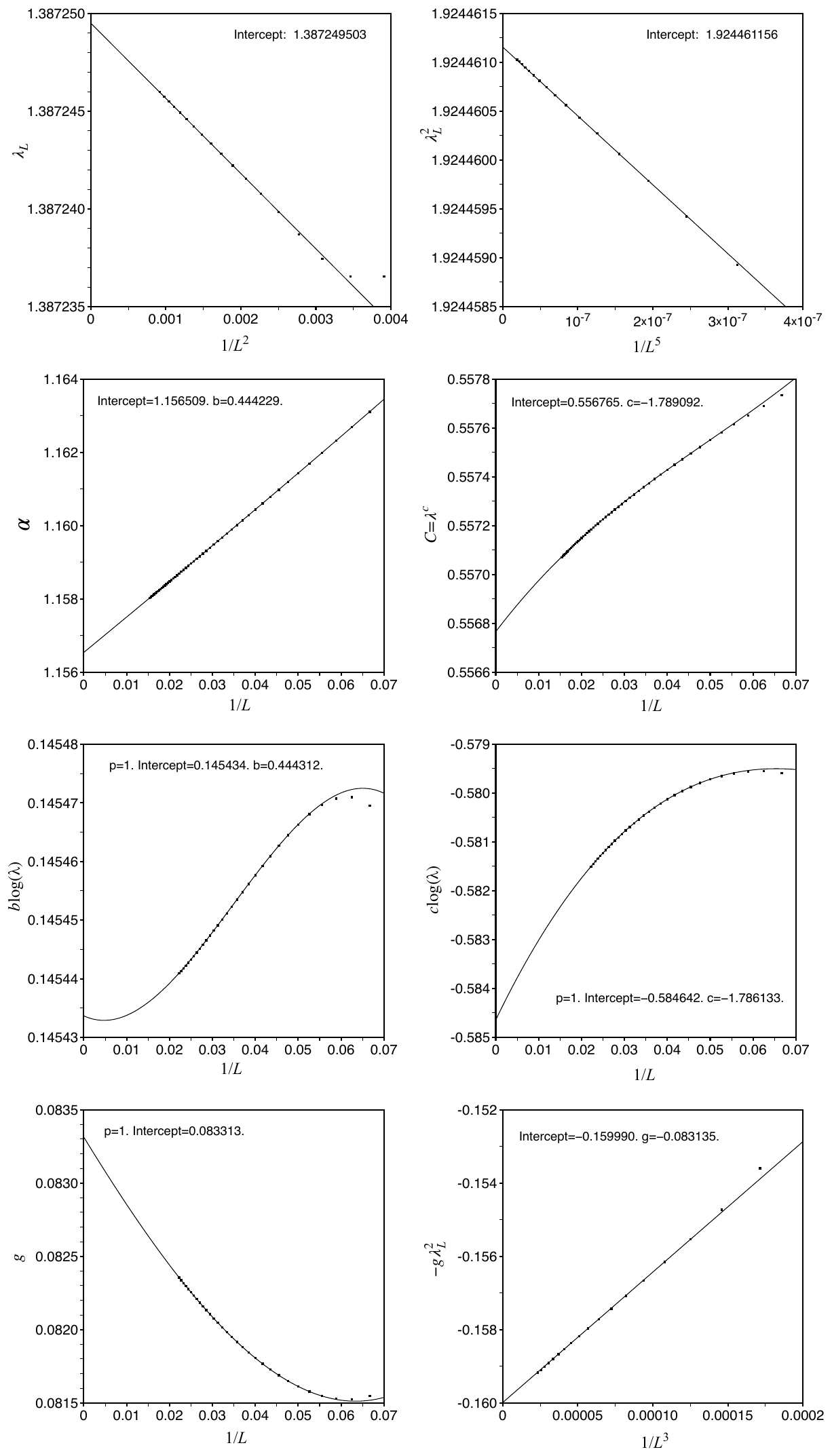} 
 \caption{\label{fig:wcah-tri-top}  
 Plots of the various estimators used in the analysis of the data for SAWs crossing a triangular domain of the hexagonal lattice while
 passing through the topmost vertex.
 }
 \end{center}
\end{figure}

\begin{figure}[ht!] 
\begin{center}
\includegraphics[width=0.78\textwidth,angle=0]{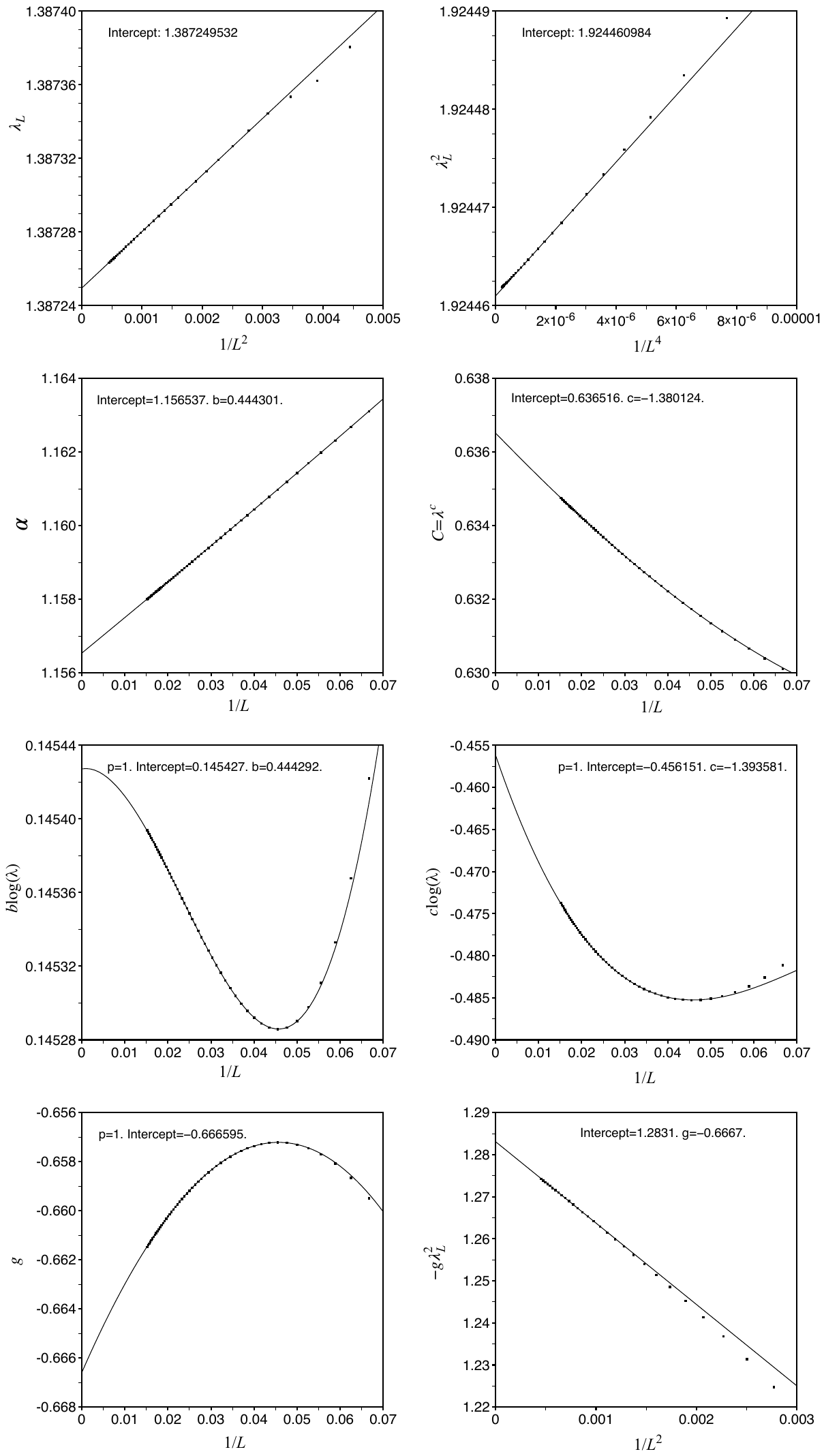} 
 \caption{\label{fig:wcah-tri-sap}  
 Plots of the various estimators used in the analysis of the data for SAPs crossing a triangular domain of the hexagonal lattice.
 }
 \end{center}
\end{figure}

\begin{figure}[ht!] 
\begin{center}
\includegraphics[width=0.78\textwidth,angle=0]{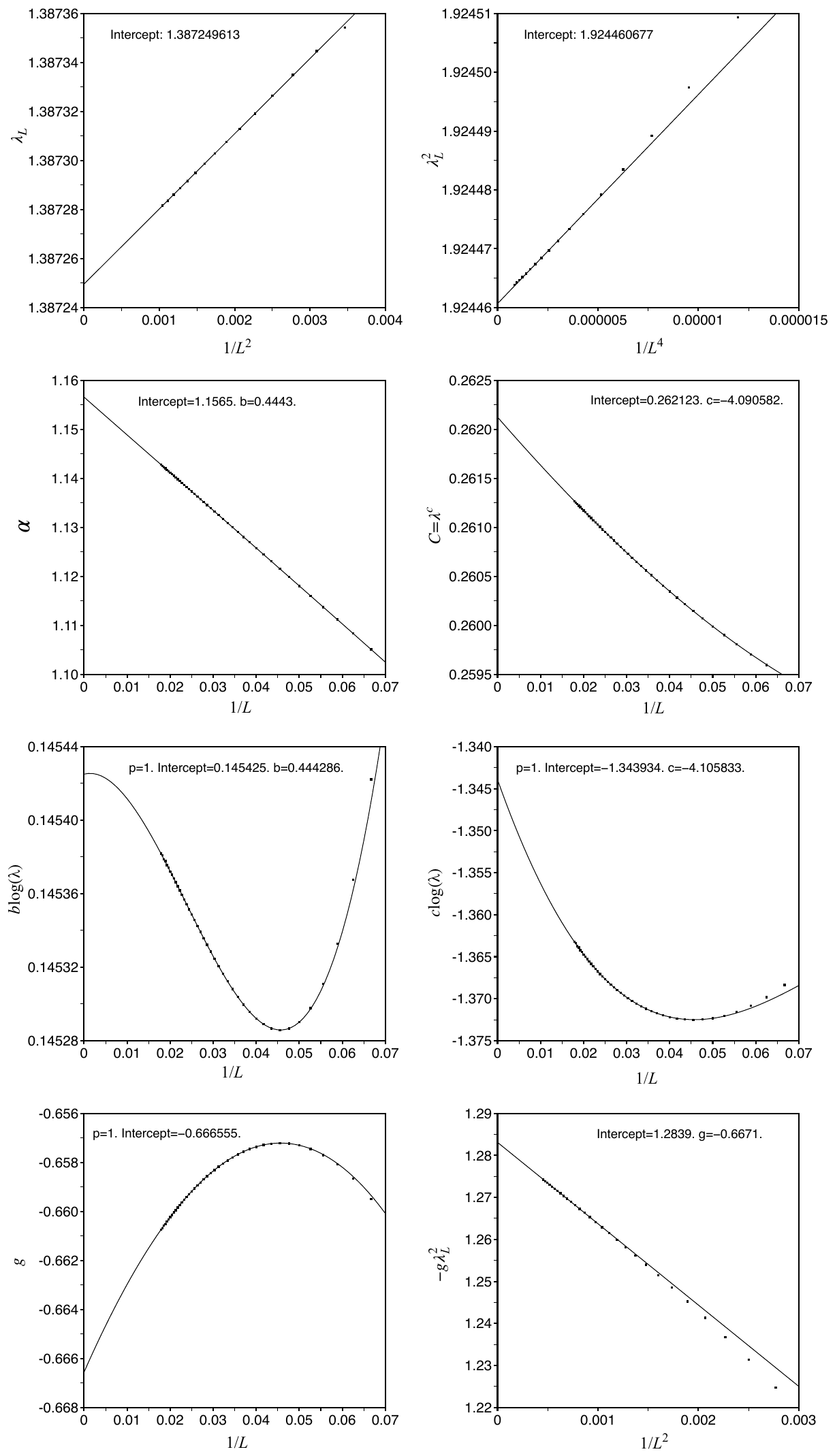} 
 \caption{\label{fig:wcah-tri-sap-top}  
 Plots of the various estimators used in the analysis of the data for SAPs crossing a triangular domain of the hexagonal lattice while
 passing through the topmost vertex.
 }
 \end{center}
\end{figure}

\begin{figure}[ht!] 
\begin{center}
\includegraphics[width=0.78\textwidth,angle=0]{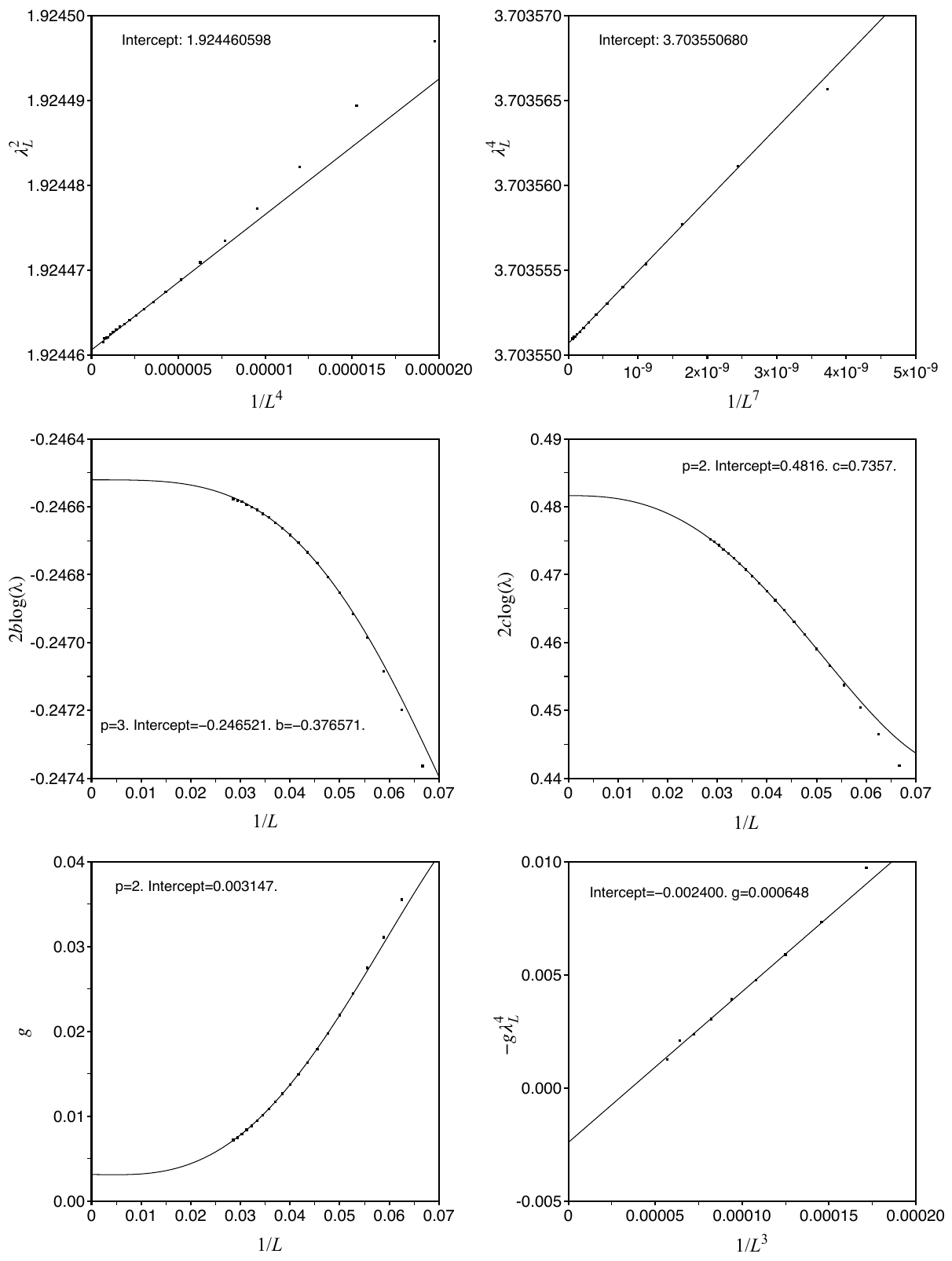} 
 \caption{\label{fig:wcah-square}  
 Plots of the various estimators used in the analysis of the data for SAWs crossing a square domain of the hexagonal lattice.
 }
 \end{center}
\end{figure}
\end{document}